\documentstyle[12pt,epsf]{article}

\catcode`\@=11
\long\def\@makefntext#1{
\protect\noindent \hbox to 3.2pt {\hskip-.9pt
$^{{\eightrm\@thefnmark}}$\hfil}#1\hfill}

\renewenvironment{thebibliography}[1]
	{\frenchspacing
	 \begin{list}{\arabic{enumi}.}
	{\usecounter{enumi}\setlength{\parsep}{0pt}
	 \setlength{\leftmargin 12.7pt}{\rightmargin 0pt}
	 \setlength{\itemsep}{0pt} \settowidth
	{\labelwidth}{#1.}\sloppy}}{\end{list}}

\def\@citex[#1]#2{\if@filesw\immediate\write\@auxout
	{\string\citation{#2}}\fi
\def\@citea{}\@cite{\@for\@citeb:=#2\do
	{\@citea\def\@citea{,}\@ifundefined
	{b@\@citeb}{{\bf ?}\@warning
	{Citation `\@citeb' on page \thepage \space undefined}}
	{\csname b@\@citeb\endcsname}}}{#1}}
\newif\if@cghi
\def\cite{\@cghitrue\@ifnextchar [{\@tempswatrue
	\@citex}{\@tempswafalse\@citex[]}}
\def\citelow{\@cghifalse\@ifnextchar [{\@tempswatrue
	\@citex}{\@tempswafalse\@citex[]}}
\def\@cite#1#2{{$\null^{#1}$\if@tempswa\typeout
	{IJCGA warning: optional citation argument
	ignored: `#2'} \fi}}

\renewcommand{\theequation}{\arabic{section}.\arabic{equation}}

\font\eightrm=cmr8

\def\today{\number\day\space
     \ifcase\month\or
       January\or February\or March\or April\or May\or June\or
       July\or August\or September\or October\or November\or December\fi
     \space\number\year}

\newcommand{\beq}{\begin{equation}}
\newcommand{\eeq}{\end{equation}}
\newcommand{\ba}{\begin{array}}
\newcommand{\ea}{\end{array}}
\newcommand{\beqa}{\begin{eqnarray}}
\newcommand{\eeqa}{\end{eqnarray}}
\newcommand{\dis}{\displaystyle}

\newcommand{\da}{^\dagger}
\newcommand{\no}{\nonumber}

\newcommand{\lsim}{\stackrel{<}{_\sim}}
\newcommand{\gsim}{\stackrel{>}{_\sim}}
\newcommand{\Gto}{\stackrel{G}{\longrightarrow}}     
   
\newcommand{\svect}[1]{\stackrel{\rightarrow}{{#1}}}   
\newcommand{\rarrow}{\longrightarrow}

\newcommand{\DD}{\bigtriangledown}
\newcommand{\Imm}{\Im m}
\newcommand{\Real}{\Re e}
\newcommand{\Ko}{{K}^0}
\newcommand{\Kob}{\bar{K}^0} 
\newcommand{\la}{\langle}
\newcommand{\ra}{\rangle}
\newcommand{\ket}[1]{\vert {#1} \rangle}
\newcommand{\bra}[1]{\langle {#1}}
\newcommand{\veps}{{\widetilde \epsilon}}
\newcommand{\wt}{\widetilde}
\newcommand{\eps}{\epsilon}
\newcommand{\epsp}{\epsilon'}

\newcommand{\sors}{\hat s}
\newcommand{\sorp}{\hat p}

\newcommand{\PL}[3]{{\it Phys. Lett.}       {\bf #1} {(19#2)} {#3}}
\newcommand{\Pre}[3]{{\it Phys. Rep.}       {\bf #1} {(19#2)} {#3}}
\newcommand{\PRL}[3]{{\it Phys. Rev. Lett.} {\bf #1} {(19#2)} {#3}}
\newcommand{\PR}[3]{{\it Phys. Rev.}        {\bf #1} {(19#2)} {#3}}
\newcommand{\NP}[3]{{\it Nucl. Phys.}       {\bf #1} {(19#2)} {#3}}
\newcommand{\ZP}[3]{{\it Z.  Phys.}         {\bf #1} {(19#2)} {#3}}
\newcommand{\AP}[3]{{\it Ann.  Phys.}       {\bf #1} {(19#2)} {#3}}
\newcommand{\ARNS}[3]{{\it Ann. Rev. Nucl. Part. Sci.}  {\bf #1} {(19#2)} {#3}}

\def\SSp{\S~sect.~}
\def\SSf{\S~fig.~}
\def\SSt{\S~tab.~}

\begin{document}    

\begin{titlepage}

\begin{flushright}
INFNNA-IV-96-29 \\
LNF-96/036(P) 
\end{flushright}

\begin{center}
\vglue 2. true cm
{\Large  CP VIOLATION IN KAON DECAYS$^*$ }
\vglue 2. true cm
{\bf Giancarlo D'Ambrosio$^1$ and Gino Isidori$^2$ }
\vglue 1. true cm
${}^{1)}$ INFN, Sezione di Napoli \\
Dipartimento di Scienze Fisiche, Universit\`a di Napoli\\
I--80125 Napoli, Italy \\[5pt]
${}^{2)}$ INFN, Laboratori Nazionali di Frascati \\ 
P.O. Box 13, I--00044 Frascati, Italy
\end{center}

\vglue 2. true cm

\begin{abstract}\noindent
We review the Standard Model 
predictions of $CP$ violation 
in kaon decays. We present
an elementary introduction to
Chiral Perturbation Theory, four--quark effective 
hamiltonians and the relation among them.
Particular attention is devoted to 
$K\to 3\pi$, $K\to 2\pi \gamma$ 
and $K\to \pi \bar{f} f$ decays.   
\end{abstract}

\vglue 2.0 true cm
\begin{center}
To appear in \\
International Journal of Modern Physics A \\
\end{center}

\vfill
\noindent {\small $^*$ Work supported in part
by HCM, EEC--Contract No. CHRX--CT920026 
 (EURODA$\Phi$NE) }

\end{titlepage}

\tableofcontents

\setcounter{equation}{0}
\setcounter{footnote}{0}
\section{Introduction.}

Since 1949, when  $K$ mesons were discovered\cite{Brown}, kaon physics has 
represented one of the richest sources of information
in the study of fundamental interactions. 

One of the first ideas, originated by the study of 
$K$ meson production and decays, was the  
Gell--Mann\cite{Gellmann53} and Pais\cite{Pais} hypothesis 
of the `strangeness' as a new quantum number. 
Almost at the same time, the famous `$\theta-\tau$ puzzle'\cite{Dalitz} 
was determinant in suggesting to  Lee and Yang\cite{LeeYang} 
the revolutionary hypothesis of parity violation in weak interactions.
Lately, in the sixties, $K$ mesons played an important role 
in clarifying global symmetries of strong interactions,
well before than QCD was proposed\cite{Gellmann}$^-$\cite{CallanT}. 
In the mean time they had a relevant role also in the formulation 
of the Cabibbo theory\cite{Cabibbo}, which unified weak interactions
of strange and non--strange particles. Finally, around  1970, the 
suppression of  flavor changing neutral currents  in kaon decays 
was one of the main reason which pushed Glashow, Iliopoulos and 
Maiani\cite{GIM} to postulate the
existence of the `charm'. Hypothesis which was lately confirmed
opening  the way to the unification of quark and lepton
electro--weak interactions.

In 1964 a completely unexpected revolution was determined by the 
Christenson, Cronin, Fitch and Turlay
observation of  $K_L \to 2\pi$ decay\cite{Cronin}, 
i.e. by the discovery of a
very weak interaction non invariant under $CP$.
Even if more than thirty years have passed by
this famous experiment, the phenomenon of $CP$ violation is still  
not completely clear and is one of the aspects which makes 
still very interesting 
the study of $K$ mesons, both from the experimental and
the theoretical point of view.

To date, in the framework of nuclear and subnuclear physics,
there are no evidences of $CP$ violation but in $K_L$ decays,
and within these processes all the observables indicate clearly
only a $CP$ violation in the mixing $\Ko-\Kob$. Nevertheless,
as shown by Sakharov\cite{Sakharov} in 1967,
also the asymmetry between matter and antimatter in the universe 
can be considered as an indication of $CP$ violation. Thus the 
study of this phenomenon has fundamental implications not 
only in particle physics but also in cosmology\cite{CKN}.
                             
As it is well known, strong and electro--weak interactions seem
to be well described within the so--called `Standard Model', i.e.
in the framework of a non--abelian gauge theory based on the 
$SU(3)_C\times SU(2)_L\times U(1)_Y$ symmetry group 
(\SSp\ref{cap:SM}). 
$CP$ violation can be naturally generated 
in this model, both in the strong and in the electro--weak sector.

$CP$ violation in the strong sector,
though allowed from a theoretical point of view\cite{CDG,JR}, 
from the experimental analysis of the 
neutron dipole moment turns out to be very suppressed.
This suppression, which does not find a natural explanation 
in the Standard Model, is usually referred
as the `strong $CP$ problem'\cite{PecceiCPstrong}.
One of the most appealing hypothesis to solve this problem is 
to extend the model including a new symmetry, which 
forbids (or drastically reduces) $CP$ violation in the 
strong sector. However, all the proposals
formulated in this direction 
have not found any experimental evidence yet\cite{PecceiCPstrong}.

 $CP$ violation in the electro--weak sector  is generated by the 
Kobayashi--Maskawa  mechanism\cite{Kobayashi}. 
This mechanism explains qualitatively the 
 $CP$ violation till now observed in the 
$\Ko-\Kob$ system but predicts also new phenomena
not observed yet: $CP$ violation in 
$|\Delta S|=1$ transitions (measured with precision 
only in  $K\to 2\pi$ decays, where turns out to be 
compatible with zero within two standard deviations\cite{PDG}) and
in $B$ decays. 

In the next years a remarkable experimental effort will be undertaken,
both in $K$ and in $B$  meson physics, in order to seriously test the 
Standard Model mechanism of $CP$ violation. Assuming there is no 
$CP$ violation in  the strong sector, in the framework of 
this model all the observables which violate $CP$ depend essentially
on one parameter (the phase of the Cabibbo--Kobayashi--Maskawa matrix). As
a consequence, any new experimental evidence of $CP$ violation 
could lead to interesting conclusions (even in minimal extensions of the 
model new phases are introduced and the relations among the observables
are modified, see e.g. Refs.\cite{Masiero,Hewett}). 
Obviously, to test the model seriously, is necessary 
to analyze with great care, from the theoretical point of view, all 
the predictions and the relative uncertainties for all
the observables which will be measured.  

In the framework of $K$ mesons
is not easy to estimate $CP$ violating observables 
with great precision, since  strong interactions are in a 
non--perturbative regime. In the most interesting case, i.e. 
in  $K\to 2\pi$ decays, this problem has been partially 
solved by combining  analytic calculations of the four--quark effective 
hamiltonian\cite{Buras2,Ciuchini2} with non--perturbative 
information on the matrix elements\cite{burasepe,Ciuchini3}.
The latter have been obtained from lattice QCD results\cite{Ciuchini3} 
or combining experimental information on $K\to 2\pi$ amplitudes 
and $1/N_c$ predictions\cite{burasepe}. Nevertheless
in other channels, like  $K\to 3\pi$ and $K\to 2\pi\gamma$ decays, 
the theoretical situation is less clear and 
there are several controversial statements in the literature.

The purpose of this review  is to analyze in detail 
all the predictions for the observables which will be measured 
in the next years, trying, where possible, to relate them 
with those of $K\to 2\pi$. The predictions
will be analyzed assuming the Cabibbo--Kobayashi--Maskawa  matrix as 
the unique source of $CP$ violation in the model (i.e. we shall
assume that exists a symmetry which forbids $CP$ violation 
in the strong sector). The tool that we shall use to relate
between each other the different observables is the so--called 
Chiral Perturbation Theory\cite{Weinberg79}$^-$\cite{GL2}
(\SSp\ref{cap:CHPT}). 
This Theory, based on the hypothesis that the eight 
lightest pseudoscalar bosons ($\pi$, $K$ and $\eta$) 
are Goldstone bosons\cite{Goldstone},
in the limit of vanishing light quark masses
($m_u=m_d=m_s=0$),
can be considered as the natural complement of  
lattice calculations. From one side, indeed, relates 
matrix elements of different processes, on the other side 
allows to calculate in a systematic way the absorptive 
parts of the amplitudes (typically not accessible from 
lattice simulations).

The paper is organized as follows: in the first section 
we shall discuss $CP$ violation in kaon decays 
in a very phenomenological way, 
outlining the general features of the problem;
in the second section the mechanism of $CP$ violation in the 
Standard Model will be analyzed, both in $K$ and in $B$ mesons,
with particular attention to the estimates of 
$K\to 2\pi$ parameters $\epsilon$ and $\epsilon'$; 
in the third section we shall introduce
Chiral Perturbation Theory. 
These first three sections represent three independent
and complementary introductions to the problem. In the following 
four sections we shall discuss the estimates of 
several $CP$ violating observables in kaon decays different 
than $K\to 2\pi$. The results will be summarized 
in the conclusions.

\setcounter{equation}{0}
\setcounter{footnote}{0}
\section{Phenomenology of $CP$ violation in kaon decays.}
\label{cap:kksystem}          

\subsection{Time evolution of the $\Ko-\Kob$ system.}
\label{sez:kksystem}
            
The state $\ket{\Psi}$ which describes a neutral kaon is in general 
a superposition of $\ket{\Ko}$ and $\ket{\Kob}$ states, with
definite strangeness, eigenstates of strong and electromagnetic 
interactions\footnote{~For excellent reviews about the 
arguments presented in this section and, more in general,
about $CP$ violation in kaon decays see 
Refs.\protect\cite{LeeW}$^-$\cite{handbook} }.

Introducing the vector $\Psi=\left[\ba{c} \Psi_1\\ \Psi_2 \ea \right]$, 
so that $\ket{\Psi}= \Psi_1\ket{\Ko}+\Psi_2\ket{\Kob}$,
the time evolution of  $\ket{\Psi}$, in the particle rest frame, 
is given by:
\beq
i {\partial \over \partial  t} \Psi(t) = H\Psi(t) = (M-{i\over 2}\Gamma)\Psi(t),
\eeq
where $M$ and  $\Gamma$ are $2\times 2$ hermitian 
matrices with positive eigenvalues. 
Denoting by $\lambda_\pm$ and $\Psi_\pm$ the eigenvalues and 
eigenvectors of $H$, respectively, we have:
\beq
\ket{\Psi(t)}=c_+ e^{-i\lambda_+ t} \ket{\Psi_+} + c_- e^{-i\lambda_- t} 
\ket{\Psi_-}.
\eeq
                                                                        
Under the discrete symmetries $P$, $C$ and $T$ (parity, 
charge conjugation and time reversal), 
strangeness eigenstates transform in the following way\cite{LeeW,TDLee}:
\beq
\ba{ll}
P\ket{\Ko}   = - \ket{\Ko},  \qquad 
&P\ket{\Kob} = - \ket{\Kob}, \\
C\ket{\Ko}   = e^{i\alpha_c} \ket{\Kob}, \qquad 
&C\ket{\Kob} = e^{-i\alpha_c} \ket{\Ko}, \\
T\ket{\Ko}   = e^{i(\theta-\alpha_c)} \ket{\Ko},  \qquad
&T\ket{\Kob} = e^{i(\theta+\alpha_c)} \ket{\Kob}, \qquad
\ea
\label{CPT1}
\eeq
where $\alpha_c$ and  $\theta$ are arbitrary phases\footnote{~Note 
that $T(\eta \ket{\Psi}) = \eta^* T \ket{\Psi}$.}. Since 
strangeness is  
conserved in strong and electromagnetic interactions, is possible to 
redefine $\ket{\Ko}$ and $\ket{\Kob}$  phases in the following way:
\beq
\ket{\Ko} \rarrow e^{-i\alpha {\hat S}}\ket{\Ko}=e^{-i\alpha }\ket{\Ko}
\qquad 
\ket{\Kob} \rarrow e^{-i\alpha {\hat S}}\ket{\Kob}=e^{i\alpha}\ket{\Kob},
\eeq
where ${\hat S}$ is the operator which define the 
strangeness\footnote{~${\hat S}\ket{\Ko}=+\ket{\Ko}$,
${\hat S}\ket{\Kob}=-\ket{\Kob}$, ${\hat S}\ket{\mbox{\rm non--strange\
particles}}=0$.}. Thus the evolution matrix  $H$ is  defined up to 
the transformation
\beq 
\left[ \ba{cc} H_{11} & H_{12} \\   H_{21} & H_{22}  \ea \right]
\rarrow \left[ \ba{cc} H_{11} & e^{2i\alpha }H_{12} \\  e^{-2i\alpha } 
H_{21}  & H_{22}  \ea \right].\label{Halpha}
\eeq 
Choosing $\alpha=(\pi-\alpha_c)/2$, the transformations of 
$\ket{\Ko}$ and $\ket{\Kob}$ under $CP$ are given by:
\beq
CP \ket{\Ko} =  \ket{\Kob}, 
\label{convfase1}
\eeq
whereas those under $\Theta=CPT$ remain unchanged:
\beq
\Theta \ket{\Ko} = - e^{i\theta} \ket{\Kob},  \qquad
\Theta \ket{\Kob} = - e^{i\theta} \ket{\Ko}. 
\eeq                                             
Using this phase choice, the transformation laws of $H$
under  $CP$ and $CPT$ are given by:
\beqa
&CP \left[ \ba{cc} H_{11} & H_{12} \\   
H_{21} & H_{22}  \ea \right] (CP)^{-1} = \left[ \ba{cc} H_{22} & H_{21} \\ 
  H_{12} & H_{11}  \ea \right],& \label{Hcp} \\
&\Theta \left[ \ba{cc} H_{11} & H_{12} \\   H_{21} & H_{22}  \ea \right] 
\Theta^{-1} = \left[ \ba{cc} H_{22} & H_{12} \\   H_{21} & H_{11}  
\ea \right].&  \label{Hcpt} 
\eeqa 

From Eqs.~(\ref{Hcp}-\ref{Hcpt}) and (\ref{Halpha})  
we can easily deduce the conditions for $H$ to be 
invariant under $CP$ and $CPT$.
The time evolution matrix is invariant under  $CPT$ if $H_{11}=H_{22}$, 
i.e.\footnote{~Note that $M_{11}$ and 
$\Gamma_{11}$ are real and positive, since 
$M$ and $\Gamma$ are hermitian and positive.} 
\beq
M_{11}=M_{22} \qquad \mbox{\rm and} \qquad \Gamma_{11}=\Gamma_{22},
\label{CPTreq}
\eeq
no matter of the phase choice in (\ref{Halpha}).
Eq.~(\ref{CPTreq}) is a necessary but not sufficient 
condition to have invariance under $CP$.
To insure $CP$ invariance
is necessary to constraint also the off--diagonal elements of $H$. 
The condition following from (\ref{Hcp}), 
i.e. $M_{12}-i\Gamma_{12}/2=M_{12}^*-i\Gamma_{12}^*/2$, 
depends on the phase choice in Eq.~(\ref{Halpha}), 
indeed we can always choose $\alpha$ in such a way to 
make $M_{12}$ or $\Gamma_{12}$ real. The supplementary condition to 
insure $CP$ invariance, independently from the phase choice in 
(\ref{Halpha}), is:
\beq
\arg\left({M_{12} \over \Gamma_{12}}\right) = 0.
\label{CPreq}
\eeq

For now on we shall consider the $\Ko-\Kob$ system  assuming $CPT$ 
invariance. In this case the matrix $H$ can be written as 
\beq
H=\left[ \ba{cc} H_{11} & H_{12} \\   H_{21} & H_{11}  \ea \right]
\eeq 
and solving the eigenvalue equation one gets:
\beq
\lambda_\pm=H_{11} \pm \sqrt{H_{12}H_{21}}, \qquad\qquad \Psi_\pm \propto \left[
\ba{c} 1 \\ \pm \sqrt{{H_{21}/ H_{12}}} \ea \right].
\label{eigenv}
\eeq
In the limit where also $CP$ is an exact symmetry 
of the  $\Ko-\Kob$ system, i.e. the $CP$ operator commutes
with $H$, from Eq.~(\ref{CPreq}) follows that
\beq
\sqrt{ {H_{12}\over H_{21}}} = 
\sqrt{  {M_{12} -i\Gamma_{12}/2 \over M_{12}^*-i\Gamma_{12}^*/2} }
\eeq
is just a phase factor and, with an opportune phase transformation
(\ref{Halpha}), the normalized eigenvectors (conventionally called 
$\ket{K_1}$ e $\ket{K_2}$) are given by: 
\beqa
\ket{K_1}&=&{1\over \sqrt{2}}\left( \ket{\Ko} + \ket{\Kob}\right),  \\
\ket{K_2}&=&{1\over \sqrt{2}}\left( \ket{\Ko} - \ket{\Kob}\right).  
\eeqa
On the other hand, if $CP$ is an approximate symmetry of the 
$\Ko-\Kob$ system, the eigenvectors of $H$ are usually written as
\beqa
\ket{K_S}&=&{1\over \sqrt{1+\vert\veps\vert^2}}
\left( \ket{K_1} + \veps\ket{K_2}\right) \no \\
 &=& {1\over \sqrt{2(1+\vert\veps\vert^2)}}\left( (1+\veps)\ket{\Ko} + 
 (1-\veps) \ket{\Kob} \right),   \no \\ && \label{klksdef}\\
\ket{K_L}&=&{1\over \sqrt{1+\vert\veps\vert^2}}
\left( \ket{K_2} + \veps\ket{K_1}\right) \no \\ &=&
{1\over \sqrt{2(1+\vert\veps\vert^2)}}\left( (1+\veps) \ket{\Ko} -
 (1-\veps) \ket{\Kob}\right),  \no
\eeqa
where $\veps$ 
is given by:
\beq
{1 + \veps \over 1 - \veps} =  
\sqrt{  {M_{12} -i\Gamma_{12}/2 \over M_{12}^*-i\Gamma_{12}^*/2} }~.
\label{vepsdef}
\eeq
The $\veps$ parameter is not an observable quantity
and indeed is phase convention dependent. On the other hand,
the quantity
\beq
{\Real(\veps)\over 1+ \vert\veps\vert^2}
= {\Imm(\Gamma_{12})\Real(M_{12})-\Imm(M_{12})\Real(\Gamma_{12})
\over 4\vert M_{12}\vert^2 +\vert \Gamma_{12}\vert^2 },
\label{realveps}
\eeq
that vanishes if Eq.~(\ref{CPreq}) is 
satisfied\footnote{~In the r.h.s.
of Eq.~(\protect\ref{realveps})
we neglect terms of order $\Real(\veps)^2$.},
is phase independent and possibly observable.

The two eigenstates of $H$ have different 
eigenvalues also if $CP$ is an exact symmetry\footnote{~In the 
r.h.s. of Eqs.~(\protect\ref{cap130}--\protect\ref{cap131}) 
we neglect the imaginary parts of
 $M_{12}$ and $\Gamma_{12}$.}:   
\beqa
\lambda_S &=& M_{11}-{i\over 2} \Gamma_{11}+
\left( M_{12} -{i\over 2} \Gamma_{12} \right)
\left( {1 + \veps \over 1 - \veps} \right) \no \\ &\simeq &
M_{11} +  \Real M_{12} -{i\over 2} \left( \Gamma_{11} + 
\Real \Gamma_{12}  \right), \label{cap130} \\
\lambda_L &=& M_{11}-{i\over 2} \Gamma_{11} - \left( M_{12} -{i\over 2} 
\Gamma_{12} \right)\left( {1 + \veps \over 1 - \veps} \right) \no \\
&\simeq & M_{11} - \Real M_{12} -{i\over 2} \left( \Gamma_{11} - 
\Real \Gamma_{12}   \right). \label{cap131}
\eeqa
From the experimental data\cite{PDG} follows:
\beq
\ba{rcccl}
M_{11} &=& (M_L + M_S )/2  &=&  (497.672 \pm 0.031)\ \mbox{\rm MeV} ,\\
-\Real M_{12} &\simeq& (M_L - M_S )/2 &=& (1.755 \pm 0.009)
\times 10^{-12}\ \mbox{\rm MeV}, \\
\Gamma_{11}+ \Real \Gamma_{12} &\simeq& 
\Gamma_S &=& (7.374 \pm 0.010)\times 10^{-12}\ \mbox{\rm MeV}, \\
\Gamma_{11}- \Real \Gamma_{12} &\simeq& 
\Gamma_L &=& (1.273 \pm 0.010)\times 10^{-14}\ \mbox{\rm MeV}. 
\label{expm}
\ea
\eeq
The big difference among $M_{11}$ and the other matrix elements
of $H$  can be simply explained: whereas $M_{11}$ is dominated by 
the strong self--interaction of a strange meson, the other terms, 
connecting states with different strangeness, are
due to weak interactions. $M_{12}$, in particular, is determined by 
the $\vert\Delta S\vert=2$ 
 transition amplitude
which connects $\Ko$ and $\Kob$, whereas $\Gamma_{11}$ and 
$\Gamma_{12}$ are given by the product of two 
$\vert\Delta S\vert=1$ weak transitions (those responsible of 
$\Ko$ and $\Kob$ decays). Assuming that  $\vert\Delta S\vert=2$
transitions are generated by the product of two $\vert\Delta S\vert=1$
transitions, then is natural to expect\footnote{~We denote by 
$G_F$, $M_P$ and  $\theta$ the Fermi constant, the proton mass and
the Cabibbo angle, respectively.} (\SSp\ref{cap:SM})
\beq
M_{12} \sim \Gamma_{11} \sim \Gamma_{12}\sim 
{ G_F^2 M_P^4 \sin^2\theta \over (2\pi)^4}M_K  \sim 10^{-12} 
\mbox{\rm MeV}.
\eeq

\subsection{$K \to 2 \pi$ decays.} 
\label{sez:2pi}

If $CP$ was an exact symmetry of the kaon system, then 
the $\ket{K_2}$ ($CP$-negative eigenstate) should not 
decay in a two--pion final state (eigenstate of $CP$ with positive 
eigenvalue). However, experiments have shown
that both eigenstates of $H$ decay into two--pion 
final states\cite{Cronin,PDG}. The $\ket{K_S}$ decays
into  $\ket{2\pi}$  almost $100\%$ of the times, 
whereas the $\ket{K_L}$ has a branching ratio in this channel 
 about 3 order of magnitude smaller.

To analyze these decays more in detail, is convenient to 
introduce the isospin decomposition of  
$K \to 2 \pi$ amplitudes. Denoting by  
$S_{\rm strong}$ the  $S$--matrix of  strong interactions
(that we assume to be invariant under isospin transformations),
we define the $S$--wave `re--scattering' phases 
of a two--pion state with definite isospin $I$ as:
\beq
_{{\rm out}}\bra{ 2\pi, I} \ket{2\pi, I}_{{\rm in}} =
\bra{2\pi, I}\vert S_{{\rm strong}} \ket{2\pi, I}  \doteq e^{i 2\delta_I}.
\label{kodec01}
\eeq
With this definition,  $\Ko \to 2 \pi$ transition amplitudes
can be written in the form
\beq
_{{\rm out}}\bra{2\pi, I}\vert H_{{\rm weak}} \ket{ 
\Ko}_{{\rm in}} \doteq 
\sqrt{3 \over 2} A_I  e^{i \delta_I},
\label{kodec}
\eeq
where $H_{{\rm weak}}$ indicates the weak hamiltonian 
of $\vert \Delta S\vert =1$  transitions. Assuming 
 $CPT$ invariance, from Eqs.~(\ref{kodec01}-\ref{kodec}) it
follows\cite{Okun}:
\beq
_{{\rm out}}\bra{2\pi, I}\vert H_{{\rm weak}} \ket{ 
\Kob}_{{\rm in}} = \sqrt{3 
\over 2} A_I^* e^{i \delta_I}.
\eeq

The two--pion isospin states allowed in 
$S$--wave are $I=0$ and $I=2$, defining 
$A(K \to 2 \pi)=\  _{{\rm out}}\bra{2\pi}\vert 
H_{{\rm weak}} \ket{K}_{{\rm in}}$
and using the appropriate Clebsh--Gordan coefficients, we 
find:\footnote{~We assume $\Delta I \leq 3/2$.}
\beqa 
A(\Ko \to \pi^+ \pi^-)&=&A_0e^{i\delta_0}+{1 \over \sqrt{2}} A_2e^{i\delta_2}, 
\no \\
A(\Ko \to \pi^0 \pi^0)&=&A_0e^{i\delta_0}- \sqrt{2} A_2e^{i\delta_2},
\label{k2pdec} \label{definA_0A_2}  \\
A( K^+ \to \pi^+ \pi^0)&=&{3 \over   2} A_2e^{i\delta_2}. \no
\eeqa
$A(\Kob \to 2 \pi)$ amplitudes are obtained 
from (\ref{definA_0A_2}) with the simple substitution 
$A_I \to A_I^*$. If $CP$ commuted with
$H_{{\rm weak}}$ then we should have, up to a phase factor,
$A(\Kob \to 2 \pi)= A(\Ko \to 2 \pi)$.
Thus, analogously to Eq.~(\ref{CPreq}), which states the 
$CP$--invariance condition of  $\Delta S=2$ amplitudes respect to 
$\Delta S=1$ ones, the $CP$--invariance condition among the
two $A_I$ amplitudes is given by:
\beq
\arg\left({ A_2 \over A_0 }\right) = 0.
\label{CPreq2}
\eeq  

At this point it is convenient to make a choice on the arbitrary phase 
of the weak amplitudes. Historically, a very popular choice 
is the famous Wu--Yang convention\cite{WuY}:  
\beq
\arg(A_0)=0.
\label{WYconv}
\eeq
Since $A_0$ is dominant with respect to the
other $A(\Ko, \Kob \to f)$ amplitudes, 
this convention is useful because from the unitarity 
relation 
\beqa
\Gamma_{12} = 2\pi \sum_{n}{ \
_{{\rm in}} \bra{ \Kob }\vert H_{{\rm weak}} \ket{n}_{{\rm out}}
 \cdot _{{\rm out}}\bra{n}\vert H_{{\rm weak}}
 \ket{ \Ko}_{{\rm in} }} \delta(M_K-E_n) \qquad\quad &&  \\
\simeq 2\pi \sum_{2 \pi (I=0)}{
_{{\rm in}} \bra{ \Kob }\vert H_{{\rm weak}} 
\ket{2\pi}_{{\rm out}} \cdot
_{{\rm out}}\bra{2\pi}\vert H_{{\rm weak}} \ket{ \Ko}_{{\rm in}}
 }\delta(M_K-E_{2\pi}),&&
\eeqa
it follows     
\beq
\arg(\Gamma_{12})\simeq 2\arg(A_0),
\eeq
thus Eq.~(\ref{WYconv}) implies also  $\arg(\Gamma_{12})\simeq 0$. 
By this way all  weak phases are `rotated' on suppressed 
amplitudes, like $A_2$ and $M_{12}$. From  
(\ref{vepsdef}--\ref{realveps}) follows  also
\beqa
\arg\left( \veps \big\vert_{{\rm  WY }}
\right) &=& - \arctan \left[ 2 \dis{ \Real
 (M_{12}) \over \Real (\Gamma_{12})}\right] \no\\
&=&  \arctan \left[ 2 \dis{ M_L-M_S \over \Gamma_L-\Gamma_S }\right]
= (43.6 \pm 0.1)^\circ.
\label{defpsw}
\eeqa
However, in the following  we shall not adopt the 
Wu--Yang phase convention, which is not the 
most natural choice in the  Standard Model (\SSp\ref{cap:SM}), 
but we shall impose only 
\beq
\arg(A_0) \ll 1,
\label{Myconv}
\eeq
in order to treat as perturbations   
weak--amplitude phases.

From the experimental information on 
$\Gamma(K_S \to 2 \pi)$ and $\Gamma(K^+ \to \pi^+\pi^0)$\cite{PDG}, 
neglecting $CP$ violating effects,
one gets\footnote{~Due to small isospin--breaking
effects, amplified by the suppression of  $A_2$ 
(\SSp\ref{sez:SMepsp}), the value of $(\delta_2-\delta_0)$ 
extracted by $K\to 2 \pi$ data is not reliable. 
We shall adopt the 
prediction of Gasser and 
Mei{\ss}ner\protect\cite{GasserM} that is in good 
agreement with the value of $(\delta_2-\delta_0)$  extrapolated from 
$\pi-N$ data\protect\cite{Ochs}.}
\beqa
\Real( A_0 ) &=& (2.716  \pm 0.007)\times 10^{-4}\ \mbox{\rm MeV}, 
\label{k2pexp1} \\
\omega^{-1}  &=& {\Real( A_0 ) \over \Real( A_2 ) } = (22.2 \pm 0.1)
\qquad  \mbox{\rm and} \qquad  (\delta_0-\delta_2) = (45 \pm 6)^\circ. 
\label{k2pexp2}
\eeqa
As anticipated, $\Real( A_0 )  \gg \Real( A_2 )$, i.e.
the $\Delta I=1/2$ transition amplitude is dominant with respect
to the  $\Delta I=3/2$ one (`$\Delta I=1/2$ rule'). 

Conventionally, in order to study  $CP$ violating effects, the following
parameters are introduced:
\beqa
\eta_{+-}\ \doteq& \dis{ A(K_L \to \pi^+ \pi^-) \over A(K_S \to \pi^+ \pi^-) }
&\doteq\ \eps + \epsp, \\
\eta_{00}\ \doteq& \dis{ A(K_L \to \pi^0 \pi^0) \over A(K_S \to \pi^0 \pi^0) }
&\doteq\ \eps -2 \epsp.
\eeqa
Using Eqs.~(\ref{klksdef}) and (\ref{k2pdec}), and
neglecting terms proportional to  $\Imm (A_I)^2/$ $\Real(A_I)^2$ 
and $\omega^2 \Imm (A_I)/\Real(A_I) $,  $\eps$ 
ed $\epsp$ are given by:
\beqa
\eps &=& \veps + i {\Imm (A_0) \over \Real(A_0) } =  \veps \big
\vert_{{\rm  WY }}
\label{eps1} \\
\epsp &=& i { e^{i(\delta_2-\delta_0)} \over \sqrt{2}}\omega \left[
{\Imm (A_2) \over \Real(A_2)}-{\Imm (A_0) \over \Real(A_0)}\right].
\label{eps2}
\eeqa
Both $\eps$ and $\epsp$ are measurable quantities with  definite phases:
\beqa
\arg(\eps)&=& \arg\left( \veps \big\vert_{{\rm  WY }} \right) =  
(43.6\pm 0.1)^\circ, \label{cpt111}\\
\arg(\epsp)&=&(\delta_2-\delta_0+\dis{\pi \over 2}) = (45\pm 6)^\circ ,
\eeqa
and both vanish in the exact-$CP$ limit: $\eps$ 
vanishes if Eq.~(\ref{CPreq}) if satisfied, 
whereas $\epsp$ vanishes if Eq.~(\ref{CPreq2}) is satisfied. 
The $\eps$ parameter  is referred as the `indirect' $CP$--violating 
parameter, since Eq.~(\ref{CPreq}) could be violated by 
a non--zero weak phase in $\vert \Delta S \vert=2$ amplitudes
only. 
On the other hand, $\epsp$ is called the  `direct' 
$CP$--violating parameter, since 
a non--zero weak phase in  $\vert \Delta S \vert=1$ amplitudes
is necessary  to violate Eq.~(\ref{CPreq2}). 

It is interesting to note that the first identity in Eq.~(\ref{cpt111})
could be violated if $CPT$ was not an exact symmetry. Thus, given that
$\arg\left( \veps \big\vert_{{\rm  WY }} \right) \simeq   
\arg(\epsp)$, a large value of $\Imm(\epsp/\eps)$ 
(~$\Imm(\epsp/\eps) \gsim \Real(\epsp/\eps)$~) would be a signal of 
$CPT$ violation.\cite{handbook} 

From the analysis of experimental data on  $K_L \to 2 \pi$
decays\cite{PDG} follows:
\beqa
\vert\eps\vert&=(2.266 \pm 0.017)\times 10^{-3}\qquad \arg(\eps)
&=(44.0 \pm 0.7)^\circ, \label{epsexp}\\
\Real\Big({\epsp \over \eps}\Big)&=(1.5 \pm 0.8)\times 10^{-3}\qquad 
\Imm\Big(\dis{\epsp \over \eps}\Big)&=(1.7 \pm 1.7)\times 10^{-2}. 
\eeqa
Whereas $\vert\eps\vert$ is definitely different from 
zero,  $\vert\epsp\vert$ 
is compatible with zero within two standard deviations.
This implies that the so--called `super--weak' model proposed by 
Wolfenstein more than 30 years ago\cite{Wolf,LeeWolf}, 
a model where $CP$ violation
is supposed to be generated by a very weak 
($\sim { G_F^2 M_p^4 \sin^2\theta/(2\pi)^4}$) new  
$\vert \Delta S \vert=2$ interaction, 
is still compatible with the experimental data. Thus there is no 
confirmation of the Standard Model mechanism (\SSp\ref{cap:SM}), which 
predicts also direct $CP$ violation. However,
as we shall see in the next section, the fact that 
$\vert\epsp\vert$ is much smaller of $\vert\eps\vert$ could
be explained also in  the Standard Model where,
for large values of the top quark mass, the weak phases of 
$A_0$ and $A_2$ accidentally tend to cancel. Then, within the Standard Model,
the fundamental condition for the observability of direct $CP$ violation
tends to lack in $K\to 2 \pi$ transitions:
\par {\bf [1]} {\it A necessary condition in order to observe
direct $CP$ violation in $(\Ko,\Kob)\to f$ transitions, is
the presence of at least two weak amplitudes 
with different (weak) phases.} 
                                        
\subsection{Semileptonic decays.}
\label{sez:semil}

Up to now $CP$ violation 
has been observed in the following processes:
$K_L \to 2\pi$, $K_L \to \pi^+\pi^-\gamma$
and  $K_L \to \pi l\nu$ $(l=\mu,e)$. 

As we shall discuss better in sect.~\ref{cap:Kppg}, the amount of 
$CP$ violation till now observed in $K_L \to \pi^+\pi^-\gamma$ 
is generated only by the bremsstrahlung of the 
corresponding non--radiative transition ($K_L \to \pi^+\pi^-$)
and does not carry any new information with respect to 
$K_L \to \pi^+\pi^-$.

In the  $K_L \to \pi l\nu$ case, the selection rule
$\Delta S=\Delta Q$, predicted by the Standard Model and 
in perfect agreement with the data, allows  the
existence of a single weak amplitude.\footnote{~$\Delta 
S=-\Delta Q$ transitions are allowed in the Standard Model 
only at $O(G_F^2)$; explicit calculations\protect\cite{Guberina2,Luke} 
show a suppression factor of about $10^{-6}-10^{-7}$. }
Consequently, from the 
condition {\bf [1]}, we deduce that within the Standard Model
is impossible to observe direct $CP$ violation in these channels. 

The selection rule $\Delta S=\Delta Q$ implies:
\beq
A(\Ko \to \pi^+ l^-  {\bar \nu}_l ) = A(\Kob \to \pi^- l^+  \nu_l)=0,
\eeq
thus defining 
\beq
A(\Ko \to \pi^- l^+  \nu_l ) \doteq A_l,
\eeq
from $CPT$ follows $A(\Kob \to \pi^+ l^-  {\bar \nu}_l )=A_l^*$ 
and the decay amplitudes for $K_L$ and $K_S$ are given by:
\beqa
A(K_S \to \pi^+ l^-  {\bar \nu}_l ) &= -
A(K_L \to \pi^+ l^-  {\bar \nu}_l ) &= {1-\veps \over 
\sqrt{ 2(1+\vert\veps\vert^2) }}
A_l^*, \\
A(K_S \to \pi^- l^+  { \nu}_l ) &=
A(K_L \to \pi^- l^+  { \nu}_l ) &= {1+\veps \over 
\sqrt{ 2(1+\vert\veps\vert^2) }}
A_l. 
\eeqa
If $CP$ was an exact symmetry of the system we should have
\beqa
&& \vert A(K_L \to \pi^+ l^- {\bar \nu}_l ) \vert =	
 \vert A(K_L \to \pi^- l^+ \nu_l ) \vert   \no\\
&& \quad = \vert A(K_S \to \pi^+ l^- {\bar \nu}_l ) \vert =
\vert A(K_S \to \pi^- l^+ \nu_l ) \vert, 
\eeqa
thus $CP$ violation in these channel can be observed 
via the charge asymmetries
\beq
\delta_{L,S} = { \Gamma( K_{L,S} \to \pi^- l^+  { \nu}_l ) -
\Gamma(K_{L,S} \to \pi^+ l^-  {\bar \nu}_l ) \over
\Gamma( K_{L,S} \to \pi^- l^+  { \nu}_l ) -
\Gamma(K_{L,S} \to \pi^+ l^-  {\bar \nu}_l ) }
= {2 \Real(\veps) \over 1 + \vert\veps\vert^2 }
\label{deltal}
\eeq
sensible to indirect $CP$ violation only.

The experimental data on $\delta_L$ is\cite{PDG}:
\beq
\delta_L = (3.27 \pm 0.12)\times 10^{-3},
\eeq
in perfect agreement with the numerical value
of $\Real(\eps)$ obtained form Eq.~(\ref{epsexp}).

\subsection{Charged--kaon decays.}
\label{sez:charged}

Charged--kaon decays could represent a privileged 
observatory for the study of direct $CP$ violation 
since there is no mass mixing between $K^+$ and $K^-$.

Denoting by $f$ a generic final state of $K^+$ decays, 
with ${\bar f}$ the charge conjugate and defining
\beqa
  _{{\rm out}}\bra{ f }\vert H_{{\rm weak}} \ket{ 
K^+}_{{\rm in}} &\doteq & A_f \\
  _{{\rm out}}\bra{ {\bar f} }\vert H_{{\rm weak}} \ket{ 
K^-}_{{\rm in}} &\doteq & A_{\bar f},
\eeqa
if $CP$ commuted with $H_{{\rm weak}}$ then 
$\vert A_f \vert = \vert A_{\bar f} \vert $. 
Thus $CP$ violation in this channels can be 
observed via the asymmetry\footnote{~Integrating over the full 
phase space, $\Delta_f$ can be related to the width 
charge asymmetry; with appropriate kinematical cuts one can 
isolate also slope asymmetries.}
\beq
\Delta_f = { \vert A_f \vert - \vert A_{\bar f} \vert \over
\vert A_f \vert + \vert A_{\bar f} \vert } 
\eeq
which, differently form (\ref{deltal}), is sensible to 
direct $CP$ violation only.

To analyze more in detail under which conditions  $\Delta_f$
can be different from zero, let's consider the 
case where  $\ket{f}$ is not an eigenstate of strong 
interactions, but a superposition of two states with different 
re--scattering phases. In this case, separating strong and weak phases
analogously to Eq.~(\ref{kodec}), we can write
\beq
A_f = a_f e^{i\delta_a} +b_f e^{i\delta_b}.
\eeq
In addition, imposing  $CPT$ invariance, it is found 
\beq 
A_{\bar f} = a_f^* e^{i\delta_a} +b_f^* e^{i\delta_b}, 
\eeq
thus the charge asymmetry  $\Delta_f$ is given by:
\beq
\Delta_f ={ 2\Imm(a_f^* b_f)\sin (\delta_a-\delta_b) \over
\vert a_f \vert^2 + \vert b_f \vert^2 +
2\Real(a_f^* b_f)\cos (\delta_a-\delta_b)}. 
\label{chargas}
\eeq
As it appears from Eq.~(\ref{chargas}), in order to have 
 $\Delta_f$ different from zero, is not only necessary 
to have two weak amplitudes ($a_f$ e $b_f$) 
with different weak phases (as already stated in {\bf [1]}), 
but is also necessary to have different strong re--scattering 
phases. 
\par {\bf [2]} {\it A necessary condition in order
to observe (direct) $CP$ violation in 
$K^\pm\to(f,{\overline f})$  transitions,
 is the presence of
at least two weak amplitudes with different (weak) phases
and different (strong) re--scattering phases.}

As we shall see in the next sections, the condition {\bf [2]} is not
easily satisfied within the Standard Model, indeed  
$CP$ violation in charged--kaon decays has not been observed yet. 
To be more specific,
the most frequent decay channels ($99,9\%$  of the branching ratio)
of $K^\pm$ mesons are the lepton channels $\ket{l\nu}$ and 
$\ket{\pi l \nu}$,  together with the non--leptonic states 
$\ket{2\pi}$ and $\ket{3\pi}$.
In the first two channels there is no strong re--scattering.
In the $\ket{2\pi}$ case the re--scattering phase is unique (is 
a pure $I=2$ state). Finally in the $\ket{3\pi}$ case there are different 
re--scattering phases but are suppressed, since the available phase 
space is small (\SSp\ref{cap:K3pi}).

\setcounter{equation}{0}
\setcounter{footnote}{0}
\section{$CP$ violation in the  Standard Model.}
\label{cap:SM}

The `Standard Model' of strong and electro--weak interactions
is a non--abelian gauge theory based on the 
$SU(3)_C\times SU(2)_L\times U(1)_Y$ 
symmetry group\footnote{~For excellent phenomenological reviews on 
gauge theories and, in particular, on the Standard Model see
Refs.\protect\cite{Abers}$^-$\cite{Donoghuebook}. }.    
               
The $SU(3)_C$ subgroup is the symmetry of strong (or color) 
interactions, realized \`a la Wigner--Weyl,
i.e. with a vacuum state invariant under $SU(3)_C$. 
On the other hand, the $SU(2)_L\times U(1)_Y$ symmetry, which rules
electro--weak interactions\cite{Glashow}$^-$\cite{Salam}, 
is realized \`a la Nambu--Goldstone\cite{Nambu,Goldstone}, 
with a vacuum state invariant only under 
the subgroup  $U(1)_Q$ of electromagnetic interactions.

The interaction lagrangian of fermion fields 
with  $SU(2)_L\times U(1)_Y$ electro--weak gauge bosons 
is given by:
\beqa
{\cal L}_{e-w} &=& \sum_\alpha \bar{Q}_L^{(\alpha)} \gamma^\mu \left(
gT_iW^i_\mu +\dis{1\over2} g'YB_\mu\right)  Q_L^{(\alpha)} \no\\
&+& \sum_\alpha \bar{\Psi}_L^{(\alpha)} \gamma^\mu \left(
gT_iW^i_\mu +\dis{1\over2} g'YB_\mu\right)  \Psi_L^{(\alpha)}\no\\
&+& \sum_f \bar{f}_R  \gamma^\mu \left( {1\over 2}
 g'YB_\mu\right) f_R,
\label{lew}
\eeqa
where $g,T_i,W^i_\mu(i=1,3)$ and $g',Y,B_\mu$ 
are the coupling constants, the generators and the 
gauge fields of  $SU(2)_L$ and 
$U(1)_Y$, respectively. The index $\alpha$ can assume 3 values,
according to the quark or lepton  family to which is referred, whereas
$f$ indicates the sum over all right--handed fermions. As anticipated,
the $SU(2)_L\times U(1)_Y$ symmetry 
is spontaneously broken in such a way 
that the photon field $A_\mu$ remains massless.
Introducing the weak angle  $\theta_W$, defined by the relation
$\tan\theta_W=g'/g$, and the field $Z_\mu$ (combination of 
$W^3_\mu$ and $B_\mu$ orthogonal to  $A_\mu$):
\beq 
\left( \ba{c} W^3_\mu \\ B_\mu \ea \right)  = \left(
\ba{cc} \sin\theta_W  & \cos\theta_W \\  \cos\theta_W & -\sin\theta_W \ea  
\right) \left(  \ba{c} A_\mu  \\  Z_\mu \ea \right),
\eeq
 Eq.~(\ref{lew}) becomes:
\beqa
{\cal L}_{e-w} &=&  e \sum_f \bar{f}  \gamma^\mu Q f A_\mu \no \\
&+& {g \over \cos\theta_W} \sum_f \bar{f}  \gamma^\mu 
\left( Q_Z^{V}- Q_Z^{A}\gamma_5\right)  f  Z_\mu \no \\
&+& {g \over 2\sqrt{2}} \left[ \sum_\alpha \bar{u}^{(\alpha)} \gamma^\mu
\left(1-\gamma_5 \right) d^{(\alpha)} W^+_\mu + \mbox{\rm h.c.} \right]
     \no \\
&+&  {g \over 2\sqrt{2}} \left[\sum_\alpha  \bar{\nu}^{(\alpha)} \gamma^\mu 
\left(1-\gamma_5 \right) l^{(\alpha)}  W^+_\mu  + \mbox{\rm h.c.} \right],
\eeqa
where $e=g'\cos\theta_W$, $Q_Z^{V}=T_3-Q\sin^2\theta_W$ and $Q_Z^{A}=T_3$.

\begin{table}[t]
\[ \ba{|c|c|c|c|c|c|c|}  \hline
 & \multicolumn{6}{c|}{\mbox{\rm  leptons}\qquad\qquad\quad} \\ 
 & \multicolumn{3}{|c|}{l-\mbox{type}} & \multicolumn{3}{|c|}
             {\nu-\mbox{type}} \\ \hline
 & e & \mu & \tau & \nu_e & \nu_\mu & \nu_\tau \\ \hline
m(\mbox{\rm MeV}) & 0.511 &105.7 & 1777 & < 7\times 10^{-6} & 
               < 0.27 & < 31 \\ \hline
Q & \multicolumn{3}{|c|}{-1} & \multicolumn{3}{|c|}{ 0 } \\ \hline
 & \multicolumn{6}{c|}{\mbox{\rm  quarks }\qquad\qquad}\\ 
 & \multicolumn{3}{c|}{d-\mbox{type}} & 
        \multicolumn{3}{|c|}{u-\mbox{type}} \\ \hline
 & d & s & b & u & c & t \\ \hline
m(\mbox{\rm MeV}) & \sim 10 & \sim 150  & \sim 4300 & \sim 5  & 
               \sim 1200  & \sim 1.7\times 10^5 \\ \hline
Q & \multicolumn{3}{|c|}{-1/3} & \multicolumn{3}{|c|}{+2/3} \\ \hline
\ea \]
\caption{Masses and electric charges 
of quarks and leptons 
(see Ref.~\protect\cite{PDG} for a discussion about
the definition of quark masses).}
\label{tab:qandl}  
\end{table}

As in the case of gauge boson masses, 
quark and lepton masses (\SSt\ref{tab:qandl}) are dynamically generated by 
the Yukawa coupling of these fields with 
the scalars that spontaneously broke 
$SU(2)_L\times U(1)_Y$ symmetry.
The effective lagrangian for quark and lepton masses 
can be written in the following way
\beq
{\cal L}^{eff}_{m} =  \bar{U}M_U U + \bar{D}M_D D +  \bar{L}M_L L,
\label{L_m}
\eeq
where
\beq
U=\left( \ba{c} u  \\ c \\ t \ea\right),
\qquad
D=\left( \ba{c} d \\ s \\ b \ea\right),
\qquad 
L=\left( \ba{c} e \\ \mu \\ \tau \ea \right),
\eeq
$M_U=\mbox{\rm diag}\{ m_u, m_c, m_t \}$,
$M_D=\mbox{\rm diag}\{ m_d, m_s, m_b \}$ and
$M_L=\mbox{\rm diag}\{ m_e, m_\mu, m_\tau \}$
(we assume massless neutrinos).
The fermion fields which appear in Eq.~(\ref{lew}) 
are not in general  eigenstates of mass matrices. 
Introducing the unitary matrices $V_U$, $V_D$ and $V_L$, so that
\beq
\left( \ba{c} u^{(1)}  \\ u^{(2)}  \\ u^{(3)}  \ea\right) =
V_U U,
\qquad
\left( \ba{c} d^{(1)}  \\ d^{(2)}  \\ d^{(3)}  \ea\right) =
V_D D,
\qquad \mbox{\rm and} \qquad 
\left( \ba{c} l^{(1)}  \\ l^{(2)}  \\ l^{(3)}  \ea\right) =
V_L L,
\eeq 
thus writing electro--weak eigenstates 
in terms of mass eigenstates, Eq.~(\ref{lew}) becomes
\beqa
{\cal L}_{e-w} &=&  e \sum_f \bar{f}  \gamma^\mu Q f A_\mu \no \\
&+& {g \over \cos\theta_W} \sum_f \bar{f}  \gamma^\mu 
\left( Q_Z^{V}+ Q_Z^{A}\gamma_5\right)  f  Z_\mu \no \\
&+& {g \over 2\sqrt{2}} \left[ \bar{U} (V_U\da V_D)  \gamma^\mu
\left(1-\gamma_5 \right) D W^+_\mu + \mbox{\rm h.c.} \right]
     \no \\
&+&  {g \over 2\sqrt{2}} \left[\sum_\alpha \bar{N} \gamma^\mu 
\left(1-\gamma_5 \right) L  W^+_\mu  + \mbox{\rm h.c.} \right],
\label{lewm}
\eeqa
where
\beq
N=V_L\da  \left( \ba{c} \nu^{(1)} \\ \nu^{(2)} \\ \nu^{(2)} \ea\right)
\doteq \left( \ba{c} \nu_e \\ \nu_\mu \\ \nu_\tau \ea\right).
\eeq
Since fermion mass matrices commute with electric charge
and with hypercharge ($Y$),
the only difference among Eqs.~(\ref{lew}) and (\ref{lewm}) is
the presence of the unitary matrix $V_U\da V_D$.

As we have seen in the previous section, a
necessary condition to induce  $CP$ violation in kaon decays
is the presence of weak amplitudes with different weak phases.
In the Standard  Model framework, this condition is equivalent 
to the requirement of complex coupling constants in ${\cal L}_{e-w}$. 
These complex couplings must not to be cancelled by a redefinition
of field phases and leave ${\cal L}_{e-w}$
hermitian. From these two conditions 
follows that the only coupling constants of ${\cal L}_{e-w}$
which can be complex are the matrix elements of $V_U\da V_D$. 
This matrix,  known as the Cabibbo--Kobayashi--Maskawa (CKM) 
matrix\cite{Cabibbo,GIM,Kobayashi}, is the only 
source of $CP$ violation in the electro--weak sector of the 
Standard Model.

As anticipated in the introduction, in this review we will not 
discuss about possible effects due to  $CP$ violation in the 
strong sector.
 
\subsection{The CKM matrix.}
\label{sez:CKM}

In the general case of $N_f$ quark families,  
$U_{CKM}\doteq V_U\da V_D$ is a
$N_f\times N_f$ unitary matrix. 
The number of independent phases, $n_P$, and real parameters, $n_R$, 
in a $N_f\times N_f$ unitary matrix is given by:
\beq
n_P= {N_f(N_f+1) \over 2} \qquad \mbox{\rm and}\qquad  n_R= {N_f(N_f-1) 
\over 2}.
\eeq
Since the $2N_f-1$ relative phases of the  $2N_f$ 
fermion fields are arbitrary, 
the non trivial phases in $U_{CKM}$ are
\beq
n^{CP}_P= {N_f(N_f+1) \over 2} -2N_f+1 = {(N_f-1)(N_f-2) \over 2}.
\eeq
In the minimal Standard Model $N_f=3$,
thus the non--trivial phase is unique.

Parametrizing $U_{CKM}$  in terms of 3 angles
and one phase, it is possible to write\cite{MaianiCKM}$^-$\cite{Chau}:
\beqa
U_{CKM} &=& \left( \ba{ccc} V_{ud} & V_{us} & V_{ub} \\
V_{cd} & V_{cs} & V_{cb} \\ V_{td} & V_{ts} & V_{tb} \ea \right) \\
&& \!\!\!\!\!\!\!\! =\ \left( \ba{ccc} 
C_\theta C_\sigma & S_\theta C_\sigma & S_\sigma e^{-i\delta}\\
-S_\theta C_\tau - C_\theta S_\sigma S_\tau e^{i\delta} & 
C_\theta C_\tau - S_\theta S_\sigma S_\tau e^{i\delta} & C_\sigma S_\tau  \\
S_\theta S_\tau - C_\theta S_\sigma C_\tau e^{i\delta} & 
-C_\theta S_\tau - S_\theta S_\sigma C_\tau e^{i\delta} & C_\sigma C_\tau \\
\ea \right),
\label{ckma}
\eeqa
where $C_\theta$ and $S_\theta$ denote  sine and 
cosine of the Cabibbo angle\cite{Cabibbo}, 
$C_\sigma=\cos\sigma$, $S_\sigma=\sin\sigma$, etc...

It is important to remark that the  CKM phase would not be
observable if the mass matrices $M_D$ and $M_U$, 
introduced in Eq.~(\ref{L_m}), had degenerate 
eigenvalues\cite{MaianiCKM,JarlCKM}. 
As an example, if $m_c$ was equal to  $m_t$, then  $V_U$ 
would be defined up to the transformation
\beq
V_U \to \left( \ba{cc} 1  & 0 \\ 0 &  U(\bar{\theta},\bar{\delta_i})   \ea 
\right) V_U,
\eeq
where $U(\bar{\theta},\bar{\delta_i} )$ is a unitary $2\times2$
matrix. Thus adjusting the  independent 
parameters of $U(\bar{\theta},\bar{\delta_i})$ 
(1 angle and 3 phases) 
the phase of $U_{CKM}$ could be rotated away. 
In other words, 
the phenomenon of $CP$ violation is intimately related not only 
to the symmetry breaking mechanism of the electro--weak 
symmetry (or better of the chiral symmetry, \SSp\ref{cap:CHPT}),
but also to the breaking mechanism of the 
flavor symmetry (the global $SU(N_f)$ symmetry that we should have if
all the quark masses were equal). The two mechanisms  
coincide only in the minimal Standard Model.

There is an empirical hierarchy
among CKM matrix elements ($S_\sigma \ll S_\tau 
\ll S_\theta \ll 1$), which let us express them in terms 
of a single `scale--parameter' $\lambda=S_\theta\simeq 0.22$ and
three coefficients of order 1 ($A$, $\rho$ and $\eta$)\cite{WolfCKM}.
Defining
\beq
S_\tau=A\lambda^2,\qquad S_\sigma=A\lambda^3\sigma, \qquad
\sigma e^{i\delta} = \rho + i\eta
\label{Wolfpar} 
\eeq
and expanding Eq.~(\ref{ckma}) in powers of
$\lambda$ up to $O(\lambda^3)$ terms, one finds:
\beq
U_{CKM}\simeq\left( \ba{ccc} 
1 -{\lambda^2 \over 2} & \lambda & A\lambda^3 (\rho-i\eta) \\
-\lambda  & 1 -{\lambda^2 \over 2} & A\lambda^2 \\
 A\lambda^3 (1-\rho-i\eta) & -A\lambda^2 & 1 \ea \right).
\label{Wparckm}
\eeq

\begin{table}[t]
\[ \ba{|c|c|c|} \hline
\hbox{\rm Process}  & \hbox{\rm Parameter} 
            & \hbox{\rm Value}  \\ \hline
K \to \pi l\nu  & \lambda  & 0.2205 \pm 0.0018  \\ \hline
b \to  c l\nu   & A & 0.83 \pm 0.08 \\ \hline
b \to u   l\nu  & \sigma & 0.36 \pm 0.14 \\ \hline
K_L \to \pi\pi & \cos\delta & 0.47 \pm 0.32 \\ \hline
\ea \]
\caption{Experimental determination of CKM matrix elements
(in the pa\-ra\-me\-tri\-za\-tion~(\protect\ref{Wolfpar})) 
and relative pro\-ces\-ses from which are extrac\-ted\protect\cite{PDG}
(see sect.~\protect\ref{sez:SMepsp}
for the determination of $\cos\delta$). } 
\label{tab:CKMpar}
\end{table}

An interesting quantity for the study of $CP$  violation is the `so--called'
$J_{CP}$ parameter\cite{JarlCKM}:
\beq
J_{CP}=\Imm \left( V_{ai} V_{bj} V_{aj}^* V_{bi}^* \right).
\eeq 
As can be easily shown, any observable which violates $CP$ must
be proportional to this quantity\footnote{~Let's consider, as an example,
the elementary process $u_a d_j \to  
u_a  d_i$. The amplitude is the superposition of at least two processes:
$u_a d_j \to W^+ d_i d_j \to u_a d_i$ and 
$u_a d_j \to u_a W^- u_b \to u_a d_i$, thus 
$A \sim \alpha V_{ai}V_{aj}^*+\beta V_{bj}^*V_{bi}$. The charge asymmetry,
analogously to Eq.~(\protect\ref{chargas}), will be proportional to 
$\Imm(V_{ai}V_{bj}V_{aj}^*V_{bi}^*)$.}.
The unitarity of $U_{CKM}$ insures that  $J_{CP}$ is 
independent of the choice of  $a,b,i$ and $j$ 
($a\not=b $ and $i\not=j$). In the parametrization~(\ref{Wolfpar}) 
$J_{CP}$ is given by:
\beq
J_{CP}=\eta A^2 \lambda^6 + O(\lambda^8) \lsim 10^{-4}.
\label{JCP}
\eeq
Form Eq.~(\ref{JCP}) we can infer two simple considerations: 
\begin{itemize}
\item{$CP$ violation is naturally suppressed in the 
Standard Model due to CKM matrix hierarchy.}
\item{Transitions where $CP$ violation
should be more easily detected are those where also the 
$CP$--conserving amplitude is suppressed 
by the matrix elements $V_{ub}$ and $V_{td}$.}
\end{itemize}

\subsection{Four--quark hamiltonians.}
\label{sez:4ferm}

Differently than  weak and electromagnetic interactions, 
strong interactions are not in a perturbative regime at low
energies  ($E \lsim \Lambda_\chi \sim 1$ GeV), 
i.e. at distances $d \gsim 1/\Lambda_\chi \sim 10^{-14}$ cm.
This happens because the strong coupling constant
has a divergent behaviour in the infrared limit and, presumably,
this is also the main reason why quarks 
and gluons are confined in hadrons.
With respect to the scale $\Lambda_\chi$ quarks can be 
grouped in two categories according to their
mass (\SSt\ref{tab:qandl}): $u$, $d$ and $s$ are `light', 
since $m_{u,d,s}<\Lambda_\chi$, whereas $c$, $b$ and $t$
are `heavy', since $\Lambda_\chi<m_{b,c,t}$.
As we shall discuss better in the next section, the lightest
hadrons with light valence quarks are 
the pseudoscalar mesons $\pi$, $K$ and $\eta$ (\SSt\ref{tab:ottetto}).

\begin{table}[t] 
\[ \ba{|c|c|c|c|c|} \hline
{\mbox{\rm meson}} & {\mbox{\rm valence quarks}} & {\mbox{\rm m(MeV)}}  
& (I,I_3) & S \\ \hline
\pi^+ & u\bar{d}          & 139.6 & (1,+1)     &  0 \Big. \\ \hline
\pi^0 & u\bar{u}-d\bar{d} & 135.0 & (1,0)      &  0 \Big. \\ \hline
\pi^- & d\bar{u}          & 139.6 & (1,-1)     &  0 \Big. \\ \hline
 K^+  & u\bar{s}          & 493.7 & (1/2,+1/2) & +1 \Big. \\ \hline
 \Ko  & d\bar{s}          & 497.7 & (1/2,-1/2) & +1 \Big. \\ \hline
\Kob  & s\bar{d}          & 497.7 & (1/2,+1/2) & -1 \Big. \\ \hline
 K^-  & s\bar{u}          & 493.7 & (1/2,-1/2) & -1 \Big. \\ \hline
\eta  & u\bar{u}+d\bar{d}-2 s\bar{s} & 547.5 & (0,0) &  0 \Big. \\ \hline
\ea \]
\caption{Valence quarks, masses, isospin and strangeness of 
the pseudoscalar octet (for simplicity we have neglected 
$\pi^0$-$\eta$-$\eta'$ mixing).}
\label{tab:ottetto}
\end{table}
 
\begin{figure}[t]
    \begin{center}
       \setlength{\unitlength}{1truecm}
       \begin{picture}(6.0,3.5)
       \epsfxsize 6.  true cm
       \epsfysize 3.5 true cm
       \epsffile{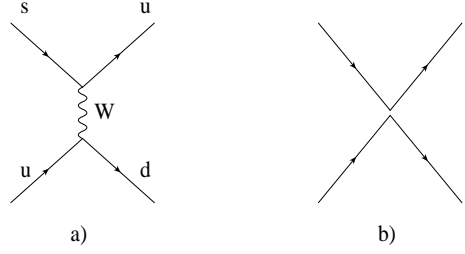}
       \end{picture}
    \end{center}
    \caption{a) Tree--level feynman diagrams 
for $|\Delta S|=1$ transitions, at the lowest order in
$G_F$ and without strong--interaction corrections. b) 
The same diagram in the effective theory $M_W \to \infty$.}
    \protect\label{fig:1}
\end{figure}

\begin{figure}[t]
    \begin{center}
       \setlength{\unitlength}{1truecm}
       \begin{picture}(10.0,2.5)
       \epsfxsize 10.  true  cm
       \epsfysize 2.5  true cm
       \epsffile{fig2a.eps}
       \end{picture}
       \centerline{a)}
       \begin{picture}(10.0,2.5)
       \epsfxsize 10.  true  cm
       \epsfysize 2.5  true cm
       \epsffile{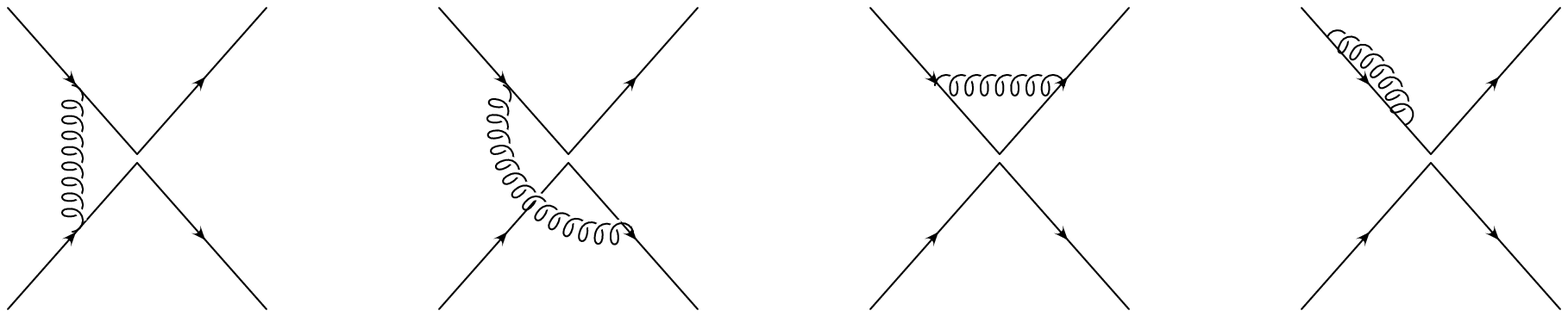}
       \end{picture}
       \centerline{b)}
    \end{center}
    \caption{QCD corrections, of order $g_s^2$, to the diagram 
in fig.~\protect\ref{fig:1}: 
      a) in the full theory; b) in the effective theory.}
    \protect\label{fig:2}
\end{figure}

Kaon decays, i.e. $\vert \Delta S\vert =1$ transitions,
being processes where initial and final state differ for the number 
of $s$ quarks (${\hat S}\ket{s}=-\ket{s}$),
involve the exchange of  at least one $W$ boson (see
Eq.~(\ref{lewm}) and fig.~\ref{fig:1}a).
%
In principle one could hope to calculate these decay amplitudes
 solving separately two problems: i)
the perturbative calculation of the weak transition  amplitude at the
quark level, ii) the non--perturbative calculation of the 
strong transition between  quark and hadron states.
Obviously this separation is not possible, at least in a trivial way,
and is necessary to manage strong interaction effects among initial and 
final state also in weak transitions (\SSf\ref{fig:2}a).
Fortunately,  both $\Lambda_\chi$ and meson masses
are much smaller than the $W$ mass and this 
helps a lot to simplify the problem.

If we neglect the transferred momenta  with respect
to  $M_W$, the  $W$--boson propagator becomes point--like
\beq
{1 \over M_W^2 - q^2 } \to  {1 \over M_W^2}   
\label{limmw}
\eeq 
and the natural scale parameter for the weak amplitudes
is given by the Fermi constant 
\beq
{ G_F \over  \sqrt{2} }={ g^2 \over 8 M^2_W } \simeq 10^{-5} \mbox{GeV}^{-2}.
\eeq 

In purely non--leptonic $|\Delta S|=1$ transitions,
neglecting  strong interaction effects, the amplitudes can be
calculated at the lowest order in $G_F$ using the four--quarks 
effective hamiltonian 
\beq
{\cal H}_{eff}^{|\Delta S|=1} = 
- {4 G_F \over  \sqrt{2}} \left[ \lambda_u (\bar{s}_L \gamma^\mu u_L) 
(\bar{u}_L \gamma^\mu d_L)  + \lambda_c (\bar{s}_L \gamma^\mu c_L) 
(\bar{c}_L \gamma^\mu d_L) + \mbox{\rm h.c.} \right]
\label{hds1},
\eeq
where $\lambda_q = V_{qs}^*V_{qd}$ and $q_L={1\over 2}(1-\gamma_5) q$
(due to their large mass, we neglect, for the moment, the effect
of $b$ and $t$). 
As it will be clear later, it is convenient  to rewrite Eq.~(\ref{hds1}) 
in the following way:
\beq
{\cal H}_{eff}^{|\Delta S|=1} = 
- {4 G_F \over  \sqrt{2}} \sum_{q=u,c} \lambda_q \left( O_q^+ +
O_q^- \right) + \mbox{\rm h.c.},
\eeq
where
\beq
O_q^\pm={1\over2} \left[ (\bar{s}_L \gamma^\mu q_L) (\bar{q}_L \gamma^\mu d_L)  
\pm (\bar{s}_L \gamma^\mu d_L) (\bar{q}_L \gamma^\mu q_L)
\right].
\label{Oqpm}
\eeq

As can be seen from fig.~\ref{fig:2}, 
QCD corrections can be calculated more easily in the
effective theory, i.e. with a point--like  $W$  propagator,
than in the full theory. Indeed, in the first case, feynman diagrams with 
four `full propagators' are reduced to 
diagrams with only three `full propagators',  simplifying 
the calculation. However, the presence of point--like propagators 
induces new ultraviolet divergences, not present in the
full theory, that must be eliminated by  an appropriate
renormalization of the four--quarks operators\cite{GL,AM}$^,$\footnote{~The 
first identity in Eq.~(\protect\ref{ope}) 
follows from Eq.~(\protect\ref{lewm}), 
defining $J_\mu(x)=\bar{U}(x) (V_U\da V_D)$
 $\gamma^\mu\left(1-\gamma_5 \right) D(x)$ and denoting by 
$D^{\mu\nu}_W(x,M_W)$ 
the $W$ propagator in  spatial coordinates.}: 
\beqa
\bra{F}| {\cal H}_{w}^{|\Delta S|=1}  \ket{I} &=& 
{ g^2 \over 8 }\int {\rm d}^4 x
D_W^{\mu\nu}(x,M_W) \bra{F}|T\left( J_\mu(x)J\da_\nu(0) \right) \ket{I} \no\\
&\to&  {G_F \over  \sqrt{2}}  \sum_i C_i(\mu) \bra{F}| O_i(\mu) \ket{I}.
\label{ope}
\eeqa
%
%
This procedure is nothing but an application of the Wilson's
`Operator Product Expansion'\cite{WilsonOPE}, the 
 technique which let us expand a non--local product of operators,
in this case the weak currents, in a series of local terms.
The scale  $\mu$, which appears in Eq.~(\ref{ope}), is a consequence
of the  renormalization procedure which eliminates the 
`artificial' divergences of the effective operators.
The requirement that the product $C_i(\mu)O_i(\mu)$,
and thus all physical observables, be independent of $\mu$,
fix unambiguously the evolution of the coefficients 
$C_i$ as a function of $\mu$:
\beq 
\left[ \delta_{ij} \left( \mu {\partial \over \partial\mu} +  
\beta_s(g_s) {\partial \over \partial g_s }\right) - 
\gamma^T_{ij}(g_s)
 \right] C_j(\mu) =0.
\label{rge1}
\eeq
Eq.~(\ref{rge1}), known as the Callan--Symanzik equation for the 
 coefficients $C_i$, takes this simple form only in a `mass--independent'
regularization scheme\cite{Weinberg73,Hooft2}.
The functions $\beta(g_s)$ and $\gamma_{ij}(g_s)$ 
are defined by
\beq
\beta_s=\mu {\partial \over \partial\mu} g_s(\mu) \qquad \mbox{and} 
\qquad  
\gamma_{ij}={\hat \gamma}_{ij}-2\gamma_{_J}\delta_{ij},  
\eeq
where ${\hat \gamma}_{ij}$ is the anomalous dimension matrix  
of the effective operators\footnote{~Given a set of operators
$O_i$, which mix each other through strong interactions, calling
$Z_{ij}$ the matrix of renormalization constants of such operators 
($O_i^{ren}=Z^{-1}_{ij}O_j$), the anomalous dimension matrix  
is defined by $\gamma_{ij} = Z^{-1}_{ik} \mu {\partial \over \partial\mu}
Z_{kj}$.} and  $\gamma_{_J}$ is the anomalous dimension of the weak 
current.
In addition to Eq.~(\ref{rge1}), which rules the evolution of the  
$C_i$ as a function of $\mu$, in order to use Eq.~(\ref{ope})
is necessary to fix the values of the $C_i$
at a given scale, imposing the identity in the last term 
(`matching' procedure). This scale is typically chosen to  be of the 
order of the  $W$ mass, where the perturbative calculations 
are  much simpler ($g_s(M_W)\ll 1$).\footnote{~For a wider discussion 
about matching conditions and about the integration of
Eq.~(\protect\ref{rge1}) see Ref.\protect\cite{Ciuchini2,Manohar,NewBuras}. } 

\begin{figure}[t]
    \begin{center}
       \setlength{\unitlength}{1truecm}
       \begin{picture}(8.0,4.0)
       \epsfxsize 8.  true cm
       \epsfysize 4.  true cm
       \epsffile{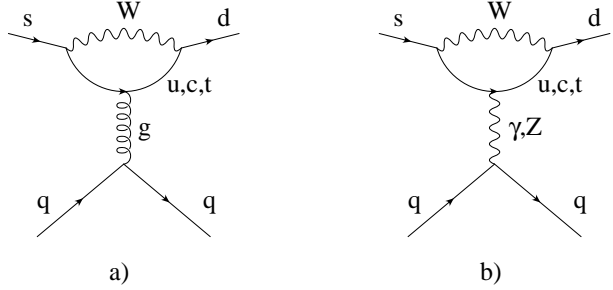}
       \end{picture}
    \end{center}
    \caption{Penguin diagrams: a) `gluon penguin'; b)
                `electromagnetic penguin'.}
    \protect\label{fig:3}
\end{figure}

If we consider only the diagrams in fig.~\ref{fig:2} 
and we neglect the effects due to non--vanishing  light quark masses, 
the operator $O_i$ which appear  in Eq.~(\ref{ope}) are just  
the $O_q^\pm$ of Eq.~(\ref{Oqpm}). These operators are renormalized
only in a multiplicative way, as a consequence 
 $\gamma_{ij}$ is diagonal:
\beq
\gamma_{ij}= 
{g_s^2 \over 4\pi^2} \left[ \ba{cc} \gamma_+ & 0 \\ 0 & \gamma_- \ea \right] =
{g_s^2 \over 4\pi^2} \left[ \ba{cc} +1 & 0 \\ 0 & -2  \ea \right].
\eeq
Since $C_q^\pm(M_W)=\lambda_q[1+O(g_s(M_W))]$, solving Eq.~(\ref{ope})
one finds\cite{GL,AM}\footnote{~For simplicity we neglect the 
effect due to the  $b$ threshold in the integration of 
Eq.~(\protect\ref{rge1}). } 
\beqa
{\cal H}_{eff}^{|\Delta S|=1} &=& 
- {4G_F \over  \sqrt{2}} \sum_{q=u,c} \lambda_q \left\{
\left[ {g_s(M_W) \over g_s(\mu)} \right]^{\gamma_+\over \beta_0} 
O_q^+(\mu)  +
\left[ {g_s(M_W) \over g_s(\mu)} \right]^{\gamma_-\over \beta_0} 
O_q^-(\mu) \right\} + \mbox{\rm h.c.} \qquad \no \\
&\simeq & -{4G_F \over  \sqrt{2}} \sum_{q=u,c} \lambda_q \left\{
\left[ 1-{g_s^2(\mu) \over 4 \pi^2}\log\left({M_W\over\mu}\right)\right]
O_q^+(\mu) \right. \no\\ 
&&\qquad\qquad\quad + \left.
\left[1+2{g_s^2(\mu) \over 4 \pi^2}\log\left({M_W\over\mu}\right)\right]
O_q^-(\mu) \right\} + \mbox{\rm h.c.}, \label{am1}
\eeqa
where $\beta_0={1\over 12}(33-2N_f)$ is defined by 
\beq
\beta_s(g_s)=- {g_s^2\over 4\pi^2}\left(\beta_0 g_s+O(g_s^3)\right).
\eeq

Eq.~(\ref{am1}) is a good approximation to the weak hamiltonian 
for  $m_c <\mu< m_b$~\cite{Bardeen,NeubertStech}. The coefficients
$C_q^\pm(\mu)$ keep track, indeed, of all 
the leading QCD corrections, i.e. of all the terms of
order $g_s(\mu)^{2n} \log(M_W/\mu)^n$.
Now we know that non-perturbative effects
start to play a role for $\mu<m_c$, however historically
people tried anyway to extrapolate the Wilson coefficients 
down to $\mu \sim m_{\rho}$. At this low scales  $C_-$  increases 
substantially ($C_- \simeq 2.2$) while $C_+$ is slightly suppressed. 
Using the factorization hypothesis\cite{Okun} for the 
evaluation of $O_-$ matrix elements, the result 
for  $K\rightarrow 2\pi$ $ \Delta I =1/2$
transitions, though improved by the gluon exchange,
still underestimates the phenomenological amplitudes by a factor 
five. 

Also, Eq.~(\ref{am1}) is not sufficient to study $CP$--violating 
effects: in this case is necessary to consider other operators.
As it is well known, an important
role is played by the so--called `penguin diagrams'\cite{SVZ}$^-$\cite{Buras00}
 (\SSf\ref{fig:3}) which,
though suppressed with respect to those in fig.~\ref{fig:2}, 
give rise to new four--quark operators with different weak phases. 
The suppression of the diagrams in fig.~\ref{fig:3} is nothing but a
particular case of the GIM mechanism\cite{GIM}: 
due to the unitarity of CKM matrix, their contribution vanishes
in the limit $m_u=m_c=m_t$.

The complete set of operators relevant to  
non--leptonic $|\Delta S|=1$ transitions, is 
given by the following 12 dimension--six
terms\cite{Ciuchini3}$^,$\footnote{~The number of independent 
operators decrease to 11 and  7, for  $\mu< m_b$ and  $\mu< m_c$,  
respectively\protect\cite{Ciuchini3}. }: 
\beqa
O_1  &=& (\bar{s}_L^\alpha \gamma^\mu d_{L\alpha}) 
         (\bar{u}_L^\beta  \gamma^\mu u_{L\beta}), \no\\
O_2  &=& (\bar{s}_L^\alpha \gamma^\mu d_{L\beta}) 
         (\bar{u}_L^\beta  \gamma^\mu u_{L\alpha}),  \no\\
O_{3,5}&=& (\bar{s}_L^\alpha \gamma^\mu d_{L\alpha}) \sum_{q=u,d,s,c}
         (\bar{q}_{L,R}^\beta  \gamma^\mu q_{L,R \beta}),  \no\\
O_{4,6}&=& (\bar{s}_L^\alpha \gamma^\mu d_{L\beta}) \sum_{q=u,d,s,c}
         (\bar{q}_{L,R}^\beta  \gamma^\mu q_{L,R \alpha}),  \no\\
O_{7,9}&=& (\bar{s}_L^\alpha \gamma^\mu d_{L\alpha}) \sum_{q=u,d,s,c}
         e_q (\bar{q}_{L,R}^\beta  \gamma^\mu q_{L,R \beta}),  \no\\
O_{8,10}&=& (\bar{s}_L^\alpha \gamma^\mu d_{L\beta}) \sum_{q=u,d,s,c}
         e_q (\bar{q}_{L,R}^\beta  \gamma^\mu q_{L,R \alpha}),  \no\\
O_1^c&=& (\bar{s}_L^\alpha \gamma^\mu d_{L\alpha}) 
         (\bar{c}_L^\beta  \gamma^\mu c_{L\beta}), \no\\
O_2^c&=& (\bar{s}_L^\alpha \gamma^\mu d_{L\beta}) 
         (\bar{c}_L^\beta  \gamma^\mu c_{L\alpha}), 
\label{ciubase}
\eeqa
where $\alpha$ and $\beta$ are the color indices and $e_q$ is the electric 
charge of the quark $q$.
Using the relation $\lambda_u+
\lambda_c+\lambda_t=0$, in the basis~(\ref{ciubase}) the 
weak hamiltonian assumes the following form:
\beqa
{\cal H}_{eff}^{|\Delta S|=1} &=&
- {4 G_F \over  \sqrt{2}}  \Big\{ \lambda_u 
\left[  C_1(\mu)O_1(\mu)+C_2(\mu) O_2(\mu)\right] \no \\  
 - \lambda_u &&
\!\!\!\!\!\!\!\!\!\!\!\!\!\! 
 \left[  C_1(\mu)O_1^c(\mu)+C_2(\mu)O_2^c(\mu)\right]  
-  \lambda_t \sum_{i=3}^{10}
 C_i(\mu) O_i(\mu) \Big\} + \mbox{\rm h.c.}
\label{ciuham}
\eeqa
In Refs.\cite{Buras2,Ciuchini2} the $10\times 10$ 
anomalous dimension matrix of the  coefficients $C_i(\mu)$ 
has been calculated at two loops, including corrections of 
order $\alpha_s^2$, $\alpha_s\alpha_{em}$ and $\alpha_{em}^2$.
Correspondingly the initial condition for the 
$C_i(\mu)$, at $\mu=M_W$, have been calculated including 
terms of order $\alpha_s(M_W)$ and $\alpha_{em}(M_W)$. Using 
these results is  possible to calculate all
the next--to--leading--order corrections to the 
Wilson coefficients of the effective hamiltonian.

After the work of Refs.\cite{Buras2,Ciuchini2} is useless to 
push further the perturbative calculation of the  
 Wilson coefficients. At this point, the main source of error in 
estimating $CP$ violation in kaon decays, is represented by
the non--perturbative evaluation of the hadronic 
matrix elements of Eq.~(\ref{ope}).
In the next subsection, following Ciuchini et al.\cite{Ciuchini3},
we shall see how this problem 
has been solved in $K\to2\pi$ 
using lattice results.

\subsection{$K\to 2 \pi$ parameters 
$\eps$ and $\epsp$.}
\label{sez:SMepsp}

\subsubsection{${\cal H}_{eff}^{|\Delta S|=2}$ 
and the estimate of $\eps$.}

In the Standard Model the value of  $\eps$ cannot be predicted 
but is an important constraint on the CKM phase.
From Eqs.~(\ref{realveps}) and (\ref{defpsw}), 
assuming $\arg(\eps)= \pi/4$ and neglecting
terms of order $|\eps|^2$, follows
\beq
\eps \simeq { e^{i\pi/4} \over 2\sqrt{2} }\left(
 {\Imm M_{12} \over \Real M_{12} }\right)
= - { e^{i\pi/4} \over \sqrt{2} }\left( 
{\Imm M_{12} \over \Delta M_K }\right). 
\label{neps1} 
\eeq
%
\begin{figure}[t]
    \begin{center}
       \setlength{\unitlength}{1truecm}
       \begin{picture}(5.0,2.6)
       \epsfxsize 5.0 true cm
       \epsfysize 2.6 true cm
       \epsffile{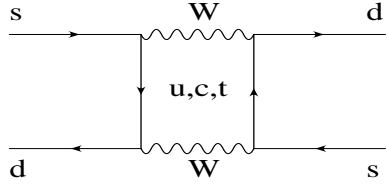}
       \end{picture}
    \end{center}
    \caption{a) Box diagram  
for $|\Delta S|=2$ transitions, without QCD corrections.}
    \protect\label{fig:4}  
\end{figure}
%
In order to calculate $M_{12}$ is necessary to determine
the effective hamiltonian responsible of 
$|\Delta S|=2$ transitions. In this case the situation is much simpler
than in the   $|\Delta S|=1$ case, previously discussed,
since there is only one relevant operator, the one 
created by the box diagram of fig.~\ref{fig:4}.
Thus the effective hamiltonian for $|\Delta S|=2$ transitions
can be written as
\beqa
{\cal H}_{eff}^{|\Delta S|=2} &=& 
{G_F^2 \over 4 \pi^2 } M_W^2 (\bar{s}_L\gamma^\mu d_L)^2 
\big[\lambda_c^2 \eta_1  F(x_c) +  \lambda_t^2 \eta_2 F(x_t) \no \\
&& \qquad\qquad + 2 \lambda_c\lambda_t \eta_3 F(x_c,x_t) \big]+ 
\mbox{\rm h.c.} \label{hds2},
\eeqa
where $F(x_q)$ and $F(x_q, x_j)$
are the Inami--Lim functions\cite{Inami},
$x_q=m_q^2 /M_W^2$, and $\eta_i=1+O(g_s^2)$ are the 
QCD corrections, calculated at the next--to--leading order
in Refs.\cite{Buras0,Herrlich}. Since
%
%
\beq
\Imm(M_{12})={1\over 2M_K} \Imm\left[
\bra{\Kob}|{\cal H}_{eff}^{|\Delta S|=2} \ket{\Ko} \right],
\label{neps2}
\eeq
from the previous equations~(\ref{neps1}--\ref{neps2}) follows 
\beqa
|\eps| &=& {G_F^2 M_W^2  \over 4 \sqrt{2}\pi^2 M_K \Delta M_K } 
A^2\lambda^6\sigma\sin\delta \bra{\Kob}| (\bar{s}_L\gamma^\mu
d_L)^2  \ket{\Ko}
\no\\ &&\qquad  \times
\left[ \eta_3 F(x_c,x_t) -\eta_1 F(x_c) 
+ A^2\lambda^4(1-\sigma\cos\delta) \eta_2 F(x_t) \right] 
\label{epsth}
\eeqa
where $A,\ \lambda,\ \delta$ and $\sigma$ 
are the CKM parameters defined in 
sect.~\ref{sez:CKM}.

Few comments about Eq.~(\ref{epsth}) before going on:
\begin{itemize}
\item{Both the $\eta_i$ and  
the matrix element $\bra{\Kob}| (\bar{s}_L\gamma^\mu d_L)^2 
\ket{\Ko}$ depend  on $\mu$, but their product is scale independent.}
\item{Due to the large value of the top mass, the term proportional 
to $F(x_t)$ is relevant even if it is
suppressed by the factor  $A^2\lambda^4$.}
\item{In order to avoid the calculation of the matrix element
 $\bra{\Kob}| (\bar{s}_L\gamma^\mu d_L)^2 \ket{\Ko}$, 
one could try to evaluate  $\Delta M_K$ using
${\cal H}_{eff}^{|\Delta S|=2}$. However, this is not convenient
since  $\Real(M_{12})$ receive also a  long distance contribution 
that is difficult to evaluate with high accuracy\cite{deRafaelBK}.  }
%
%
\end{itemize}
In Ref.\cite{Ciuchini3} the matrix element has been parametrized in the
following way:
\beq 
\bra{\Kob}| (\bar{s}\gamma^\mu(1-\gamma_5) d)^2  \ket{\Ko}=
{8 \over 3} f_K^2 M_K^2 B_K\alpha_s(\mu)^{6/25},
\label{bkdef}
\eeq
where $f_K=\sqrt{2}F_K=160$ MeV is the $K$--meson decay constant 
(\SSp\ref{cap:CHPT}) and  $B_K$ is a $\mu$--independent 
parameter.\footnote{~If one considers also the next--to--leading 
terms in the $\eta_i$, then Eq.~(\protect\ref{bkdef}) 
must be modified, in order to preserve the $\mu$ invariance
of $B_K$.}  Lattice estimates 
of this matrix element at scales $\mu\sim 2-3$ GeV 
imply\cite{LatticeBK} $B_K=0.75\pm0.15$.
With this result and the  experimental value of  $\eps$, 
Eq.~(\ref{epsth}) imposes two possible solutions for $\cos\delta$, 
with different signs (\SSf\ref{fig:ciufig}).
The negative solution can be eliminated imposing 
additional conditions (coming from lattice
estimates the  $\bar{B}_d-B_d$ mixing) and the
final result of Ref.\cite{Ciuchini3} is:
\beq
\cos\delta=0.47\pm0.32.
\label{cosdeltas}
\eeq
                
\subsubsection{The estimate of $\epsp$.}

As shown in sect.~\ref{sez:2pi}, assuming
the isospin decomposition of $K\to 2 \pi$ amplitudes, 
it follows
\beq
\epsp=i {e^{i(\delta_2-\delta_0)} \over \sqrt{2} } {\omega\over \Real A_0}
 \left[ \omega^{-1} \Imm(A_2) - \Imm(A_0)\right].
\label{epspvvv1} 
\eeq
Actually the decomposition (\ref{definA_0A_2}) is not exactly true,
it receives small corrections due to the mass difference
 $m_u-m_d\not=0$, which breaks isospin symmetry.

The main effect generated by the mass difference among 
$u$ and  $d$ quarks, is to induce a mixing between
the $\pi^0$ ($I=1$) and the two isospin--singlet $\eta$ and $\eta'$.
As a consequence the  transition $\Ko \to \pi^0\pi^0$ can 
occur also through the intermediate state  $\pi^0\eta(\eta')$  
($\Ko \to \pi^0\eta(\eta') \to \pi^0\pi^0$). 
Due to the hierarchy of weak amplitudes
($\Real A_0 \gg \Real A_2$), we can safely neglect  
 isospin--breaking terms in $A_0$. Furthermore, we know that 
$K^0 \to \pi^0 \eta (\eta')$ amplitudes
are $\Delta I=1/2$ transitions, thus the global 
effect of isospin breaking in $\epsp$ can be simply reduced 
to a correction of $\Imm A_2$ proportional 
to $\Imm A_0$\cite{Donoghuek3p}. Following 
Refs.\cite{Donoghuek3p,BurasGerard} we define
\beq
\Imm A_2= \Imm A_2'+\Omega_{IB}\omega \Imm A_0,
\label{ibparam_1}
\eeq
where $A_2'$ is the `pure' $\Delta I=3/2$ 
amplitude (without isospin--breaking terms).
In order to estimate $\Omega_{IB}$ is necessary
to evaluate $\pi^0-\eta-\eta'$ mixing and 
the imaginary parts of 
$K^0 \to \pi^0 \eta (\eta')$ amplitudes. The first 
problem is connected with  the relation among quark and 
meson masses, and can be partially solved in the framework of
Chiral Perturbation Theory (\SSp \ref{subsez:LS}).
On the other hand, the second problem requires the non--perturbative 
knowledge of weak matrix elements. Evaluating these elements 
in the large $N_c$ limit, 
Buras and Gerard\cite{BurasGerard} 
estimated $\Omega_{IB}\simeq 0.25$.
Due to the large uncertainties which affect this estimate, in the 
following we shall assume\cite{LMMR} $\Omega_{IB}=0.25\pm 0.10$.   

Using Eq.~(\ref{ibparam_1}), the expression of
$\epsp$ becomes
\beq
\epsp=i {e^{i(\delta_2-\delta_0)} \over \sqrt{2} } {\omega\over \Real A_0}
 \left[ \omega^{-1} \Imm(A_2')-(1-\Omega_{IB}) \Imm(A_0)\right],
\label{epspvvv}
\eeq
where  $A_0$ and $A_2'$ can be calculated using the 
effective hamiltonian (\ref{ciuham}) 
in Eq.~(\ref{kodec}).\footnote{~Bertolini et al.\cite{Gabrielli}
pointed out that a non--negligible contribution to 
$\epsp$ could be generated by the dimension--5 
gluonic--dipole operator, not included in the basis~(\ref{ciubase}). 
The matrix element of this operator is however suppressed in the chiral 
expansion (see the discussion about the electric--dipole operator
in sect.~\ref{sez:kppgstimecp}) and was overestimated in 
Ref.~\cite{Gabrielli}.  }  
Analogously to the case of $\eps$, is convenient 
to introduce opportune $B$--factors to parametrize 
${\cal H}_{eff}^{|\Delta S|=1}$ matrix elements. 
Following again Ref.\cite{Ciuchini3}  we define: 
\beq
\bra{2\pi,I}| O_i(\mu)  \ket{\Ko} = B_i^{1+I\over 2}(\mu) 
\bra{2\pi,I}| O_i \ket{\Ko}_{VIA},
\label{bfactordef}
\eeq
where $\bra{2\pi,I}| O_i \ket{\Ko}_{VIA}$ indicates the 
matrix element calculated in the  vacuum insertion approximation\cite{Okun}.

\begin{table}[t]
\[ \ba{|c|c|c|} \hline
     &  \sqrt{1 \over 24}\bra{2\pi,0}|O_i\ket{\Ko}_{VIA} &
        \sqrt{1 \over 12}\bra{2\pi,2}|O_i\ket{\Ko}_{VIA}  \\ \hline
O_3  &  +X/3 \big. &  0   \\ \hline
O_4  &  +X                &  0   \\ \hline
O_5  &  -Z/3 \big. &  0   \\ \hline
O_6  &  -Z                &  0   \\ \hline
O_7  &  +2Y/3+Z/6+X/2  &
        +Y/3-X/2 \big.   \\ \hline
O_8  &  +2Y+Z/2+X/6  &
        +Y-X/6 \big.  \\ \hline
O_9  &  -X/3 \big. &  +2X/3 \\ \hline
O_1^c&  +X/3 \big. &  0 \\ \hline
O_2^c&  +X \big. &  0 \\ \hline
\ea \]
\caption{Matrix elements of the four--quarks operators
of ${\cal H}_{eff}^{|\Delta S|=1}$ in the vacuum insertion 
approximation; $X=f_\pi(M_K^2-M_\pi^2)$,
$Y=f_\pi M_K^4/(m_s+m_d)^2$ and $Z=4Y(f_K-f_\pi)/f_\pi$. }
\label{tab:OinVIA}
\end{table}

VIA results for matrix elements which 
contribute\footnote{~$O_1$ and $O_2$ do not
contribute since  $\Imm(\lambda_u)=0$; $O_{10}$ has 
been  eliminated through the relation 
 $O_{10}=O_9+O_4-O_3$ which holds for $\mu<m_b$.}    to
$\Imm(A_0)$ and $\Imm(A_2')$
are reported in table~\ref{tab:OinVIA}, in 
tables~\ref{tab:Ci} and \ref{tab:Bi} we report
 Wilson coefficients and  corresponding $B$--factors
at   $\mu=2$ GeV. The two column of table~\ref{tab:Ci}    
correspond to different regularization schemes:
the 't Hooft--Veltman scheme  (HV)  and the 
naive--dimensional--regularization scheme (NDR). The differences 
among the $C_i$ values in the two tables give an estimate of the
next--to--next--to--leading--order corrections  
which have been neglected.

The dominant contribution to the real parts of $A_0$ and $A_2$  
is generated by $O_1$ and $O_2$. The lattice estimates 
of the corresponding $B$--factors, which must be 
substantially different from one in order to reproduce the observed
$\Delta I=1/2$ enhancement, are affected by large uncertainties 
and are not reported in table~\ref{tab:Ci}. Fortunately this 
uncertainty does not affect the imaginary parts, which in the 
basis~(\ref{ciubase}) are dominated  by 
$O_6$ and $O_8$:
\beqa
\Imm(A_0) &\simeq& {G_F \over \sqrt{2} } \lambda^5 A^2 \sigma\sin\delta
(C_6 B_6^{1/2} Z), \no\\
\Imm(A_2) &\simeq& - G_F \lambda^5 A^2 \sigma\sin\delta
(C_8 B_8^{3/2} Y ).
\label{approxImm}
\eeqa
Using these equations we can derive a simple and interesting 
phenomenological expression (similar to the one
proposed in Ref.\cite{Buras1}) for $\Real(\epsp/\eps)$:
\beq
\Real\left({\epsp\over \eps}\right)\simeq \left(3.0\times10^{-3}\right)
\left[B_6^{1/2}-{\widetilde r}B_8^{3/2}\right]A^2\sigma\sin\delta
=(2.6\pm2.3)\times 10^{-4},
\label{epsfen}
\eeq
where
\beq
{\widetilde r}={\sqrt{2} \over (1-\Omega_{IB})\omega }\left\vert
{C_8 \over C_6} \right\vert\simeq (0.6\pm 0.2).
\label{rtilde}
\eeq
From Eq.~(\ref{epsfen}) it is clear that the weak phases of
$A_0$ and $A_2'$ accidentally tends to cancel each other.
As anticipated in the previous section,
this cancellation is due to the large value $m_t$, which enhance
 $C_8$ (for $m_t\sim 200$ GeV we found  ${\widetilde r}\sim 1$,
whereas for $m_t\sim 100$ GeV, ${\widetilde r}\sim 10^{-1}$). 
Nevertheless, the other essential ingredient of this
cancellation is the `$\Delta I=1/2$ rule', i.e. the dynamical
suppression of  $\Delta I=3/2$ amplitudes with respect to  
$\Delta I=1/2$ ones (the $\omega^{-1}$ factor in Eq.~(\ref{rtilde}) 
is essential for the enhancement of  ${\widetilde r}$).

\begin{table}[t]
\[ \ba{|c|c|c|} \hline
     & \mbox{\rm HV} & \mbox{\rm NDR} \\ \hline
C_1  &  (-3.29\pm0.37\pm0.00)\times10^{-1} & (-3.13\pm0.39\pm0.00)
\times10^{-1}\\ \hline
C_2  &  (104.13\pm0.54\pm0.00)\times10^{-2} & (11.54\pm0.23\pm0.00)
\times10^{-1}\\ \hline
C_3  &  (1.73\pm0.26\pm0.00)\times10^{-2} & (2.07\pm0.33\pm0.00)
\times10^{-2}\\ \hline
C_4  &  (-3.82\pm0.44\pm0.01)\times10^{-2} & (-5.19\pm0.71\pm0.01)
\times10^{-2}\\ \hline
C_5  &  (1.20\pm0.11\pm0.00)\times10^{-2} & (10.54\pm0.16\pm0.02)
\times10^{-3}\\ \hline
C_6  &  (-5.08\pm0.72\pm0.03)\times10^{-2} & (-0.72\pm0.13\pm0.00)
\times10^{-1}\\ \hline
C_7  &  (0.01\pm0.00\pm0.18)\times10^{-3} & (0.01\pm0.04\pm0.20)
\times10^{-3}\\ \hline
C_8  &  (0.77\pm0.12\pm0.12)\times10^{-3} & (0.81\pm0.16\pm0.14)
\times10^{-3}\\ \hline
C_9  &  (-6.71\pm0.27\pm0.68)\times10^{-3} & (-7.49\pm0.15\pm0.75)
\times10^{-3}\\ \hline
\ea \]
\caption{Wilson coefficients of    
${\cal H}_{eff}^{|\Delta S|=1}$ operators,  at $\mu=2$ GeV, 
calculated including next--to--leading--order
corrections in two different regularization 
schemes\protect\cite{Ciuchini3}. The first error is  
due to the uncertainty on $\alpha_s(\mu)$, the second to 
the uncertainty on $m_t$. }
\label{tab:Ci}
\end{table}
\begin{table}
\[ \ba{|c|c|c|c|c|c|} \hline
B_{1c,2c}^{1/2}  & B_{3,4}^{1/2}  & B_{5,6}^{1/2}  & 
B_{7,8,9}^{1/2} &  B_{7,8}^{3/2} & 
B_{9}^{3/2}    \\ \hline
0-0.15^{(*)}   &   1-6^{(*)}  & 1.0\pm 0.2 & 1^{(*)} & 1.0\pm 0.2 &
0.62 \pm 0.10  \\ \hline
\ea \]
\caption{$B$--factors, defined in Eq.~(\protect\ref{bfactordef})
at $\mu=2$ GeV. The entries marked with `*'
are pure `theoretical guesses', whereas the others are obtained by 
lattice simulations\protect\cite{Ciuchini3}.}
\label{tab:Bi}
\end{table}

\begin{figure}
\begin{center}
\epsfxsize=1.00\textwidth
\leavevmode\epsffile{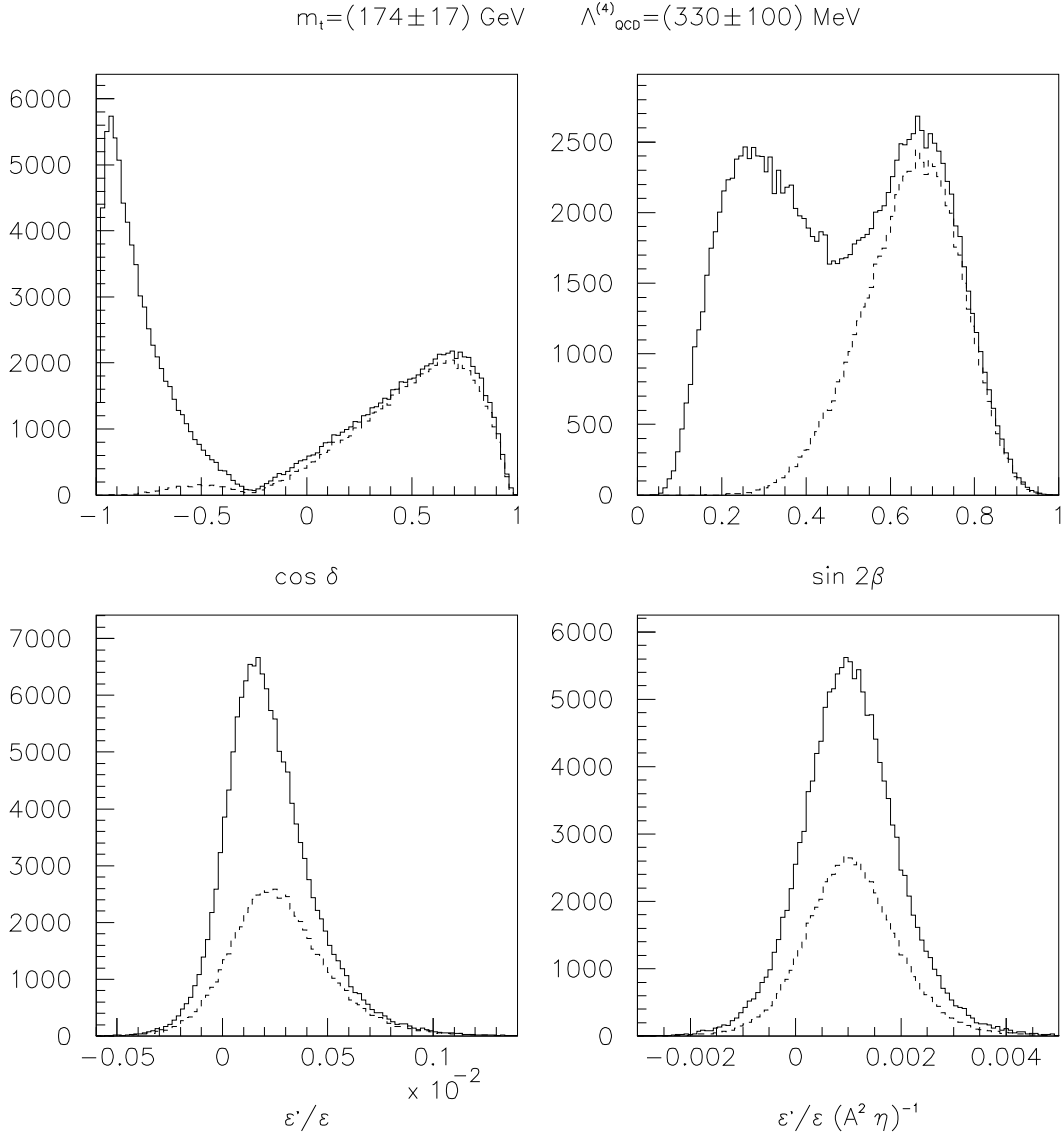}
\caption{ Distributions (in arbitrary units) of $\cos \delta$,
$\sin 2 \beta=2\Imm(V_{td})\Real(V_{td})/|V_{td}|^2$, 
$\epsp/\eps$ and $\epsp/\eps (A^2\eta)^{-1}$ as obtained 
by Ciuchini et al.\protect\cite{Ciuchini3}. Dotted histograms 
have been obtained adding the additional constraint 
coming from $\bar{B}_d-B_d$ mixing. }
\label{fig:ciufig}
\end{center}
\end{figure}

An accurate statistical analysis of the theoretical estimate of
$\Real(\epsp/\eps)$ has been recently carried out by Ciuchini et 
al.\cite{Ciuchini3}. In fig.~\ref{fig:ciufig} we report 
some  results of this analysis. 
Histograms have been obtained by varying, according to their errors, 
all quantities involved in the calculation of 
$\eps$ and $\epsp$: Wilson coefficients, 
$B$--factors, experimental values of  $\alpha_s$ and $m_t$, 
CKM parameters, $\Omega_{IB}$ and $m_s$
(the latter is extracted by lattice simulations). 
The final estimate of  $\Real(\epsp/\eps)$ thus obtained 
is\cite{Ciuchini3}:
\beq
\Real\left({\epsp\over \eps}\right)
=(3.1\pm 2.5)\times 10^{-4},
\label{epspthciu}    
\eeq
in agreement with previous and more recent 
analyses\cite{LMMR,burasepe,NewBuras,Bertolini}. 

Actually, in Ref.\cite{NewBuras} the final error on $\Real(\epsp/\eps)$ is
larger since the various uncertainties have been combined linearly and not 
in a  gaussian way, like in Ref.\cite{Ciuchini3}. A large uncertainty 
has been obtained also in Ref.\cite{Bertolini}, where
the matrix elements have been estimated in a completely different approach.
Nevertheless, all analyses agree on excluding a value of
$\Real(\epsp/\eps)$ substantially larger than $1 \times 10^{-3}$. 

\subsection{$B$ decays.}
\label{sez:Bdecays}
                                    
$CP$ violation in kaon decays is strongly suppressed 
by CKM--matrix  hierarchy: $\epsilon$,  as an example,  is
of order $O(\lambda^4) \sim O(10^{-3})$.
On the other hand, in the decays of $B_d$ and $B_s$ 
mesons
this suppression is avoidable in several cases\cite{Bigi1,Bigi2}
(see the discussion at the end of sect.~\ref{sez:CKM})
and the study of $CP$ violation 
turns out to be more various and promising.
                 
With respect to the $\Ko-\Kob$ system, 
$B_d-{\bar B}_d$ and  $B_s-{\bar B}_s$ systems 
have the following interesting differences:
\begin{itemize}
\item{Due to the large number of initial and final states, 
both even and odd under $CP$, 
the difference of the decay widths is much smaller than the 
 mass difference, i.e.
$|\Gamma_{12}| \ll |M_{12}|$.
As a consequence, according to Eq.~(\ref{vepsdef}), 
the mixing parameters 
$\epsilon_{B_d}$ and $\epsilon_{B_s}$  have  very small  real 
parts.}       
\item{Mass differences are originated by box diagrams 
(similar to the one of fig.~\ref{fig:4}) which, differently from the 
 $\Ko-\Kob$ case, are dominated by top--quark exchange not only in the 
imaginary part but also in the real part.
In the CKM phase convention of Eq.~(\ref{ckma}) one finds: 
\beqa
{1-\epsilon_{B_d} \over 1+\epsilon_{B_d}} &\simeq & {V_{td} \over
V^*_{td} } \doteq e^{-2i\beta}, \label{defbeta} \\
{1-\epsilon_{B_s} \over 1+\epsilon_{B_s}} &\simeq & {V_{ts} \over
V^*_{ts} } \simeq 1.
\eeqa  }     
\end{itemize}
Since real parts of  $\epsilon_{B_d}$ and 
$\epsilon_{B_s}$ are small,  the study of  
$CP$ violation via charge asymmetries of 
semileptonic decays (\SSp\ref{sez:semil}) is not convenient in   
neutral--$B$--meson systems. The best way to observe a $CP$ 
violation in these channels\cite{Bigi1,Bigi2}   is to compare the 
time evolution of the states 
$\ket{B_q(t)}$ and $\ket{\bar{B}_q(t)}$ (states which 
represent, at  $t=0$, $B_q$ and $\bar{B}_q$ mesons)
 in a   final $CP$--eigenstate
 $\ket{f}$ ($CP\ket{f} = \eta_f\ket{f}$):
\beqa
\Gamma( B_q(t) \to f ) & \propto e^{-\Gamma_{B_q} t }
\left[ 1 -\eta_f \lambda_f \sin (\Delta M_{B_q} t) \right], \\
\Gamma( \bar{B}_q(t) \to f ) &\propto  e^{-\Gamma_{B_q} t }
\left[ 1 + \eta_f \lambda_f \sin (\Delta M_{B_q} t) \right].
\eeqa
An evidence of  $\lambda_f\not=0$ necessarily 
implies $CP$ violation, either in the  mixing or in the 
decay amplitudes. As we shall see in the following,
for particular states  
$\ket{f}$,  the CKM mechanism predicts $\lambda_f = O(1)$.

If $CP$ violation was originated only 
at the level of $B_q -\bar{B}_q$ mixing, 
$\lambda_f$ would not depend on the decay channel. On the contrary,
in the Standard  Model $\lambda_f$ 
assumes different values according to the channel. 
In all transitions where only one weak amplitude is
dominant, is possible to factorize strong interaction 
effects and to extract CKM matrix--elements 
{\it independently from the knowledge of hadronic matrix elements}. 
According to the dominant process
at the quark level, 
$\lambda_f$ assumes the following values\cite{Peccei}:
\beq \ba{ll}  
\lambda_f= \sin 2\beta          &\qquad B_d,\ b\to c \\
\lambda_f= \sin 2(\beta+\delta)\doteq \sin 2\alpha &\qquad B_d,\ b\to u \\
\lambda_f= 0        &\qquad B_s,\ b\to c \\
\lambda_f= \sin 2\delta &\qquad B_s,\ b\to u \ea
\label{abc}
\eeq
where $\delta$ is the  CKM phase in the 
parametrization (\ref{ckma}) and $\beta$, already introduced in
Eq. (\ref{defbeta}), is given by:
\beq
\tan \beta = {\sigma \sin\delta \over 1 - \sigma \cos\delta } =
{\eta \over 1 - \rho}.
\eeq

\begin{figure}[t]
    \begin{center}
       \setlength{\unitlength}{1truecm}
       \begin{picture}(10.0,6.0)
       \epsfxsize 10.0  true cm
       \epsfysize 6.0 true cm
       \epsffile{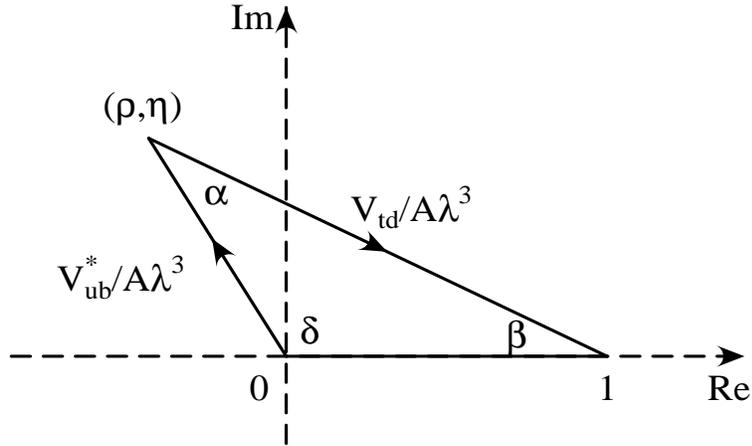}
       \end{picture}
    \end{center}
\caption{Unitarity triangle in the complex plane.} 
\label{fig:triangle}
\end{figure}

The phases $\alpha$, $\beta$ and $\delta$, which appear in 
Eq.~(\ref{abc}), 
have an interesting phenomenological interpretation:
are the angles of the so--called `unitarity  triangle'
(\SSf\ref{fig:triangle}). Indeed, from  CKM--matrix unitarity, 
which implies
\beq
V_{ub}^* V_{ud} + V_{cb}^* V_{cd} +  V_{tb}^* V_{td} = 0, 
\label{notdegent}
\eeq
i.e., at the leading order in $\lambda$, 
\beq
V_{ub}^* +  V_{td} = A\lambda^3, 
\eeq
follows the relation
\beq
\alpha+\beta+\delta = \pi.
\eeq
Obviously, CKM--matrix unitarity impose also other constraints,
in addition to Eq.~(\ref{notdegent}), which can be re--formulated in terms of
different triangles. However, the one in  fig.~\ref{fig:triangle}
is the most interesting one since
the three sizes are of the same order in $\lambda$
and thus the triangle is not degenerate.
 
Limits on  CKM--parameter $\rho$ and $\eta$ 
(\SSt\ref{tab:CKMpar}), coming both from  $K$ and $B$ physics, 
put some constraints on the  angles 
$\alpha$, $\beta$ and $\delta$. Recent correlated 
analyses of such limits\cite{Ciuchini3,PecceiW}
lead to the conclusion  that, whereas  $\sin 2\delta$ and  
$\sin 2\alpha$ can vanish, 
$\sin 2\beta$ is necessarily different from zero and possibly  
quite large:
\beq
0.23 \leq \sin 2\beta \leq   0.84.
\eeq
Fortunately, the measurement of  $\sin 2\beta$ 
is also the most accessible from the experimental point of view. 
Indeed, the decay  $B_d \to \Psi K_{S,L}$ is 
dominated\footnote{~Actually there is 
also a small contribution coming from the 
penguin diagram ${\bar b}\to s q{\bar q}$, however this term has the
same weak phase (zero in the standard 
CKM parametrization) of the dominant one.} 
by the tree--level process ${\bar b}\to c{\bar c}s$, thus, according to
Eq.~(\ref{abc}), from this decay is possible to extract in a clean way 
$\sin 2\beta=\lambda_{\Psi K}$.
The measurement of $\lambda_{\Psi K}$ is one of the main goal of
next--generation high--precision experiments on  $B$
decays and will represent  a
fundamental test for the 
CKM mechanism of $CP$ violation.

The measurements of the other two phases ($\alpha$ and $\delta$), 
very interesting from the theoretical point of view, both to exclude new  
`super--weak' models (models where $CP$ is generated only by 
$|\Delta S|=2$ and $|\Delta B|=2$ interactions) and to 
test CKM--matrix unitarity (limiting the presence of 
new quark families), are much more difficult.
In the case of $\alpha$, for instance,  
the most promising channel is the decay
$B_d \to \pi\pi$, dominated by  the tree--level transition
${\bar b}\to u{\bar u}d$, but the very small branching ratio and
the  contamination of penguin diagrams with different weak 
phases\cite{Nir}, makes the 
measurement of $\lambda_{\pi\pi}$ quite difficult  and the 
successive extraction of  $\sin 2\alpha$ not very clean.
A detailed analysis of all the processes that can be 
studied in order to measure these phases is beyond the 
purpose of this article, we refer the reader to the numerous 
works on this subject which are present in the literature  
(see e.g. Refs.\cite{Winstein,Peccei,Nir}$^-$\cite{Ali} 
and references cited therein).

\setcounter{equation}{0}
\setcounter{footnote}{0}
\section{Chiral Perturbation Theory.}
\label{cap:CHPT}

As already stated in the previous section,  
color interactions between quarks and gluons are  
non perturbative at low energies, and the confinement 
phenomenon is probably the most evident consequence of this
behaviour. Nevertheless, from the experimental point of view  
is known  that at very low energies 
 pseudoscalar--octet mesons (\SSt \ref{tab:ottetto}) 
interact weakly, both among themselves and with nucleons.
Therefore it is reasonable to expect that with a suitable
choice of degrees of freedom
QCD can be treated perturbatively even at low energies.
Chiral Perturbation Theory\cite{Weinberg79}$^-$\cite{GL2} (CHPT), using  
the pseudoscalar--octet mesons as degrees of freedom,
has exactly this goal.\footnote{~For excellent reviews on 
CHPT and, in  particular, for its applications in kaon dynamics see
Refs.\protect\cite{deRafael2,handbook,Donoghuebook,Burasbook}$^-$\protect\cite{Ecker}. }

Neglecting light quark masses, the QCD lagrangian 
\beq
{\widetilde {\cal L}}_{QCD}=\sum_{q=u,d,s} {\bar q} \gamma^\mu \left(
i\partial_\mu - {g_s} {\lambda_a \over 2} G_\mu^a  \right) q
 -{1\over 4} G^a_{\mu\nu} G^{a\mu\nu}+O(\mbox{\rm heavy quarks}),
\eeq
a part from the local invariance under $SU(3)_C$, possesses a global 
invariance under 
 $SU(N_{ql})_L\times SU(N_{ql})_R\times U(1)_V\times U(1)_A$,
where $N_{ql}=3$ is the number of massless quarks. The 
$U(1)_V$ symmetry, which survives also in the case of non--vanishing
quark masses, is exactly conserved and its generator is the 
barionic number. On the other hand, the $U(1)_A$  symmetry is 
explicitly broken at the quantum level by the abelian 
anomaly\cite{Hooft76,CDG}. 
 $G=SU(3)_L\times SU(3)_R$ is the group of chiral transformations:
\beq
\psi_{_{L,R}} \Gto g_{_{L,R}} \psi_{_{L,R}}, 
\qquad \mbox{\rm where}\qquad
\psi=\left( \ba{c} u \\ d \\ s \ea\right)\qquad \mbox{\rm and}\qquad
g_{_{L,R}}\in G,
\eeq
spontaneously broken by the quark condensate 
($\bra{0}|\bar{\psi}\psi\ket{0}\not=0$). 
The subgroup which remains unbroken after the 
breaking of  $G$ is $H=SU(3)_V\equiv SU(3)_{L+R}$ (the famous
$SU(3)$ of the `eightfold way'\cite{Gellmann}); as a consequence 
the coset space $G/H$ is isomorphic to $SU(3)$.

The fundamental idea of CHPT is that, in the 
limit $m_u=m_d=m_s=0$
(chiral limit), pseudoscalar--octet mesons are 
Goldstone bosons generated by the 
spontaneous breaking of $G$ into $H$. Since 
Goldstone fields can be always re-defined in such a way
that interact only through derivative 
couplings\cite{Goldstone}, this 
hypothesis justify the soft  behaviour of  
 pseudoscalar interactions at low  energies. If these mesons 
were effectively  Goldstone bosons, they should be massless, 
actually this is not the case due to the light--quark--mass
terms
which  explicitly  break  $G$. Nevertheless, since
$m_{u,d,s} < \Lambda_\chi$, is natural to expect that these 
breaking terms can be treated as small 
perturbations. The fact that pseudoscalar--meson 
masses are much smaller  
than the typical hadronic scale ($M^2_\pi/\Lambda_\chi^2 \ll 1$)  
indicates that also this hypothesis is reasonable.  
Summarizing, the two basic assumptions of CHPT are\cite{Ecker}:

{\bf [1]} {\it In the chiral limit, 
pseudoscalar--octet mesons are Goldstone bosons
originated by the spontaneous breaking of $G$ into $H$. }

{\bf [2]} {\it The mass terms of  light quarks 
can be treated as perturbations.}  

According to these hypotheses, to describe QCD interactions 
of pseudoscalar mesons is necessary to 
consider the most general lagrangian  invariant under $G$, 
written in terms of Goldstone--boson fields, and add to it
the breaking terms,  which transform linearly under 
$G$.\cite{Weinberg79}
The problem of this approach is that the lagrangian  built in this
way is non renormalizable and thus contains an infinite number 
of operators. Nevertheless, as we shall see in the following, 
in the case of low energy processes  ($E<\Lambda_\chi$), 
the error done by considering only a finite number of such
operators is under control (of order $(E/\Lambda_\chi)^n$) .

\subsection{Non--linear realization of $G$.}
\label{sez:nonlinear}

Goldstone--boson fields parametrize the 
coset space $G/H$ and thus do not transform linearly under  $G$. 
The general formalism to construct invariant  operators, or
operators which transform linearly, in terms of the Goldstone--boson
fields of a spontaneously broken (compact, connected and semisimple) 
symmetry, has been analyzed in detail 
by Callan, Coleman, Wess and 
Zumino\cite{CWZ,CCWZ}. In this subsection
we shall only illustrate the application of this formalism to 
the chiral symmetry case. 

First of all is possible to define a unitary matrix 
$u$ ($3\times 3$), which depends on the Goldstone--boson fields ($\phi_i$),
and which transforms in the following way: 
\beqa 
u(\phi_i) &\Gto& g_{_R} u(\phi_i) h^{-1}(g,\phi_i) =
h(g,\phi_i) u(\phi_i) g_{_L}^{-1} \no\\
u(\phi_i)\da &\Gto& g_{_L} u(\phi_i)\da h^{-1}(g,\phi_i) =
h(g,\phi_i) u(\phi_i)\da g_{_R}^{-1}, 
\eeqa
where $h(g,\phi_i)$, the so--called `compensator--field', is an 
element of the subgroup $H$. If $g\in H$, $h$ 
is a unitary matrix, independent of the
$\phi_i$, which furnishes a linear representation of $H$:
if $\Psi$ is a  matrix  which transforms linearly under 
$H$, then
\beq
\Psi_i \Gto  h(g,\phi_i) \Psi_i  h^{-1}(g,\phi_i).
\eeq
There are different parametrizations of $u$ in terms of
the fields  $\phi_i$, which correspond to different choices
of coordinates in the coset space $G/H$.
A convenient parametrization is the exponential 
parametrization:\footnote{~We denote by $\eta_8$ the  
 octet component of the $\eta$ meson.}              
\beqa
u^2 = U &=& e^{i\sqrt{2}\Phi/F}, \no\\
  \Phi &=& {1\over \sqrt{2} } \sum_i\lambda_i\phi^i 
= \left[ \ba{ccc}\dis{\pi^0\over \sqrt{2}}+\dis{\eta_8 \over \sqrt{6}}&\pi^+ &
 K^+ \\ 
\pi^- & -\dis{\pi^0\over\sqrt{2}}+{\eta_8 \over \sqrt{6}} & \Ko \\ 
K^- & \Kob & -\dis{2\eta_8 \over \sqrt{6}} 
\ea \right],
\eeqa
where $F$ is a dimensional constant (dim$[F]$=dim$[\Phi]$)
that, as we shall see, can be related to the pseudoscalar--meson
decay constant. Note that $U \Gto g_R U g_L^{-1}$. 

Successively, it is convenient to introduce the following derivative 
operators:
\beq       
u_\mu \doteq i(u\da \partial_\mu u  - u \partial_\mu u\da) = iu\da\partial_\mu Uu\da
= u_\mu\da    
\qquad\qquad   u_\mu \Gto h u_\mu h\da,   
\eeq  
which transforms like $\Psi$, and 
\beq
\Gamma_\mu \doteq {1\over 2}(u\da \partial_\mu u  + u \partial_\mu u\da) 
= -\Gamma_\mu\da \qquad\qquad\qquad  \Gamma_\mu \Gto h \Gamma_\mu h\da +
h \partial_\mu h\da,  
\eeq
which let us build the covariant derivative of  $\Psi$: 
\beq
\DD_\mu \Psi = \partial_\mu  \Psi - [\Gamma_\mu, \Psi]. 
\eeq 
With these  definitions is very simple to  construct the 
operators we are interested in:
if $A$ is any operator which  transforms linearly under 
$H$ (like $\Psi$, $u_\mu$ and their covariant derivatives), 
then $u A u\da$ and $u\da A u$ transforms linearly
under $G$, whereas their trace is invariant:
\beqa 
u A u\da &\Gto&  g_{_R} (u A u\da) g_{_R}^{-1}, \no\\ 
u\da A u &\Gto&  g_{_L} (u\da A u) g_{_L}^{-1}.  
\eeqa  

\subsection{Lowest--order lagrangians.}
\label{sez:Lp2}

In absence of external fields, the invariant operator which contains the 
lowest number of derivatives
is unique:\footnote{~We denote by $\la A\ra$ 
the trace of $A$.}
$\la u_\mu u^\mu \ra=\la \partial_\mu U \partial^\mu U\da \ra$. 
Fixing the coupling constant of this operator 
in order to get the kinetic term of spin--less fields, 
leads to: 
\beq
{\widetilde {\cal L}}_S^{(2)} = {F^2 \over 4} 
\la \partial_\mu U \partial^\mu U\da \ra= {1\over 2} \partial_\mu \Phi   
\partial^\mu \Phi + O(\Phi^4)  .
\eeq
This lagrangian is the chiral realization, at the lowest 
order in the derivative expansion, of
${\widetilde {\cal L}}_{QCD}$.

To include explicitly breaking terms, 
and to generate in a systematic way Green functions of quark currents,  
is convenient to modify  ${\widetilde {\cal L}}_{QCD}$,
i.e. the QCD lagrangian in the chiral limit,                         
coupling  external sources to quark currents.
Following the work of  Gasser and  Leutwyler\cite{GL1,GL2},  
we introduce the sources $v_\mu$, $a_\mu$, $\sors$ and $\sorp$,
 so that 
\beqa
r_\mu = v_\mu + a_\mu  &\Gto &  g_{_R} r_\mu g_{_R}^{-1}, \no \\
l_\mu = v_\mu - a_\mu  &\Gto &  g_{_L} l_\mu g_{_L}^{-1}, \no \\ 
\sors + i\sorp   &\Gto &  g_{_R} (\sors+i\sorp)g_{_L}^{-1}, \no \\ 
\sors - i\sorp   &\Gto &  g_{_L} (\sors-i\sorp)g_{_R}^{-1},
\label{trasfGglob}  
\eeqa
and we consider the lagrangian
\beq 
{\cal L}_{QCD}(v,a,\sors,\sorp) = {\widetilde {\cal L}}_{QCD}
+\bar{\psi} \gamma^\mu (v_\mu +a_\mu \gamma_5) \psi 
-\bar{\psi} (\sors - i \sorp \gamma_5) \psi. 
\label{lqcdvasp}
\eeq
By this way we achieve two interesting results\cite{Ecker}:

\begin{itemize}
\item{} The generating functional 
\beq
e^{i Z(v,a,\sors,\sorp) } = \int {\cal D}q  
{\cal D}\bar{q}{\cal D}G \ e^{ i  \int {\rm d}^4x
{\cal L}_{QCD}(v,a,\sors,\sorp) } 
\label{zvasp}
\eeq
is explicitly invariant under chiral transformations, but 
the explicit breaking of $G$  can be nonetheless obtained  
by calculating the  Green functions, i.e. the functional derivatives of 
$Z(v,a,\sors,\sorp)$, at 
\beq
v_\mu=a_\mu=\sorp=0 \qquad\qquad \sors=M_q=\mbox{\rm diag}(m_u,m_d,m_s).
\eeq
\item{} The global symmetry $G$ can be promoted to a local 
one modifying the transformation laws of 
$l_\mu$ and $r_\mu$ in
\beqa
r_\mu = v_\mu + a_\mu  &\Gto &  g_{_R} r_\mu g_{_R}^{-1}
+i g_R\partial_\mu g_R^{-1}, \no \\
l_\mu = v_\mu - a_\mu  &\Gto &  g_{_L} l_\mu g_{_L}^{-1}
+i g_L\partial_\mu g_L^{-1}.  
\label{trasfGloc}  
\eeqa
By this way the gauge fields  of electro--weak interactions
 (\SSp\ref{cap:SM})
are automatically included in 
$v_\mu$ and $a_\mu$: 
\beqa
v_\mu &=& -eQA_\mu - {g\over \cos\theta_W }\left[ Q\cos(2\theta_W)
-{1\over6} \right]  Z_\mu                          
- {g \over 2\sqrt{2} }\left( T_+ W^+_\mu +
\mbox{\rm h.c.}\right), \no \\ 
a_\mu &=& + {g\over \cos\theta_W } \left[ Q -{1\over6} \right]  Z_\mu
+ {g \over 2\sqrt{2} }\left( T_+ W^+_\mu +
\mbox{\rm h.c.}\right), 
\label{AWfields}   
\eeqa
\beq
\mbox{\rm with}\qquad
Q={1\over 3}\left(\ba{ccc} 2 & 0 & 0 \\ 0 & -1 & 0 \\ 0 & 0 & -1 \ea \right)
\qquad\mbox{\rm and}\qquad
T_+=\left(\ba{ccc} 0 & V_{ud} & V_{us} \\ 0 & 0 & 0 \\ 0 & 0 & 0 \ea \right). 
\eeq
As a consequence, Green functions for processes with
external photons, $Z$ or $W$ bosons,
can be simply obtained as 
functional derivatives of $Z(v,a,\sors,\sorp)$.
\end{itemize}
The chiral realization of ${\cal L}_{QCD}(v,a,\sors,\sorp)$, at the 
lowest order in the derivative expansion, is obtained by 
${\widetilde {\cal L}}_S^{(2)}$ including  external sources
in a chiral invariant way. Concerning spin--1 sources
this is achieved by means of the  `minimal substitution':
\beq
\partial_\mu U \to  D_\mu U = \partial_\mu U -ir_\mu U +i Ul_\mu.
\eeq
Non--minimal couplings, which could be built with 
the tensors
\beqa
F_L^{\mu\nu} &=& \partial^\mu l^\nu - \partial^\nu l^\mu -i\left[
l^\mu, l^\nu \right], \no \\                                    
F_R^{\mu\nu} &=& \partial^\mu r^\nu - \partial^\nu r^\mu -i\left[
r^\mu, r^\nu \right], 
\eeqa                                   
are absent at the lowest order since\footnote{~From now on, we
shall indicate with $O(p^n)\sim O(\partial^n\phi)$
terms of order $n$ in the derivative expansion.}
\beqa
U       &\qquad\qquad& O(p^0),  \no \\
u^\mu,a^\mu,v^\mu &\qquad\qquad& O(p^1), \no \\
F^{\mu\nu}_{L,R} &\qquad\qquad& O(p^2).
\label{count1}
\eeqa

Regarding spin--0 sources, it is necessary to  
establish which is the order, in the derivative expansion, of 
$\sors$ and $\sorp$. The most natural choice  is given by\cite{GL1,GL2}:
\beq
\sors,\sorp     \qquad\qquad O(p^2).
\label{sceltasp}
\eeq
As we shall see in the following, this choice is well justified a 
posteriori by the Gell--Mann--Okubo relation.

\subsubsection{The strong lagrangian.}
\label{subsez:LS}

Now we are able to write down the most general lagrangian
invariant under $G$, of order $p^2$, which includes
pseudoscalar mesons and external sources:
\beq
{\cal L}_S^{(2)} = {F^2 \over 4} 
\la D_\mu U D^\mu U\da + \chi U\da + U\chi\da \ra, \qquad
\quad \chi \doteq 2B(\sors+i\sorp).
\label{ls2}
\eeq

The two  arbitrary constants (not fixed by the symmetry) $F$ and $B$,  
which appear in ${\cal L}_S^{(2)}$, are related to two fundamental 
quantities: the pion decay constant $F_\pi$, defined by
\beq
\bra{0}| \bar{\psi} \gamma^\mu \gamma_5 \psi \ket{\pi^+(p)} \doteq  
i \sqrt{2} F_\pi p^\mu,
\eeq
and the quark condensate $\bra{0}|\bar{\psi} \psi \ket{0}$.
Indeed, differentiating with respect to the external sources, we obtain
\beqa
F_\pi &=& -i {p_\mu \over \sqrt{2} p^2 } 
\bra{0}| {\delta {\cal L}_S^{(2)} \over \delta a_\mu }  \ket{\pi^+(p)} 
= F, \label{chiralF}  \\
\bra{0}|\bar{\psi} \psi \ket{0} &=& -
\bra{0}| {\delta {\cal L}_S^{(2)} \over \delta \sors }  \ket{0} 
= -F^2 B.\label{chiralB}
\eeqa
It is important to remark that relations  (\ref{chiralF}--\ref{chiralB}) 
are exactly valid only in the chiral limit, 
in the real case ($m_q\not=0$) are modified at order $p^4$
(\SSp \ref{sez:Fgenp4}).  

The pion decay constant is experimentally  known from the process
 $\pi^+\to \mu^+ \nu$: $F_\pi=92.4$ MeV, on the other hand 
the products $Bm_u$, $Bm_d$ and $Bm_s$  are fixed by 
the following identities:
\beqa
M_{\pi^+}^2 &=& (m_u+m_d)B, \no\\
M_{K^+}^2 &=& (m_u+m_s)B,  \no\\
M_{\Ko}^2 &=& (m_d+m_s)B.  \label{masseP} 
\eeqa
The analogous equation for $M_{\eta_8}^2$ contains no free parameter 
and give rise to a consistency relation: 
\beq
3M_{\eta_8}^2=4M_{K}^2-M_{\pi}^2,
\label{GOkuborel}
\eeq
the well--known  Gell--Mann--Okubo relation\cite{Gellmann,Okubo}.   

Assuming that the quark condensate does not vanish in the chiral 
limit, i.e. that  $B(m_q\to 0)\not= 0$, from  relations
(\ref{masseP}) it is easier to understand why we have chosen 
$\sors\sim O(p^2)$.  This choice is justified a posteriori 
by Eq.~(\ref{GOkuborel}) and a priori by  
lattice calculations of the ratio $B/F$ (see references 
cited in Ref.\cite{Ecker}), nevertheless it is  important to remark that 
it is an hypothesis which go beyond the fundamental assumptions of 
CHPT. Eq.~(\ref{sceltasp}) has also the big advantage 
to facilitate the power counting in the derivative expansion 
(this choice avoids Lorentz--invariant terms of order $p^{2n+1}$). 
The approach of Stern and Knecht\cite{Stern},  
i.e. the hypothesis that the quark condensate could be very 
small, or even vanishing, in the chiral limit  (so that 
$O(m_q^2)$ corrections to Eqs.~(\ref{masseP}) cannot be neglected)
gives rise to a large number 
of new operators for any fixed power of $p$ and strongly reduces
the predictive power of the theory.\footnote{~See 
Ref.\protect\cite{Ecker} for a complete discussion 
about the  relation between  `standard' CHPT 
($\sors \sim O(p^2)$) and `generalized' CHPT  ($\sors \sim O(p)$).} 

\subsubsection{The  non--leptonic  weak lagrangian.}
\label{subsez:LW}

The lagrangian (\ref{ls2}) let us to calculate at order $p^2$ 
Green functions for weak and  electromagnetic transitions, 
beyond the strong ones, in processes with external gauge fields, like 
semileptonic kaon decays. However, the lagrangian  ($\ref{ls2}$) 
is not sufficient to describe non--leptonic decays of  $K$ 
mesons, since, as shown in sect.~\ref{sez:4ferm}, in this case 
is not possible to trivially factorize strong--interaction effects.
The correct procedure to describe these processes, 
is to build the chiral realization of the effective hamiltonian
(\ref{ciuham}).
                      
Under $SU(3)_L\times SU(3)_R$ transformations, the operators of 
Eq.~(\ref{ciubase}) transform linearly in the following way:
\beq \ba{lcl}
O_1, O_2, O_9 &  & (8_L,1_R) + (27_L,1_R),   \\	
O_1^c, O_2^c, O_3, O_4, O_5, O_6 & \qquad & (8_L,1_R),  \\	
O_7, O_8  &  & (8_L,8_R).  \ea
\label{trasfOi}	
\eeq
Analogously to the case of light--quark mass terms,
chiral  operators which transform like the  $O_i$ can be built 
introducing appropriate scalar  sources. 
As an example, to build the  $(8_L,1_R)$ operators, 
we introduce the  source  
\beq
{\hat \lambda}  \Gto   g_{_L} {\hat \lambda}  g_{_L}^{-1}
\eeq
and we consider all the operators, invariant under $G$, linear  
in $\lambda$ (operators bilinear  in $\lambda$ correspond  to terms 
of order  $G_F^2$ in the effective hamiltonian). 
Successively, fixing the source 
to the constant value 
\beq
{\hat \lambda}  \to \lambda={1\over 2}(\lambda_6-i\lambda_7),
\eeq
we select the $\Delta S=1$ component of all
 possible $(8_L,1_R)$ operators. For 
$(27_L,1_R)$ terms the procedure is very similar,
the only change is the source structure. 
On the other hand for $(8_L,8_R)$ operators,
generated by electromagnetic--penguin diagrams
(\SSp\ref{sez:4ferm}), is necessary to introduce two sources,
corresponding to charged and  
neutral currents. The lowest order operators obtained by this procedure
are\cite{Cronin2,GRW}:
\beq 
\ba{lcll}
W^{(2)}_8    = \la \lambda L_\mu L^\mu \ra  & \qquad & (8_L,1_R) \qquad 
& O(p^2), \\
W^{(2)}_{27} = (L_\mu)_{23} (L^\mu)_{11} 
   + {2\over 3}(L_\mu)_{21} (L^\mu)_{13}    &  & (27_L,1_R)
& O(p^2),  \\
W^{(0)}_{\underline{8}} = F^2 \la \lambda U\da Q U \ra  &   & (8_L,8_R)
& O(p^0),  \ea
\eeq  
where $L_\mu=u\da u_\mu u$. Whereas   singlets under $SU(3)_R$ are
of order $p^2$, the  $(8_L,8_R)$ operator is of order $p^0$.
This however is not a problem, since  
electromagnetic--penguin operators at the quark level ($O_7$ and 
$O_8$ in the basis (\ref{ciubase})) 
are suppressed by a factor $e^2$ with respect to the dominant terms.
Thus the chiral lagrangian for $|\Delta S|=1$ non--leptonic transitions,
at order  $(G_Fp^2e^0)+(G_Fp^0e^2)$, is given by:  
\beq
{\cal L}_W^{(2)} = {G_F\over \sqrt{2}} \lambda_u  F^4 \left[ 
\sum_{i=8, 27 }  g_i W_i^{(2)} +  g_{\underline{8}} W_{\underline{8}}^{(0)}
\right ] + \mbox{\rm h.c.}
\label{lagr2w}
\eeq

The three constants $g_i$ which appear in ${\cal L}_W^{(2)}$ are not
fixed by chiral symmetry but is natural to expect them to be 
of the order of the Wilson coefficients of table~\ref{tab:Ci}.
The $g_i$ are real in the  limit where  $CP$ is an 
exact symmetry.

In principle, the  $g_i$ could be determined either by 
comparison with experimental data, on
$K\to 2\pi$ or $K\to 3\pi$, or by  
by comparison with theoretical estimates, coming from 
lattice QCD or other non--perturbative 
approaches\cite{deRafaelBK,Bardeen}$^-$\cite{Bertoldelta}.
In practice, the choice is reduced because:  i) 
there are no experimental information on the imaginary 
parts of the weak amplitudes;
ii) lattice calculations for the real parts of $K\to 2\pi$
amplitudes are still not reliable (the estimates are
dominated by the large errors on the $B$-factors of 
$O_1$ and $O_2$); iii) all the other non--perturbative 
approaches are affected 
from large theoretical uncertainties.
As a consequence, in our opinion the best  choice
to determine the $g_i$ is
to fix the real parts by comparison with 
experimental data (those on $K\to 2 \pi$ for simplicity),
and to fix the imaginary parts by comparison with 
lattice calculations\cite{IMP}.
   
Using the lagrangian~(\ref{lagr2w}) at tree level,
from Eqs.~(\ref{definA_0A_2}) and~(\ref{ibparam_1})  
follows:\footnote{~In the 
following we will neglect isospin--breaking effects but in 
$\epsp/\eps$ (\SSp\protect\ref{sez:SMepsp}), since  there are not 
sufficient data to systematically analyze isospin breaking beyond 
$K\to 2\pi$\protect\cite{Burasbook,KMWlett}.}
\beqa
A_{0}  &=& \sqrt{2} F \left[ \left(M_K^2 -M_\pi^2\right)\left(G_8+{1\over 9} 
G_{27} \right) - {2\over 3} F^2 G_{\underline{8}} \right], \label{A_01}\\
A_{2}' &=& F \left[ {10 \over 9} G_{27} \left(M_K^2-M_\pi^2 \right) 
- {2\over 3} F^2 G_{\underline{8}} \right],
\label{A_21}
\eeqa  
where, for simplicity, we have introduced the dimensional
couplings  $G_i=G_F
\lambda_u g_i/$ $\sqrt{2}$. 
Neglecting the contribution of the  
$(8_L, 8_R)$ operator,\footnote{~As can be seen 
from Eq.~(\protect\ref{ciuham}), the contribution of $(8_L,8_R)$
operators in the real parts is completely negligible.}
the comparison with the experimental data~(\ref{k2pexp1}--\ref{k2pexp2})
leads to:  
\beqa
|G_8| &=& 9.1 \times 10^{-6}\ \mbox{\rm GeV}^{-2}, \label{G_8xx} \\
g_{27}/g_8  &=& {9 \sqrt{2} \omega  \over 10}  = 5.7 \times 10^{-2}.
\eeqa 
Note that the 
value of $g_8$ corresponding to (\ref{G_8xx}) is about five times 
larger than the one obtained from (3.27) in the factorization  
hypothesis\cite{deRafael2}.

Regarding the imaginary parts, the comparison with 
the results shown in sect.~\ref{sez:SMepsp} 
leads to\cite{IMP}:
\beqa
\Imm g_8 &=& \Imm\left( {\lambda_t \over \lambda_u }\right) 
\left[ {C_1\over 3} B_{1c}^{1/2} + {C_2} B_{2c}^{1/2} + 
       {C_3\over 3} B_{3}^{1/2} + {C_4} B_{4}^{1/2} \right. \no\\
 && \left.
\qquad\qquad\qquad  -\left( {C_5\over 3} B_{5}^{1/2} + {C_6} B_{6}^{1/2} \right)
{Z\over X} -  {C_9\over 3} B_{9}^{1/2} \right],    \no\\
\Imm g_{27} &=& \Imm\left( {\lambda_t \over \lambda_u }\right) 
\left[ {6 C_9\over 9} B_{9}^{3/2} \right],  \no\\
\Imm g_{\underline{8}} &=& \Imm\left( {\lambda_t \over \lambda_u }\right) 
\left[ -C_7 B_{7}^{3/2} -3C_8 B_{8}^{3/2} \right]\left(
{Y \over \sqrt{2} F_\pi^3 }\right).  
\eeqa
                                                   
Once fixed  the $g_i$, by comparison with 
 $K\to 2\pi$ amplitudes, the theory is absolutely predictive 
in all other non--leptonic channels: 
the comparison between these predictions  and the 
experimental data leads to useful indications about the 
convergence of the derivative (or chiral) expansion. 
In  table~\ref{tab:k3piconf} we report
the results of a fit\cite{KMWlett} on the experimental data, 
together with  
the predictions of  ${\cal L}_W^{(2)}$, for the dominant 
$K\to 3\pi$ amplitudes (\SSp\ref{cap:K3pi}). As can be noticed,
the discrepancy between lowest order (order $p^2$)
chiral predictions  and data
is about $30\%$. To obtain a better agreement 
is necessary to  consider next--order 
(order $p^4$) corrections\cite{KMWlett}.
As we shall see in sect.~\ref{cap:Kppg}, the need of considering
O($p^4$) terms is even more evident in the 
case of radiative decays, where the lowest order predictions  
vanish except for the  bremsstrahlung amplitudes.

\begin{table}[t]
\[ \ba{|c|c|c|} \hline
\mbox{amplitude} & \mbox{exper. fit}  & O(p^2)-\mbox{prediction} \\ \hline
 a_c & -95.4 \pm 0.4 & -76.0 \pm 0.3  \\ \hline
 a_n & +84.4 \pm 0.6 & +69.9 \pm 0.6  \\ \hline
 b_c & -26.9 \pm 0.3 & -17.5 \pm 0.1  \\ \hline
 b_n & -28.1 \pm 0.5 & -18.0 \pm 0.2  \\ \hline
 b_2 &  -3.9 \pm 0.4 & -3.1  \pm 0.4 \\ \hline
\ea \]
\caption{Comparison between experimental data
and  lowest order CHPT predictions for the  
dominant $K\to 3 \pi$ amplitudes\protect\cite{KMWlett}
 (\SSp\protect\ref{cap:K3pi}). } 
\label{tab:k3piconf} 
\end{table}

\subsection{Generating functional at order $p^4$.}
\label{sez:Fgenp4}

In the previous subsection  we have seen how to build 
the chiral realization of ${\cal L}_{QCD}$ and of the 
non--leptonic effective hamiltonian at the lowest order 
in the chiral expansion. At this order   
Green functions can be calculated using the above lagrangians 
at tree level. On the other hand, at the next order,  
is necessary to calculate the whole generating functional 
to obtain  Green functions in terms of  meson fields.
 
In the case of strong interactions we can rewrite  
the generating functional (\ref{zvasp}) in the following way:
\beq
e^{i Z(v,a,\sors,\sorp) } = \int {\cal D}U(\Phi)  
 \ e^{ i  \int {\rm d}^4x
{\cal L}_S (U,v,a,\sors,\sorp) } 
\eeq
where ${\cal L}_S(U,v,a,\sors,\sorp)$ is a local function 
of  meson fields and external sources. 
Since  $Z(v,a,\sors,\sorp)$  is locally invariant  
for chiral  transformations, except for the anomalous term\cite{Adler,Bell}, 
it is natural to expect that also 
${\cal L}_S(U,v,a,\sors,\sorp)$ be locally invariant\cite{Weinberg79}.
Indeed, it has been shown by Leutwyler\cite{Leutwyler94} that 
the freedom  in the definition of  $U$ and  
${\cal L}_S$ let us always to put the latter  
in a locally--chiral--invariant  
form. Only the anomalous part
of the functional cannot be written in terms of 
locally--invariant operators\cite{Wess,Witten}.

The expansion of ${\cal L}_S$  in powers of $p$, 
by means of the power counting rules
(\ref{count1}--\ref{sceltasp}),
\beq
{\cal L}_S={\cal L}_S^{(2)}+{\cal L}_S^{(4)}+...,
\eeq
induces a corresponding expansion of the generating functional.
At the lowest order we have 
\beq
Z^{(2)}(v,a,\sors,\sorp)  = \int {\rm d}^4x
{\cal L}_S^{(2)}(U,v,a,\sors,\sorp).
\label{Za2}
\eeq 
At the next order is necessary to  consider both  
one--loop amplitudes generated by ${\cal L}_S^{(2)}$ and  
local terms of ${\cal L}_S^{(4)}$.
As anticipated at the beginning of this section,
${\cal L}_S^{(2)}$ is non renormalizable, however, 
by symmetry arguments,
all  one--loop divergences generated by 
${\cal L}_S^{(2)}$  which cannot be re--absorbed in a re--definition
of ${\cal L}_S^{(2)}$ coefficients have
exactly the same structure of ${\cal L}_S^{(4)}$
local terms. 
The same happens at order $p^6$:  non--re--absorbed
divergences generated at two loop by ${\cal L}_S^{(2)}$ 
and at one loop by  ${\cal L}_S^{(4)}$
have the structure of  ${\cal L}_S^{(6)}$ local terms. Thus 
the theory is  renormalizable order by order 
in the chiral expansion. 

It is important to remark that loops play a fundamental 
role: generating the imaginary parts of the amplitudes
let us to implement the 
unitarity of the theory in a perturbative way.

Furthermore, the loop expansion suggests 
a natural scale for the expansion in powers of $p$,
i.e. for the scale  $\Lambda_\chi$ which rules
the suppression  ($p^2/\Lambda^2_\chi$) of  $O(p^{n+2})$ terms
with respect to  $O(p^n)$ ones. Since any 
loop carries a  factor  $1/ (4\pi F)^2$ 
($1/F^2$ comes from the expansion of $U$ and $1/16\pi^2$ 
from the integration on loop variables), the naive 
expectation is\cite{GeorgiMan}
\beq
\Lambda_\chi \sim 4\pi F_\pi = 1.2\ \mbox{\rm GeV}.
\label{naivelc}
\eeq
Obviously Eq.~(\ref{naivelc}) is just and indicative estimate of
$\Lambda_\chi$, more refined analysis suggests that is lightly 
in excess (see the discussion  in Ref.\cite{Ecker}),
but it is sufficient to understand that in   
kaon decays, where $|p|\leq M_K$, the
convergence might be slow, as shown in the previous subsection.

At order $p^4$, beyond loops and local terms of ${\cal L}_S^{(4)}$
there is also the anomalous term, the so--called 
Wess--Zumino--Witten functional\cite{Wess,Witten} ($Z_{WZW}$). Thus
the complete expression of $Z^{(4)}$ is:
\beq
Z^{(4)}(v,a,\sors,\sorp)  = \int {\rm d}^4x
{\cal L}_S^{(4)}(U,v,a,\sors,\sorp) + Z^{(4)}_{1-loop}+Z_{WZW}.
\label{Za4}
\eeq
Whereas $Z_{WZW}$ is finite  and is not renormalized
(\SSp\ref{subsez:WZW}), $Z^{(4)}_{1-loop}$ is 
divergent and is necessary to regularize it. 
Using dimensional regularization,  chiral power counting insures that 
the divergent part of 
$Z^{(4)}_{1-loop}$ is of order $p^4$ and, as already stated, has the
structure of ${\cal L}_S^{(4)}$ local terms.
In $d$ dimension we can write
\beq 
Z^{(4)}_{1-loop}= -\Lambda(\mu)\sum_i\gamma_i O_i^{(4)} +
Z^{(4)fin}_{1-loop}(\mu),
\eeq
where
\beq
\Lambda(\mu)={\mu^{d-4} \over (4\pi)^2}\left\lbrace {1\over d-4} - {1\over 2}
\left[ \ln(4\pi) +1 +\Gamma'(1) \right] \right\rbrace,
\eeq
$\gamma_i$ are appropriate coefficients, independent of $d$, and
$Z^{(4)fin}_{1-loop}(\mu)$ is finite in the limit  $d\to 4$. 
By this way,
calling $L_i$ the coefficients of ${\cal L}_S^{(4)}$ operators:
\beq
{\cal L}_S^{(4)}=\sum_i L_i  O_i^{(4)},
\eeq
and defining 
\beq
L_i = L_i^r(\mu) + \gamma_i \Lambda(\mu),
\label{mudepL}
\eeq
the sum of the first two terms in Eq.~(\ref{Za4}) is renormalized:
\beq
\int {\rm d}^4x
{\cal L}_S^{(4)}(L_i) + Z^{(4)}_{1-loop}
=\int {\rm d}^4x
{\cal L}_S^{(4)}(L_i^r(\mu)) + Z^{(4)fin}_{1-loop}(\mu).
\label{Za4r}
\eeq

\subsubsection{$O(p^4)$ Strong counterterms.}
\label{subsez:LS4}

The most general lagrangian of order $p^4$, invariant under  
local--chiral transformations, Lorentz  transformations,
$P$, $C$ and $T$, consists of  12 operators\cite{GL1}:
\beqa
{\cal L}_S^{(4)} &=& L_1 \la D_\mu U\da D^\mu U \ra^2   
+ L_2 \la D_\mu U\da D_\nu U \ra  \la D^\mu U\da D^\nu U \ra  \no\\ &+&\!\! 
L_3 \la D_\mu U\da D_\mu U  D^\nu U\da D^\nu U \ra 
+ L_4 \la D_\mu U\da D^\mu U \ra \la \chi\da U+  U\da\chi \ra \no\\ &+&\!\! 
L_5 \la D_\mu U\da D^\mu U ( \chi\da U+ U\da\chi)  \ra
+ L_6 \la \chi\da U+  U\da\chi \ra^2 
+ L_7 \la \chi\da U-  U\da\chi \ra^2 \no \\&+&\!\! 
L_8 \la \chi\da U\chi\da U +  U\da\chi U\da\chi \ra 
-iL_9 \la F_R^{\mu\nu} D_\mu U D^\nu U\da +F_L^{\mu\nu} D_\mu U\da D^\nu U 
         \ra\no \\ &+&\!\! 
L_{10} \la U\da F_R^{\mu\nu}  U  F_{L\mu\nu} \ra 
+ L_{11} \la F_R^{\mu\nu}F_{R\mu\nu} + F_L^{\mu\nu}F_{L\mu\nu} \ra 
+ L_{12} \la \chi\chi\da \ra .
\label{ls4gl}
\eeqa 
Since at order  $p^4$ this lagrangian operate only at the 
tree level, the equation of motion of  $L_S^{(2)}$,
\beq
\Box U U\da - U\Box U\da = \chi U\da - U\chi\da -{1\over 3} \la
\chi U\da - U\chi\da \ra, \label{eqmotion}
\eeq
has been used to reduce the number
of independent terms\cite{GL1}.

The constants  
$L_{1}\div L_{10}$ of Eq.~(\ref{ls4gl}) are not determined by the theory
alone and must be fixed by experimental data. 
The value of the renormalized
constants, defined by Eq.~(\ref{mudepL}), together  
with the corresponding scale factor $\gamma_i$ and the  processes  
used to fix them are reported  
in table~\ref{tab:L_i} at 
$\mu=M_\rho\simeq 770$ MeV.
  To obtain the $L^r_i(\mu)$ at different scales, 
using Eq.~(\ref{mudepL}) we find
\beq
L^r_i(\mu_1) = L^r_i(\mu_2) + {\gamma_i \over (4\pi)^2} 
\ln{\mu_2 \over \mu_1}.
\eeq
It is important to remark that the processes where the
$L_i^r(\mu)$ appear are more than those used to fix them,
thus the theory is predictive 
(see e.g. Refs.\cite{Ecker,handbook} for a discussion on  
CHPT tests in the sector of strong--interactions).

\begin{table}[t]
\[ \ba{|c|c|c|c|} \hline
i & L^r_i(M_\rho)\times 10^3  & \mbox{\rm process} 
	& \gamma_i 	\\ \hline
1 &  0.4 \pm 0.3 & K_{e4},\ \pi\pi \to \pi\pi 	& 3/32  	\\ 
2 & 1.35 \pm 0.3 & K_{e4},\ \pi\pi \to \pi\pi 	& 3/16  	\\ 
3 & -3.5 \pm 1.1 & K_{e4},\ \pi\pi \to \pi\pi 	& 0     	\\ 
4 & -3.5 \pm 0.5 & \mbox{Zweig\ rule}  		& 1/8   	\\ 
5 &  1.4 \pm 0.5 & F_K/F_\pi			& 3/8   	\\ 
6 & -0.2 \pm 0.3 & \mbox{Zweig\ rule} 		& 11/144   	\\ 
7 & -0.4 \pm 0.2 & \mbox{\rm Gell-Mann-Okubo}, L_5, L_8 & 0  	\\ 
8 &  0.9 \pm 0.3 & M_{K^0}-M_{K^+},L_5		& 5/48 	 	\\ 
9 &  6.9 \pm 0.7 & \la r^2 \ra_V^\pi  		& 1/4 	 	\\ 
10& -5.5 \pm 0.7 & \pi\to e\nu\gamma		& -1/4 	 	\\ \hline
11&  & 	& -1/8 	 	\\ 
12&  & 	& 5/24 	 	\\ \hline
\ea \]
\caption{Values of the  $L^r_i(M_\rho)$, processes used to fix them,
and relative scale factors\protect\cite{Bijnens}.}
\label{tab:L_i} 
\end{table}

The constants $L_{11}$ and $L_{12}$ are not measurable 
because the corresponding operators are 
contact terms of the external--field, 
necessary to renormalize the theory but 
without any physical meaning.

Finally, using  the $L^r_i(M_\rho)$ fixed by  data, 
we can verify the reliability of the naive estimate of
 $\Lambda_\chi$ (\ref{naivelc}). Using, as an example, 
$L^r_9(M_\rho)$ (the largest value in  
table~\ref{tab:L_i}), from the tree--level calculation of the 
electromagnetic pion form factor, follows
\beq
f^{e.m.}_\pi(t) \doteq 1 + \la r^2 \ra_V^\pi t + O(t^2) =
1+ {2 L^r_9(M_\rho) \over F_\pi^2 }  t + O(t^2),
\eeq
which implies  
\beq
\Lambda^2_\chi \gsim {F^2_\pi \over 2 L^r_9(M_\rho)} \simeq  M^2_\rho.
\eeq

\subsubsection{The WZW functional.}
\label{subsez:WZW}

The generating functional which reproduces the 
QCD chiral anomaly in terms of  meson fields 
was originally built by Wess and Zumino\cite{Wess}, 
successively has been re--formulated by Witten\cite{Witten} 
in the following way:
\beqa
Z_{WZW}(l,r) &=& -{i N_c \over 240 \pi^2 } \int_{M^5} \mbox{d}^5 x
\epsilon^{ijklm} \la U\da \partial_i U\partial_j U\da 
\partial_k U\partial_l U\da \partial_m U \ra \no \\
             &&  -{i N_c \over 48 \pi^2 } \int \mbox{d}^4 x
\epsilon^{\mu\nu\rho\sigma} \left[ W(U,l,r)_{\mu\nu\rho\sigma} -
W(1,l,r)_{\mu\nu\rho\sigma} \right],
\eeqa                 
where
\beqa
&& W(U,l,r)_{\mu\nu\rho\sigma} \ =\ \la
  U l_\mu l_\nu l_\rho U\da r_\sigma 
+ {1\over 4} U l_\mu U\da r_\nu U l_\rho U\da r_\sigma 
+ i U \partial_\mu l_\nu l_\rho U\da r_\sigma 			\no\\ &&\qquad
+ i \partial_\mu r_\nu U l_\rho U\da r_\sigma  
- iU\da \partial_\mu U  l_\nu  U\da r_\rho U l_\sigma 
- \partial_\mu U\da\partial_\nu r_\rho U l_\sigma  		\no\\ &&\qquad
+ \partial_\mu U\da \partial_\nu U U\da r_\rho U l_\sigma 
+ U\da \partial_\mu U l_\nu \partial_\rho l_\sigma
+ U\da \partial_\mu U \partial_\nu l_\rho l_\sigma  		\no\\ &&\qquad
-i U\da \partial_\mu U  l_\nu l_\rho l_\sigma 
+ {1\over 2} U\da \partial_\mu U l_\nu U\da \partial_\rho U l_\sigma
+i U\da \partial_\mu U\partial_\nu U\da\partial_\rho Ul_\sigma  \no\\ &&\qquad
- (U \leftrightarrow U\da,  l_\mu \leftrightarrow r_\mu )\ra . 
\eeqa
The $Z_{WZW}$ functional let us to compute all the contributions generated by
the  chiral anomaly to  electromagnetic and semileptonic decays
of pseudoscalar mesons. This does not mean that there are no 
other contributions to these decays. 
However, since $Z_{WZW}$ satisfies the anomalous
Ward identities,  contributions not generated by $Z_{WZW}$
must be locally invariant under chiral transformations. 

Chiral power counting 
insures that  $Z_{WZW}$ coefficients are not renormalized
by next--order contributions (for a detailed discussion about the 
odd--intrinsic--parity sector at $O(p^6)$ see 
Refs.\cite{Bijnens90,Ecker}).

\subsubsection{$O(p^4)$ Weak counterterms.}
\label{subsez:LW4}

Also in the case of non--leptonic transitions, in order to calculate 
the  Green functions at order $p^4$ is convenient to introduce 
an appropriate generating functional.
Since we are  interested only in  contributions of order $G_F$, 
we proceed analogously to the strong interaction case
 (\SSp\ref{sez:Fgenp4}) with the simple substitution
\beqa
{\cal L}_S^{(2)} &\to &{\cal L}_S^{(2)}+ {\cal L}_W^{(2)},\no\\
{\cal L}_S^{(4)} &\to &{\cal L}_S^{(4)}+ {\cal L}_W^{(4)},
\eeqa 
where ${\cal L}_W^{(4)}$ is an $ O(p^4)$ lagrangian that
transforms linearly  under  $G$ like ${\cal L}_S^{(2)}$ 
and consequently absorbs all one--loop
divergences generated by ${\cal L}_S^{(2)}\times{\cal L}_W^{(2)}$.

The  operators of order $p^4$ which  transforms like 
$(8_L,1_R)$ and $(27_L,1_R)$ under $G$, have been classified for the 
first time by Kambor, Missimer and Wyler\cite{KMWnucl}: 
the situation is worse than in the strong case because the
number of independent operators is much larger. 
For this reason, since $\Delta I=3/2$ amplitudes are experimentally very
suppressed, following Ecker, Kambor and Wyler\cite{EckerWyler}
we shall limit to consider only $(8_L,1_R)$ operators. 

\begin{table}
\[ \ba{|c|c|c|c|} \hline
i & W^{(4)}_i  & \mbox{\rm decay channel} & \gamma_i 	\\ \hline
1 & \la \lambda u\da u_\mu u^\mu u_\nu u^\nu u\ra & \geq 3\pi & 2 
  \\ \hline
2 & \la \lambda u\da u_\mu u_\nu u^\nu u^\mu u\ra & \geq 3\pi & -1/2 
  \\ \hline
3 & \la \lambda u\da u_\mu u_\nu u \ra\la u^\mu  u^\nu \ra & \geq 3\pi & 0
  \\ \hline
4 & \la \lambda u\da u_\mu u \ra\la u^\mu u_\nu  u^\nu \ra & > 3\pi & 1
  \\ \hline
5 & \la \lambda u\da \{ \chi_+, u_\mu u^\mu \} u \ra & \geq 2\pi & 3/2
  \\ \hline                                                              
6 & \la \lambda u\da u_\mu u \ra \la \chi_+ u^\mu  \ra & \geq 2\pi & -1/4
  \\ \hline                                                              
7 & \la \lambda u\da \chi_+ u \ra \la u_\mu u^\mu  \ra & \geq 2\pi & -9/8
  \\ \hline                                                              
8 & \la \lambda u\da u_\mu u^\mu u \ra \la  \chi_+ \ra & \geq 2\pi & -1/2
  \\ \hline                                                              
9 & \la \lambda u\da \left[ \chi_-, u_\mu u^\mu \right] u \ra &\geq 2\pi &3/4
  \\ \hline                                                              
10 & \la \lambda u\da \chi_+^2 u \ra & \geq 2\pi & 2/3
  \\ \hline                                                              
11 & \la \lambda u\da \chi_+ u \ra\la  \chi_+ \ra &\geq 2\pi & -13/18
  \\ \hline                                                              
12 & \la \lambda u\da \chi_-^2 u \ra & \geq 2\pi & -5/12
  \\ \hline                                                              
13 & \la \lambda u\da \chi_- u \ra\la  \chi_- \ra & \geq 2\pi & 0
  \\ \hline                                                      
14 & i \la \lambda u\da \{ f_+^{\mu\nu}, u_\mu u_\nu \} u \ra & \geq \pi\gamma\ 
  (E) & 1/4 \\ \hline                                                      
15 & i \la \lambda u\da u_\mu f_+^{\mu\nu}  u_\nu u \ra & \geq \pi\gamma\ (E)
  & 1/2 \\ \hline                                                      
16 & i \la \lambda u\da \{ f_-^{\mu\nu}, u_\mu u_\nu \} u \ra & \geq 2\pi\gamma
  \ (E) & -1/4 \\ \hline                                                      
17 & i \la \lambda u\da u_\mu f_-^{\mu\nu} u_\nu u \ra & \geq 2\pi\gamma\ (E)
  & 0 \\ \hline                                                      
18 & \la \lambda u\da \left(f_{+\mu\nu}^2 - f_{-\mu\nu}^2 \right) u \ra & 
  \geq \pi\gamma\gamma\ (E) & -1/8 \\ \hline
28 & i\epsilon_{\mu\nu\rho\sigma} \la \lambda u\da u^\mu u \ra\la 
  u^\nu u^\rho u^\sigma \ra & \geq 3\pi\gamma\ (M) & 0 \\ \hline 
29 & \la \lambda u\da \left[ {\tilde f}_{+\mu\nu} - {\tilde f}_{-\mu\nu}, u^\mu 
  u^\nu \right] u \ra & \geq 2\pi\gamma\ (M) & 0 \\ \hline 
30 & \la \lambda u\da u^\mu u \ra\la 
  u^\nu {\tilde f}_{+\mu\nu} \ra & \geq \pi\gamma\ (M) & 0 \\ \hline 
31 & \la \lambda u\da u^\mu u \ra\la 
  u^\nu {\tilde f}_{-\mu\nu} \ra & \geq 2\pi\gamma\ (M) & 0 \\ \hline 
\ea \]
\caption{$W^{(4)}_i$ operators, in the 
basis of Ref.\protect\cite{EckerWyler}, relevant to
non--leptonic kaon decays  at $O(G_F)$, with  relative  
scale factors. In the third column are indicated the  
processes  which the operators can contribute to: the symbol $>$ 
indicates  that $\pi$ or $\gamma$ can be added, whereas  $(E)$
and $(M)$ indicate electric and magnetic transitions, respectively;
no distinction is made for real or virtual photons. }
\label{tab:W_i} 
\end{table}

In  Ref.\cite{EckerWyler} the number of independent $(8_L,1_R)$ 
operators has been reduced to 37 using the lowest order 
equation of motion for $U$ and the Cayley--Hamilton theorem. 
Successively, terms that contribute only to processes with 
external $W$ bosons (i.e. terms which generate
 $O(G_F^2)$ corrections to semileptonic decays) 
and contact terms have been isolated.
By this way, the number of independent operators 
relevant to non--leptonic kaon decays at $O(G_Fp^4)$
turns out to be only 22. 

In the basis of Ref.\cite{EckerWyler} the 
$(8_L,1_R)$ component of the $O(p^4)$ 
weak lagrangian is  
written in the following way 
\beq
{\cal L}_W^{(4)} = G_8 F^2 \sum_{i=1}^{37} N_i W_i^{(4)} +\mbox{\rm h.c.},
\eeq
where the $N_i$ are adimensional constants. The 22 relevant--operator  
$W_i^{(4)}$ are reported 
in table~\ref{tab:W_i}, where, for
simplicity, has been introduced the fields
\beqa
f^{\mu\nu}_\pm &=&  u F^{\mu\nu}_L u\da  \pm u\da F^{\mu\nu}_R u, 
\qquad  {\tilde f}_{\pm\mu\nu} = 
\epsilon_{\mu\nu\rho\sigma}f_{\pm}^{\rho\sigma}, \no\\
\chi_\pm &=&  u\da \chi u\da  \pm u \chi u. 
\eeqa

Analyzing the effects of the 
 $W_i^{(4)}$ in 
$K\to 2\pi$, $K\to 3\pi$, $K\to \pi\gamma^*$, 
$K\to \pi\gamma\gamma$, $K\to 2\pi\gamma$ and $K\to 3\pi\gamma$
decays, some interesting consequences (which we shall 
discuss more in detail
in the next sections)  can be deduced:
\begin{itemize} 
\item{It is not possible to fix  
the coefficients $N_{5}\div N_{13}$: 
their effect is just to renormalize the value of 
$G_8$ fixed at $O(p^2)$ (in principle, some combinations could be fixed by 
off--shell processes, like $K\to \pi\pi^*$).}
\item{Two combinations of  $N_{1}\div N_{3}$ can be fixed by widths 
and linear slopes of $K\to 3\pi$, then is possible to 
make predictions for the quadratic slopes of these 
decays\cite{Kamborslopes} (\SSp\ref{cap:K3pi}).
As shown in Ref.\cite{DEIN}, radiative non--leptonic processes
do not add further information about $N_{1}\div N_{13}$. }
\item{The coefficients  $N_{14}\div N_{18}$ and three 
independent combinations of $N_{28}\div N_{31}$ 
could in principle fixed by the analysis of radiative  
non--leptonic decays (unfortunately present data
are too poor). Then, also in this case 
several predictions could be made\cite{Eckerrep} 
(\SSp\ref{cap:Kppg}).}
\end{itemize} 

Obviously, the above statements are valid only for
the real parts of the coefficients $N_i$. 
For what concerns the imaginary parts, related to 
$CP$ violation, up to now there are neither useful 
experimental informations  nor lattice results.
In order to make definite predictions
is necessary to implement an hadronization model. Nevertheless,
as we shall see in the following, chiral symmetry alone 
is still very useful to 
relate each other different $CP$--violating observables.

\subsection{Models for counterterms.}
\label{sez:VMD}

Due to the large number of $O(p^4)$ counterterms, 
both in the strong and expecially  in the non--leptonic weak
sector, it is interesting to consider theoretical  models
which let us to predict the value of counterterms at a given 
scale. By construction these models have nothing to do with 
the chiral constraints, already implemented, but
are based on additional less--rigorous assumptions dictated by the
phenomenology of strong interactions at low energy.

There are different classes of such  models (for an 
extensive discussion see Ref.\cite{deRafael2,handbook}); one of the 
most interesting hypothesis is the idea  that
counterterms  are saturated,
around $\mu = M_\rho$, by the contributions coming from  
low--energy--resonance ($\rho$, $\omega$, $\eta'$, 
etc...) exchanges\cite{Donoghueres,EGLR}. In the framework of this 
hypothesis  (known as `chiral duality') 
it is assumed that the dominant contribution is generated by 
 spin--1 mesons, in agreement with the old idea of
`vector meson dominance'. 

In order to calculate the  resonance contribution to counterterms,
is necessary: i) to consider the most general chiral--invariant 
lagrangian containing both  resonance  and pseudoscalar meson
fields; ii) to integrate over the resonance degrees of freedom,
in order to obtain a non--local effective action for 
pseudoscalar mesons only; iii) to  expand this action in terms of
local operators. Since strong and electromagnetic 
coupling constants of resonance fields are  
experimentally known, in the case of
${\cal L}_S^{(4)}$ this procedure leads to 
interesting unambiguous predictions\cite{EGLR}. 
                                   
As an example, to calculate spin--1 resonance effects, 
we can introduce two antisymmetric  tensors  
$V^{\mu\nu}$ and $A^{\mu\nu}$, 
which describe the  $SU(3)_{L+R}$ octets of 
$1^{--}$ and $1^{++}$ resonances, and which  under 
 $G$ transform  in the following way:
\beq
R^{\mu\nu} \Gto h(g,\phi_i) R^{\mu\nu} h^{-1}(g,\phi_i) 
\qquad\quad R^{\mu\nu}=V^{\mu\nu},\ A^{\mu\nu}.
\label{GlawforVA}
\eeq
The lowest--order chiral lagrangian describing 
$V^{\mu\nu}$ and $A^{\mu\nu}$ interactions  
with  pseudoscalar mesons and   
gauge fields is:\footnote{~Actually,
the choice of Eqs.~(\protect\ref{GlawforVA}--\protect\ref{LkinR}) 
to describe spin--1 resonances couplings is not unique, there
exist different formulations which  however lead to 
equivalent  results\protect\cite{EGLRb}.}                      
\beqa
{\cal L}^{(2)}_{V,A} &=& {\cal L}_{kin}(V) + {\cal L}_{kin}(A) +
{F_V \over 2\sqrt{2} } \la V_{\mu\nu} f^{\mu\nu}_+\ra  \no\\
&& + {iG_V \over \sqrt{2} } \la V_{\mu\nu} u^\mu u^\nu \ra +
{F_A \over 2\sqrt{2} } \la A_{\mu\nu} f^{\mu\nu}_-\ra, 
\label{LintR}
\eeqa
where
\beq
{\cal L}_{kin}(R) = -{1\over 2} \la \DD^\mu R_{\mu\nu}  
\DD_\sigma R^{\sigma\nu}\ra  + {1\over 4} M_R^2 
\la R_{\mu\nu} R^{\mu\nu} \ra. 
\label{LkinR}
\eeq
Integrating over 
 resonance degrees of freedom and expanding up to $O(p^4)$, 
leads to identify the following contribution to the $L_i$:
\beq
\ba{lll} 
{L}^V_1 = \dis{G_V^2 \over 8 M_V^2}, \quad  & 
{L}^V_2 = -L^V_3/3 = 2L^V_1, \quad  & 
{L}^V_9 = \dis{F_VG_V \over 2 M_V^2}, \\
{L}^V_{10} = \dis{F_A^2 \over 4 M_A^2}- {F_V^2 \over 4 M_V^2},\quad &
{L}^V_4 = {L}^V_5 = {L}^V_6 = {L}^V_7 = 
{L}^V_8 = 0. & 
\ea \label{L_iVMD}
\eeq
The constants $G_V$, $F_V$ and $F_A$ 
can be experimentally fixed by the measurements of
 $\Gamma(V \to \pi\pi)$, 
$\Gamma(V \to e^+e^-)$ and  $\Gamma(A \to \pi\gamma)$, respectively. 
The results obtained by this procedure for the non--vanishing
 ${L}^V_i$ are reported in the second column of
table~\ref{tab:L_iVMD}:
as can be noticed the agreement with the fitted $L_i$ is very good. 
For the constants $L_{4-8}$, which do not receive any contribution 
from  spin--1 resonances, is necessary to calculate 
the contribution of scalar resonances. At any rate, 
as can be noticed from table~\ref{tab:L_i},
these constants have smaller values respect to the dominant ones
($L_3$, $L_9$ and $L_{10}$) to which  
spin--1 resonance contribute. We finally note that 
imposing on the lagrangian (\ref{LintR}) 
Weinberg sum rules\cite{Weisum}
and the so--called KSFR relations\cite{KSFRa,KSFRb}
(which are in good agreement with experimental data)
then $F_V$, $G_V$, $F_A$ and $M_A$ satisfy
the following identities:
\beq
F_V=2G_V=\sqrt{2} F_A=\sqrt{2} F_\pi, \qquad  M_A=\sqrt{2} M_V.
\label{minimalsetF_V}  
\eeq
As a consequence, in this case the  ${L}^V_i$ 
can be expressed in term of a single parameter: $M_V$. 
The values of the 
${L}^V_i$ thus obtained are 
reported in the third column of  
table~\ref{tab:L_iVMD}: in spite of the simplicity of the model,
even in this case the agreement is remarkable.

\begin{table}[t]
\[ \ba{|c|c|c|c|} \hline
i & L^r_i(M_\rho)\times 10^3  & {L}^V_i \times 10^3 & 
{L}^V_i\times 10^3  \ (*)  \\ \hline
1 &  0.4 \pm 0.3 & 0.6\pm 0.2	& 0.9  	\\  \hline
2 & 1.35 \pm 0.3 & 1.2\pm 0.3	& 1.8	\\  \hline
3 & -3.5 \pm 1.1 & -3.0\pm 0.7	& -4.9  \\  \hline
9 &  6.9 \pm 0.7 &  7.0\pm  0.4	& 7.3	\\  \hline
10& -5.5 \pm 0.7 & -6.0\pm  0.8	& -5.5	\\  \hline
\ea \]
\caption{Comparison between the fitted $L^r_i(M_\rho)$ (first column)
and the vector--meson--dominance  predictions
(\protect\ref{L_iVMD}).
The values in the second column have been obtained 
with  $F_V$, $G_V$, $F_A$, $M_V$ and $M_A$
fixed by  experimental data, the corresponding errors are related 
to the different possibilities to fix 
these constants ($F_V$, as an example, can be 
fixed either from  $\Gamma(\rho \to e^+e^-)$  or from 
$\Gamma(\omega \to e^+e^-)$). The values reported in the last column 
have been obtained using the relations (\protect\ref{minimalsetF_V})
and fixing $M_V=M_\rho$.  }
\label{tab:L_iVMD} 
\end{table}

\subsubsection{The factorization hypothesis of ${\cal L}_W$.}  

Clearly, in the sector of non--leptonic weak interactions
the situation is more complicated 
since there are no experimental information about  weak
resonance couplings.
To make predictions is necessary to add further assumptions, 
one of these is the factorization 
hypothesis\cite{deRafaeletal,IP,Bruno}.
Since the dominant terms of the four--quarks hamiltonian 
are factorizable as the product of two  left--handed currents,
we assume that also the  
chiral weak lagrangian has the same structure.

As we have seen in sect.\ref{sez:Lp2}, the lowest--order  chiral  
realization of the  left--handed current
 $\bar{q}_L \gamma^\mu q_L$ is given by the  
functional derivative of $Z^{(2)}$ 
with respect to the external source $l_\mu$:
\beq
J^{(1)}_\mu = {\delta  Z^{(2)}(l,r,\sors,\sorp) 
\over \delta l_\mu} = - {1\over 2} F^2 L_\mu.
\eeq 
Furthermore, since the lowest--order  weak lagrangian 
can be written as 
\beq 
{\cal L}_W^{(2)} =  4 G_8 \la \lambda  J_\mu^{(1)} J^{\mu(1)} \ra
+ \mbox{h.c.}, 
\eeq                                        
the factorization hypothesis of ${\cal L}_W^{(4)}$
consists of assuming the following structure:
\beq
{\cal L}_{W-fact}^{(4)} = 4 k_f G_8 \langle \lambda \left\lbrace J_\mu^{(1)}, 
J^{\mu(3)} \right\rbrace\rangle  + \mbox{h.c.}, 
\eeq
where $k_f$ is a positive parameter of order 1 and $J^{\mu(3)}$
is the chiral realization of the left--handed  current  
at order $p^3$. In general  $J^{\mu(3)}$ can be 
expressed as functional derivative of  ${\cal L}_S^{(4)}$,
with respect to the source $l_\mu$, and in this case depends
on the value of the $L^r_i(\mu)$.
In order to make the model more predictive, relating it to the 
vector--meson--dominance hypothesis previously discussed,
one can assume $L^r_i(M_\rho)={L}^V_i$.

To date, experimental data on weak $O(p^4)$ counterterms
are very poor, not sufficient to draw quantitative conclusions  
about the validity of the factorization hypothesis.
At any rate, in the only channels where there are useful and
reliable experimental data, i.e.  $K\to 3\pi$ and
 $K^+\to \pi^+e^+e^-$ decays,  the estimates of the sign and of the 
order of magnitude of counterterms, calculated within this
model, are more or less 
correct\cite{EckerWyler}. Only in the next years,
when new high--statistics data on both neutral and charged kaon  
decays will be available, it will be possible to make an accurate 
analysis of the non--leptonic sector. With the expected new data
it will be possible not only to test
the factorization model, but
also to study in 
general  the convergence of the chiral 
expansion at $O(p^4)$ in the non--leptonic sector.

\setcounter{equation}{0}
\setcounter{footnote}{0}
\section{$K\to 3 \pi$ decays.}
\label{cap:K3pi}

\subsection{Amplitude decomposition.}
\label{sez:K3pidec}

There are four distinct channels for 
$K\to 3\pi$ decays:
\beq \ba{rll}
K^\pm &\to \pi^\pm \pi^\pm \pi^\mp \qquad & I=1,2, \\
K^\pm &\to \pi^0 \pi^0 \pi^\pm \qquad & I=1,2, \\
\Ko(\Kob) &\to \pi^\pm \pi^\mp \pi^0 \qquad & I= 0,1, 2 ,  \\
\Ko(\Kob) &\to \pi^0 \pi^0 \pi^0 \qquad & I= 1. \\
\ea \label{k3pi1}
\eeq 
Near each channel we have indicated the final--state isospin
assuming $\Delta I \leq 3/2$. 

In order to write the transition amplitudes, it is convenient 
to introduce the following kinematical variables:
\beq
s_i\doteq (p_K-p_i)^2,
\qquad \mbox{and} \qquad s_0\doteq {1\over 3} (s_1+s_2+s_3) = 
{1\over3} M^2_K + {1\over3}\sum_{i=1}^3 M^2_{\pi_i}, 
\eeq 
where $p_K$ and $p_i$ denote 
kaon and  $\pi_i$ momenta ($\pi_3$ indicates 
the odd pion in the first three channels). 
With these definition, the isospin decomposition of $K\to 3 \pi$ 
amplitudes  is given by\cite{Zemach}$^-$\cite{DIPP}: 
\beqa 
A_{++-}&=& A(K^+\to \pi^+\pi^+\pi^-) =  
  2A_c(s_1,s_2,s_3)+B_c(s_1,s_2,s_3)+B_2(s_1,s_2,s_3), \no\\
A_{+00}&=&  A(K^+\to \pi^0\pi^0\pi^+) =  
  A_c(s_1,s_2,s_3)-B_c(s_1,s_2,s_3)+B_2(s_1,s_2,s_3), \no\\
A_{+-0}&=& \sqrt{2} A(\Ko \to \pi^+\pi^-\pi^0) =  
  A_n(s_1,s_2,s_3)-B_n(s_1,s_2,s_3)+C_0(s_1,s_2,s_3) \no\\
  &&\qquad\qquad\qquad\qquad\qquad +2[B_2(s_3,s_2,s_1)-B_2(s_1,s_3,s_2)]/3, 
\no\\ A_{000}&=& \sqrt{2}  A(\Ko \to \pi^0\pi^0\pi^0) =  3A_n(s_1,s_2,s_3).
\label{superdec} 
\eeqa 
The amplitudes 
$A_i$, $B_i$ ($i=c,n,2$) and $C_0$  transform  in the following way
 under $s_i$ permutations: 
the $A_i$ are completely symmetric, $C_0$ is antisymmetric  
for any exchange  $s_i \leftrightarrow s_j$, 
finally the $B_i$ are symmetric in the exchange 
 $s_1 \leftrightarrow s_2$ and obey to the relation
\beq               
B_i(s_1,s_2,s_3)+B_i(s_3,s_2,s_1)+B_i(s_1,s_3,s_2)=0.
\eeq
For what concerns isospin, $A_{c,n}$ and $B_{c,n}$ belong 
to transitions in  $I=1$, whereas
 $B_2$ and $C_0$
belong to $I=2$ and $I=0$, respectively.

Differently than in $K \to 2 \pi$, in the first three
channels  of Eq.~(\ref{superdec}) there are two 
amplitudes, which differ for the transformation property under
$s_i$--permutations, that lead to the same final state ($I=1$).
For this reason is convenient to introduce the two matrices
\beq
T_c=\left(\begin{array}{rr} 2 &1 \\ 1 &-1 \end{array}\right) 
\qquad
T_n=\left(\begin{array}{rr} 1 &-1 \\ 3 &0 \end{array}\right) ,
\label{tc}
\eeq
which project the  symmetric and  the antisymmetric components of 
the $I=1$ state in the physical 
channels for charged and neutral kaon decays:
\beq
\left(\begin{array}{c} A_{++-}^{(1)} \\ A_{+00}^{(1)} \end{array}\right)=
T_c \left(\begin{array}{c} A_c(s_i) \\ B_c(s_i) \end{array}\right)
\qquad 
\left(\begin{array}{c} A_{+-0}^{(1)} \\ A_{000}^{(1)} \end{array}\right)=
T_n \left(\begin{array}{c} A_n(s_i) \\ B_n(s_i) \end{array}\right).
\label{tctn}
\eeq

\begin{table}[t] 
\[ \ba{|c|c|c|c|c|c|} \hline
\mbox{\rm channel} & \Gamma(s^{-1}) \times 10^{-6} 
& g\times 10^1 & j\times 10^3 & h\times 10^2  & k\times 10^2 \\ \hline
\pi^+ \pi^+ \pi^- & 4.52  \pm 0.04  & -2.15 \pm 0.03 
               & / & 1.2 \pm 0.8 & -1.0  \pm 0.3  \\ \hline
 \pi^- \pi^- \pi^+ & 4.52  \pm 0.04 & -2.17 \pm 0.07 & /
                   & 1.0 \pm 0.6 & - 0.8  \pm 0.2 \\ \hline
 \pi^0 \pi^0 \pi^\pm~ ^a & 1.40 \pm 0.03 & 5.94 \pm 0.19 & /
                   & 3.5 \pm 1.5 &  - \\ \hline
(\pi^+ \pi^- \pi^0)_L & 2.39 \pm 0.04 & 6.70 \pm 0.14 & 1.1 \pm 0.8 
                   & 7.9 \pm 0.7 & 1.0 \pm 0.2 \\ \hline
(\pi^+ \pi^- \pi^0)_S  & 4.5^{+2.6}_{-1.5} \times 10^{-3}
                  & - & -  & - & - \\ \hline
(\pi^0 \pi^0 \pi^0)_L & 4.19 \pm 0.16 & / & /
                   & -0.33 \pm 0.13 & - \\ \hline
(\pi^0 \pi^0 \pi^0)_S  & < 0.4 & / & / 
                   & -  & - \\ \hline
\ea \]
\caption{Experimental data for widths and slopes
in $K\to 3\pi$ decays\protect\cite{PDG,E621_cons,CPLEAR_cons}. The symbol
/ indicates terms forbidden by Bose symmetry.}
\label{tab:K3piexp} 
\end{table}

Experimentally, the event distributions in 
 $K\to 3 \pi$ transitions are analyzed in terms of two 
adimensional and independent variables:
\beq
X= { s_1 - s_2 \over M_\pi^2 } \qquad \mbox{and}  \qquad 
Y= { s_3 - s_0 \over M_\pi^2 }, 
\eeq 
the so--called Dalitz variables. Since the three--pion phase space
is quite small ($M_K-3M_\pi < 100$ MeV), terms with 
elevate powers of  $X$ and $Y$, corresponding to states
with high angular momenta, are very suppressed (see
Ref.\cite{Amelino} and references cited therein).
Until now the  distributions have been analyzed 
including up to 
quadratic terms in $X$ and $Y$
\beq
\left| A(K\to 3 \pi) \right|^2 \propto 1 + g Y +j X +hY^2 +kX^2.
\label{slopes}
\eeq
The parameters $g\div k$ are the `Dalitz Plot slopes'. In 
table~(\ref{tab:K3piexp}) we report  the  
experimental data for the different 
channels.\footnote{~Serpukhov-167\protect\cite{Serpukhov}
presented at ICHEP '96 some preliminary data
on $K^+ \to  \pi^0 \pi^0 \pi^+$ slopes which differ 
substantially from those reported 
in the table: $g=0.736\pm 0.014\pm 0.012$,
$h=0.137\pm 0.015\pm 0.024$ and
$k=0.0197\pm 0.0045\pm 0.003$ ($\chi^2= 1.5 /ndf$).}

To relate the decomposition  (\ref{superdec}) with 
experimental data is necessary to expand  $A_i$,
$B_i$ and $C_0$ in terms of $X$ and $Y$. According to the 
transformation properties under 
$s_i$--permutations follows: 
\beqa
A_i(s_1,s_2,s_3) &=&a_i+c_i(Y^2+X^2/3)+..., \no \\
B_i(s_1,s_2,s_3) &=&b_iY+d_i(Y^2-X^2/3)+e_iY(Y^2+X^2/3)+..., \no \\
C_0(s_1,s_2,s_3) &=&f_0X(Y^2-X^2/9)+..., 
\label{k3pexpan1}
\eeqa
where dots indicate  terms at least quartic  in 
$X$ and $Y$. The parameters $a_i,b_i,...f_0$ are
 real if  strong re--scattering is neglected and  
$CP$ is conserved.

Since we are interested only in 
 $CP$ violating effects, we shall limit to consider only 
the  dominant terms in each amplitude and we will 
neglect completely the $C_0$ amplitude that
 is very suppressed. 
With this assumption, the decomposition~(\ref{superdec}) 
contains at most linear terms in $X$and $Y$:
\beqa
A_{++-} &=& 2a_c+(b_c+b_2)Y, \no \\
A_{00+} &=& a_c - (b_c - b_2)Y, \no \\
A_{+-0} &=& a_n  - b_n Y + {2\over 3} b_2 X, \no \\
A_{000} &=& 3 a_n.  
\label{superdec2}
\eeqa

\subsection{Strong re--scattering.}
\label{sez:K3piphases}

As we have seen in sect.~\ref{cap:kksystem}, to estimate   
$CP$ violation in charged--kaon decays is fundamental to know  strong
re--scattering phases of the final state.

Differently than in $K\to 2\pi$,  
$K\to 3\pi$  re--scattering phases are not constants but
depend on the kinematical variables $X$ and $Y$. Furthermore,
in the  $I=1$ final state, the two  
amplitudes with different symmetry  are mixed by 
re--scattering\cite{DIPP}. Projecting, by means of 
 $T_c$ and $T_n$, $I=1$ physical amplitudes    
in the basis of  amplitudes with definite symmetry, is 
possible 
to introduce a unique re--scattering matrix
$R$, relative to the $I=1$ final state, so that 
\beqa
\left( \ba{c}  A_{++-}^{(1)} \\ A_{+00}^{(1)}  \ea \right)_R 
= T_c R \left( \ba{c}   A_c \\ B_c \ea \right) =
  T_c R  T_c^{-1} \left( \ba{c}   A_{++-}^{(1)} \\ 
A_{+00}^{(1)} \ea \right), \\
\left( \ba{c}  A_{+-0}^{(1)} \\ A_{000}^{(1)}  \ea \right)_R 
= T_n R \left( \ba{c}   A_c \\ B_c \ea \right) =
  T_n R  T_n^{-1} \left( \ba{c}   A_{+-0}^{(1)} \\ 
A_{000}^{(1)} \ea \right). 
\eeqa
The matrix $R$ has diagonal elements which preserve
the symmetry under $s_i$--permutations
as well as  off--diagonal elements which transform
symmetric amplitudes into antisymmetric ones 
(and vice versa).
Since the phase space is limited, 
we expect re--scattering phases to be small,
i.e. that  $R$ can be expanded in the following way:
\beq
R=1+i\left( \ba{cc} \alpha(s_i) & \beta'(s_i) \\
\alpha'(s_i) & \beta(s_i) \ea \right),
\eeq 
with $\alpha(s_i),\beta(s_i),\alpha'(s_i),\beta'(s_i)\ll 1$.
Analogously, for the re--scattering in $I=2$
we can introduce a phase  $\delta(s_i) \ll 1$, so that
\beq
B_2(s_i)_R = B_2(s_i)\left[1+i\delta(s_i)\right].
\eeq
Moreover, from the  transformation properties of the 
amplitudes follows\cite{DIPP}
\beq 
\ba{rcl}
\alpha(s_i)  &=& \alpha_0 + O(X^2,Y^2),  \\
\alpha'(s_i) &=& \alpha'_0 Y + O(X^2,Y^2),  \\
\beta(s_i)   &=& \beta_0 + O( X, Y), \\
\beta'(s_i)  &=& \beta'_0 (Y^2+X^2/3)/Y + O(X^2,Y^2), \\
\delta(s_i)  &=& \delta_0 + O(X,Y).\ea 
\eeq
With these definitions, the complete 
re--scattering of Eq.~(\ref{superdec2}), including 
up to linear terms in $X$ and $Y$, is given by: 
\beqa
(A_{++-})_R &=& 2 a_c[1+i\alpha_0 +i \alpha'_0Y/2 ] + 
b_c Y  [1+i \beta_0 ]  + b_2 Y [1+i\delta_0]  \no \\
 &=& 2 a_c[1+i\alpha_0 ] + 
b_c Y \left[1+i\left( \beta_0 + {a_c\over b_c} \alpha'_0
 \right) \right]  + b_2 Y [1+i\delta_0], \no \\
(A_{00+})_R &=& a_c[1+i\alpha_0  - i \alpha'_0 Y ] -
b_c Y  [1+i  \beta_0 ]  + b_2 Y [1+i\delta_0]  \no \\
 &=& a_c[1+i\alpha_0 ] - 
b_c Y \left[1+i \left( \beta_0 + {a_c\over b_c} \alpha'_0 
  \right) \right]  
+ b_2 Y [1+i\delta_0], \no \\
(A_{+-0})_R  &=& a_n[1+i\alpha_0 - i \alpha'_0 Y ] 
- b_n Y  [1+i  \beta_0 ] + {2\over 3} b_2 X [1+i\delta_0]  \no \\
 &=& a_n[1+i\alpha_0 ] 
- b_n Y \left[ 1+i\left( \beta_0 + {a_n\over b_n} \alpha'_0 
  \right) \right]  
+ {2\over 3} b_2 X [1+i\delta_0], \no \\
(A_{000})_R  &=& 3 a_n[1+i\alpha_0 ].  
\label{superdec3}  
\eeqa
The first three amplitudes have been 
expressed in two different ways to stress that 
$Y$--dependent imaginary parts receive contributions 
from the  re--scattering of both  
 symmetric amplitudes ($a_{c,n}$) and 
antisymmetric  ones ($b_{c,n}$).

\subsection{$CP$--violating observables.}
\label{sez:K3piCP}

Considering only widths and  linear 
slopes (as can be noticed from table~\ref{tab:K3piexp},
quadratic slopes have large errors), we can 
define the following $CP$--violating observables in 
$K\to 3\pi$ transitions:
\beqa
\eta_{+-0} & \doteq & \left.
 { A^S_{+-0} \over  A^L_{+-0} }\right|_{X=Y=0} \doteq 
\epsilon +\epsilon'_{+-0}, \\
\eta_{000} & \doteq & \left.
 { A^S_{000} \over  A^L_{000} }\right|_{X=Y=0} \doteq 
\epsilon +\epsilon'_{000}, \\
\eta^X_{+-0} & \doteq & 
 \left( { \partial A^L_{+-0} /\partial X \over 
\partial A^S_{+-0} /\partial X } \right)_{X=Y=0}  \doteq 
\epsilon +\epsilon^X_{+-0}, \\
\left( \delta_g \right)_\tau & \doteq & { g^{++-} - g^{--+} \over
  g^{++-} + g^{--+} }, \label{deltagt}\\
\left( \delta_g \right)_{\tau'} & \doteq & { g^{+00} -g^{-00} \over
  g^{+00} + g^{-00} }. \label{deltagt1}
\eeqa
The first three observables belong to 
neutral kaons and, as explicitly shown, have an indirect
 $CP$--violating component. On the other hand 
$\left( \delta_g \right)_{\tau}$ and  
$\left( \delta_g \right)_{\tau'}$ are pure  indices of direct $CP$ 
violation. In principle, analogously to Eqs.~(\ref{deltagt}--\ref{deltagt1}),  
also the asymmetries of  charged--kaon
widths can be considered. However, since the integral over the  Dalitz Plot
of the terms linear in $Y$ is zero, the
width  asymmetries are very suppressed 
respect to the slope asymmetries\cite{IMP} and we will not
consider them.
\begin{figure}[t]
    \begin{center}
       \setlength{\unitlength}{1truecm}
       \begin{picture}(6.0,4.0)
       \epsfxsize 6.  true cm
       \epsfysize 4. true cm
       \epsffile{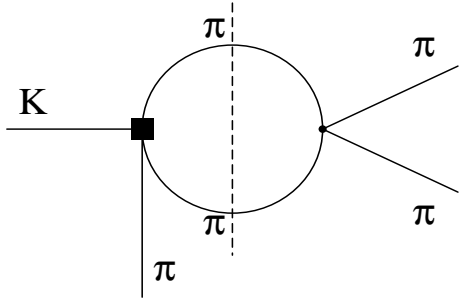}
       \end{picture}
    \end{center}
    \caption{One--loop diagram for 
$K\to 3 \pi$ re--scattering phases. }
    \protect\label{fig:k3piloop}
\end{figure}

Using the definitions of  $K_S$ and $K_L$, and 
applying 
$CPT$ to the  decomposition (\ref{superdec3}), leads to
\beqa
\epsilon'_{+-0} & = & \epsilon'_{000} \ = \  
i \left ({\Imm a_n \over \Real a_n} - {\Imm A_0 \over \Real A_0}\right) , 
\label{k3pcp1a} \\ \epsilon^X_{+-0} & = &
i \left ({\Imm b_2 \over \Real b_2} - {\Imm A_0 \over \Real A_0}\right) , \\ 
\left( \delta_g \right)_\tau & = & 
{ \Imm(a_c^* b_c) (\alpha_0 - \beta_0) + \Imm(a_c^* b_2) (\alpha_0-\delta_0)
 \over \Real(a_c^* b_c) + \Real(a_c^* b_2) },
\label{k3pcpz} \\  
\left( \delta_g \right)_{\tau'} & = & 
{ \Imm(a_c^* b_c) (\alpha_0 - \beta_0) - \Imm(a_c^* b_2) (\alpha_0-\delta_0)
 \over \Real(a_c^* b_c) - \Real(a_c^* b_2) }.
\label{k3pcp1}
\eeqa
where $A_0$ is the $K \to 2\pi$ decay amplitude in  $I=0$.

\subsection{Estimates of $CP$  violation.}
\label{sez:K3piampl}

The lowest order ($p^2$) CHPT results for  
the weak amplitudes of Eqs.~(\ref{k3pcp1a}--\ref{k3pcp1})
are:
\beqa
a_c & = & - {M_K^2 \over 3} \left[ G_8 +{2\over 3} G_{27} 
+ {3F^2 \over M_K^2}
G_{\underline 8} \right], 
\label{k3pcp2a} \\
a_n & = & + {M_K^2 \over 3} \left[ G_8 -G_{27} \right], \\
b_c & = & + M_\pi^2 \left[ G_8 -{7\over 12}G_{27}\left(1-{15\over7}
	\rho_\pi\right)
+ {3F^2 \over 4M_K^2}G_{\underline 8} \left(1+\rho_\pi\right)\right], \\
b_2 & = & - M_\pi^2 \left[ {15\over 4}G_{27}\left(1+{1\over3}\rho_\pi\right)
+ {3 F^2 \over 4M_K^2}G_{\underline 8} \left(1+\rho_\pi\right)\right], 
\label{k3pcp2}
\eeqa
where $\rho_\pi = M^2_\pi /(M_K^2-M_\pi^2) \simeq 1/12$.   

To estimate re--scattering phases at the lowest non--vanishing
order in CHPT, is necessary to calculate the imaginary 
part of one--loop diagrams of fig.~\ref{fig:k3piloop}. 
The complete analytical 
results for the phases introduced in sect.~\ref{sez:K3piphases}
can be found in  Refs.\cite{IMP,DIPP}, for what concerns the 
parameters which enter in Eqs.~(\ref{k3pcp1a}--\ref{k3pcp1}) 
we have:
\beqa
\alpha_0 & = & { \sqrt{1-4M_\pi^2/s_0 } \over 32\pi F^2 }(2s_0 +M_\pi^2)
\simeq 0.13 , \\
\beta_0 & = & -\delta_0 \ = \ { \sqrt{1-4M_\pi^2/s_0 } \over 32\pi F^2 }
(s_0 - M_\pi^2) \simeq 0.05. 
\label{k3pcp3}
\eeqa

Using   Eqs.~(\ref{k3pcp2a}--\ref{k3pcp3})
and the estimates of the imaginary parts of  
$L_W^{(2)}$ coefficients (\SSp\ref{subsez:LW}), we can finally 
predict the value of the observables (\ref{k3pcp1a}--\ref{k3pcp1})
within the Standard Model\cite{Nellohand}.

\begin{figure}[t]
    \begin{center}
       \setlength{\unitlength}{1truecm}
       \begin{picture}(8.0,8.0)
       \epsfxsize 8. true cm
       \epsfysize 8. true cm
       \epsffile{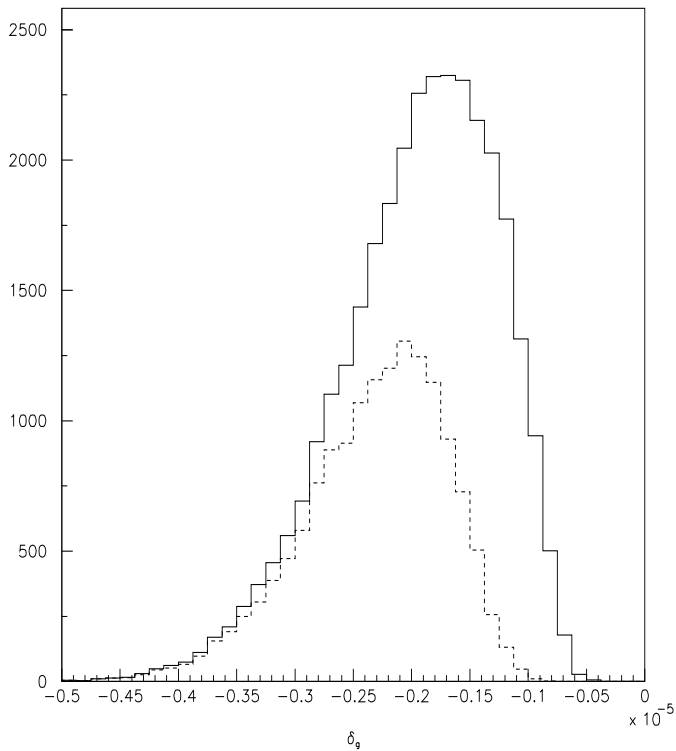}
       \end{picture}
    \end{center}
    \caption{Predictions for the 
charge asymmetry $(\delta_g)_\tau$ 
at the lowest non--vanishing order in CHPT. The two histograms
represent the probability distribution in arbitrary units,
like in fig.~\protect\ref{fig:ciufig}; the 
dashed one has been obtained adding the 
supplementary conditions coming from $B_d-{\bar B}_d$  mixing .}    
\protect\label{fig:6} 
\end{figure}

\subsubsection{Charge asymmetries.}
\label{subsez:deltag}

\begin{figure}[t]
    \begin{center}
       \setlength{\unitlength}{1truecm}
       \begin{picture}(8.0,8.0)
       \epsfxsize 8. true cm
       \epsfysize 8. true cm
       \epsffile{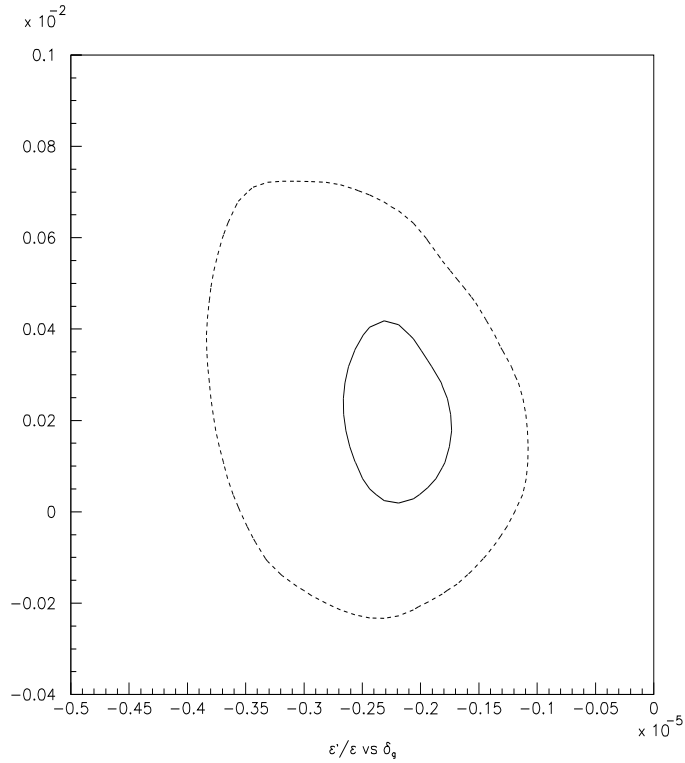}
       \end{picture}
    \end{center}
 \caption{Comparison between  $(\delta_g)_\tau$ 
and $\epsp/\eps$. Full and dotted lines 
indicate  $5\%$  and $68\%$ contours around the 
central value, respectively.}
    \protect\label{fig:7}
\end{figure}

In figs.~\ref{fig:6} and~\ref{fig:7} we show the results of a 
statistical analysis of  
$(\delta_g)_\tau$ and $\epsp$, obtained implementing 
in the  program of Ref.\cite{Ciuchini3} 
(\SSf\ref{fig:ciufig}) the calculation of $(\delta_g)_\tau$.\footnote{~For
the real parts of the amplitudes  $a_c$, $b_c$ and $b_2$ we have used the 
experimental data.}
The most interesting aspect of this analysis, as already 
stressed in Ref.\cite{IMP}, is that in $(\delta_g)_\tau$, 
differently than in $\epsp$,  
the interference between  weak phases of 
$(8_L,1_R)$ and $(8_L,8_R)$ operators is constructive.
Thus, within the Standard  Model,  charge asymmetries
in $K^\pm\to(3\pi)^\pm$ decays could be more interesting than
$\epsp$ in order to observe direct $CP$ violation. Unfortunately,
the theoretical estimates of these asymmetries are far from the 
expected sensitivities of next--future
experiments (at KLOE\cite{Alosio} 
 $\sigma [ (\delta_g)_\tau ]$ is expected to be\cite{handbook}
$\sim 10^{-4}$).

\begin{table}[t] 
\centering
\begin{tabular}{|c|c|c|} \hline
\mbox{asimmetry}& \mbox{exp. limit}&\mbox{th. estimate}\\ \hline
$\left(\delta _g\right)_{\tau}$
&$-(0.70\pm 0.53)\times 10^{-2}$&$-(2.3\pm 0.6)\times 10^{-6}$\\ \hline
$\left(\delta _{\Gamma}\right)_{\tau}$
&$(0.04\pm 0.06)\times 10^{-2}$&$-(6.0\pm 2.0)\times 10^{-8}$\\ \hline
$\left(\delta _g\right)_{\tau^\prime}$
& - &$(1.3\pm 0.4)\times 10^{-6}$\\ \hline
$\left(\delta _{\Gamma}\right)|_{\tau^\prime}$
&$(0.0\pm 0.3)\times 10^{-2}$&$(2.4\pm 0.8)\times 10^{-7}$\\ \hline
\end{tabular}
\caption{Experimental limits and theoretical estimates for the 
charge asymmetries in  $K^\pm \to \pi^\pm \pi^\pm \pi^\mp (\tau)$
and $K^\pm \to \pi^\pm \pi^0 \pi^0 (\tau')$ decays, calculated at the 
 lowest non--vanishing order in  CHPT.}
\label{tab:k3priass}
\end{table}

The results for $(\delta_g)_{\tau'}$ 
are very similar to those of $(\delta_g)_\tau$,
a part from the sign which is opposite\cite{IMP}, 
we will not show them in detail since
$(\delta_g)_\tau$ is more interesting from the experimental 
point of view.
The  mean value of  
$(\delta_g)_\tau$ and $(\delta_g)_{\tau'}$,
together with the corresponding width asymmetries, 
are reported in table~\ref{tab:k3priass}. 
As anticipated, the width asymmetries are 
definitively  suppressed with respect to the
slope asymmetries. Analogous results to those 
reported in table~\ref{tab:k3priass} have been obtained also 
by other authors\cite{Cheng0,Shabalin}$^,$\footnote{~Actually, in 
Ref.\protect\cite{Cheng0}, as well as in Ref.\protect\cite{IMP}, 
also some isospin breaking effects  
have been included. We prefer to 
neglect these effects for two reasons: i) there are not sufficient 
data to analyze systematically isospin breaking 
in all $K\to 3\pi$ channels; ii) as we will discuss in the following, 
these effects are completely negligible with respect to possible 
next--order CHPT corrections.}.

It is important to remark that the previous analysis
has been obtained using the
lowest--order CHPT results for the weak amplitudes
and, differently than in   $K\to 2\pi$,
could be sensibly modified by next--order corrections.
The difference between  $K\to 2\pi$ and $K\to 3\pi$ is that in 
former the $CP$--violating interference 
is necessarily between  a
$\Delta I=1/2$ and a  $\Delta I=3/2$ amplitude, whereas in
the latter the interference is between two $\Delta I=1/2$ 
amplitudes ($a_c$ and $b_c$). 
At  the lowest order there is only 
one dominant $(8_L,1_R)$  operator, thus the 
phase difference between  $a_c$ and  $b_c$  is determined 
by the suppressed $(27_L,1_R)$ operator. At the 
next order, $O(p^4)$, there are different $(8_L,1_R)$ operators and
we can expect that the  interference is no more suppressed by the 
 $\omega$ factor\cite{Donoghuek3p}. Unfortunately in this case is not easy 
to make definite statements, since there are no reliable
information about $O(p^4)$--operator weak phases. 
Nevertheless, according to general considerations,
is still possible to put an interesting limit\cite{DIP} on 
 $(\delta_g)_\tau$.  
  
From Eq.~(\ref{k3pcpz}), neglecting  $\Delta I=3/2$ amplitudes, follows
\beq
\left( \delta_g \right)_\tau = { (\alpha_0 - \beta_0) \over \Real a_c}
\left[ \Imm b_c {\Real a_c \over \Real b_c }- \Imm a_c \right]  
\label{k3pcpz1};
\eeq
since
\beq
\Imm a_c^{(2)} -
{\Real a_c^{(2)} \over \Real b_c^{(2)}} \Imm b_c^{(2)} 
\simeq  \omega \Imm a_c^{(2)},
\eeq  
expanding  imaginary parts at order $p^2$  we obtain,
as anticipated, a result proportional to $\omega$. 
On the other hand,
expanding the imaginary parts up to
$O(p^4)$, and neglecting  $O(\omega)$ terms, leads to
\beq
\left( \delta_g \right)_\tau = (\alpha_0 - \beta_0)\left[
{ \Imm b_c^{(4)}  \over \Real b_c^{(2)} }-{\Imm a_c^{(4)} \over \Real a_c^{(2)}
} \right].  
\eeq
In the more optimistic case we can expect
that the two $O(p^4)$
phases are of the same order of $\Imm A_0/\Real A_0$ and that
 their interference is constructive, thus\cite{DIP}
\beq
|(\delta_g)_\tau| < 2(\alpha_0 - \beta_0)\left\vert
{ \Imm A_0  \over \Real A_0 } \right\vert <  10^{-5}.
\label{finallim}
\eeq
       
A final comment on the value of $(\delta_g)_\tau$ before going on. 
The limit   (\ref{finallim})  is proportional to the 
phase difference
$(\alpha_0-\beta_0)\simeq 0.08$ which {\it is not} equal to the 
difference  between constants and $Y$--dependent 
 re--scattering terms, as explicitly shown in 
Eq.~(\ref{superdec3}). In the literature this subtle
difference has been sometimes ignored (probably due to numerical 
analysis of the  re--scattering) and, as a consequence, 
overestimates of  $(\delta_g)_\tau$ have been obtained\cite{Belkov}.

\subsubsection{The parameters $\epsp_{+-0}$ and $\epsilon^X_{+-0}$.}
\label{subsez:k3pneutri}

Defining the  `weak phases' 
\beq
\phi_8 = {\Imm G_8 \over \Real G_8}, \qquad
\phi_{27} = {\Imm G_{27} \over \Real G_{27}} \qquad \mbox{and} \qquad
\phi_{\underline 8} = {F^2 \over M_K^2 }
  {\Imm G_{\underline 8} \over \Real G_{27}},
\eeq
we can write
\beqa
\epsilon'_{+-0} &=& i \omega \sqrt{2} \left[ \phi_8 - 
  \phi_{27} +{3\over 5} \phi_{\underline 8} +O(\omega,\rho_\pi)\right], \no \\
\epsilon^X_{+-0} &=& i \left[ \phi_{27} -  \phi_8
  +{1\over 30} \phi_{\underline 8} +O(\omega,\rho_\pi) \right].  
\eeqa
For  $\epsilon'_{+-0}$ the situation is exactly the same as 
for $(\delta_g)_\tau$, i.e. the $\omega$--suppression
could be removed by next--order CHPT corrections\cite{Donoghuek3p}. 
On the other hand for $\epsilon^X_{+-0}$,
which is necessarily proportional  to the phase difference 
between a $\Delta I=3/2$  ($b_2$) and a $\Delta I=1/2$ amplitude, 
the lowest--order prediction is definitively more stable with respect to  
next--order corrections.

At the leading order in CHPT  there is a simple 
 relation between  $\epsilon'_{+-0}$ and $\epsilon'$:
\beq
\epsilon'_{+-0} = -2i|\epsp|[1+O(\Omega_{IB},\omega,\rho_\pi)],
\label{LiWolfrel}
\eeq
It is interesting to note that this relation, obtained   
many years ago by Li and Wol\-fen\-ste\-in\cite{LiW}, who considered
only  $(8_L,1_R)$ and $(27_L,1_R)$ operators, is still valid  
in presence of the lowest--order $(8_L,8_R)$ operator.

For what concerns  next--order corrections, analogously to the case 
of  $(\delta_g)_\tau$, we can estimate the upper limit for the
enhancement of  $\epsilon^X_{+-0}$ and $\epsilon'_{+-0}$
with respect to $\epsilon'$. In the more optimistic case, 
we can assume to avoid the $\omega$--suppression and 
the accidental cancellation between 
$B_6$ and $B_8$ in (\ref{epsfen}), without `paying' anything 
for having considered next--to--leading--order terms in  CHPT. 
According to  this hypothesis, from Eq.~(\ref{epsfen}) follows
\beq
|\epsilon'_{+-0}|, |\epsilon^X_{+-0}| \lsim 3\times 10^{-3}\omega^{-1}
 |\eps| A^2\sigma \sin\delta \sim  5\times 10^{-5}.
\eeq                                        

\subsection{Interference measurements for $\eta_{3\pi}$ parameters.}
\label{subsez:k3int}

The parameters $\eta_{000}$, $\eta_{+-0}$ and $\eta^X_{+-0}$, 
being defined as the ratio of two amplitudes 
(analogously to $\eta_{+-}$ and $\eta_{00}$ of $K\to2\pi$), 
can be directly measured only by the analysis 
of the interference term in the  time evolution 
of neutral kaons. This kind of measurement, achievable 
by several experimental apparata\cite{E621eta,CPLEAReta},
assumes a particular relevance 
in the case of the 
$\Phi$--factory\cite{Amelino,DamPav,handbook}. Since this method is very 
general and is useful for instance also in  $K_{L,S}\to 2\pi\gamma$
decays, we will briefly discuss it  (see Ref.\cite{handbook}
for a more detailed  discussion). 

The  antisymmetric  $\Ko-\Kob$ state,  
produced by the  $\Phi$ decay,
can be written as 
\beq
\phi \rightarrow 
\frac{N}{\sqrt{2}}\left[ K_{ S }^{(\svect{q})}K_{ L }^{(-\svect{q})} - 
K_{L}^{(\svect{q})}K_{ S }^{(-\svect{q})} \right],
\eeq
where $\svect{q}$ denotes the spatial momenta of one of the 
two kaons and  $N$ is a normalization factor.
The  decay amplitude in the final state
 $|a^{(\svect{q})}(t_1)$, $b^{(-\svect{q})}(t_2)\rangle$
is thus given by:
\beqa
A\left(a^{(\svect{q})}(t_1),b^{(-\svect{q})}(t_2)\right)&=
&{N \over {\sqrt 2}}\Big[ A(K_S\to 
a) e^{-i\lambda_S t_1} A(K_L\to b)e^{-i\lambda_L t_2} \nonumber \\ 
& &-A(K_L\to a)e^{-i\lambda_L t_1}A(K_S\to b)e^{-i\lambda_S t_2} \Big].
\eeqa
Integrating the modulus square of this amplitude  with respect
to $t_1$ and $t_2$, keeping fixed the difference  $t=t_1-t_2$,
and integrating with respect to all possible directions of
 $\svect{q}$,  leads to
\beqa
 && I(a,b;t) = \int \mbox{d}\Omega_q  \mbox{d}t_1 \mbox{d}t_2 
            |A (a(t_1),b(t_2))|^2 \delta(t_1-t_2-t) \qquad \qquad \no \\
  &&\qquad \propto { e^{-\Gamma|t|} \over 2\Gamma} \Big\lbrace 
|A_{S}^{a}|^2|A_{L}^{b}|^2 e^{-{\Delta\Gamma\over 2}t} \no\\
 &&\qquad\quad  +|A_{L}^{a}|^2|A_{S}^{b}|^2 e^{+{\Delta\Gamma\over 2}t}
  -2\Re\left[ A_S^a A_L^{a*} A_L^b A_S^{b*}
e^{+i\Delta m t}\right] \Big\rbrace, 
\label{intt}
\eeqa
where 
\beq
\Gamma={\Gamma_S+\Gamma_L\over 2},\qquad\qquad \Delta\Gamma=\Gamma_S-\Gamma_L 
\qquad \mbox{\rm and}\qquad \Delta m=m_L-m_S. 
\eeq
$I(a,b;t)$ represents the probability to have in the final state
$K_{S,L}\to a$ and $K_{L,S}\to b$ decays
separated by a time interval $t$. 

 \begin{figure}
     \begin{center}
      \setlength{\unitlength}{1truecm}
       \setlength{\unitlength}{1truecm}
       \begin{picture}(8.0,8.0)
       \epsfxsize 14. true  cm
       \epsfysize 20. true cm
       \epsffile{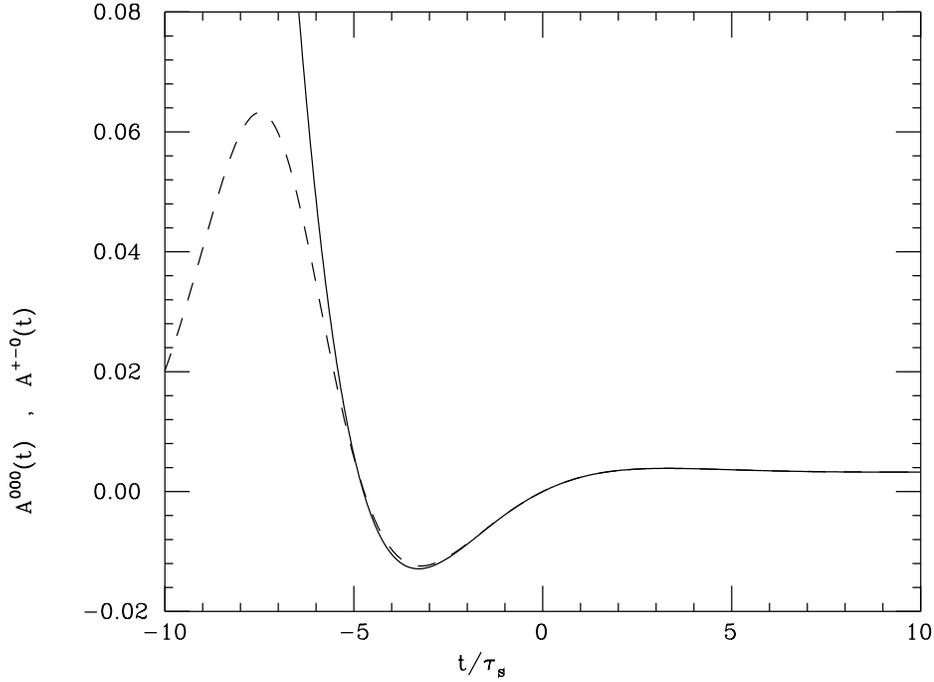}
       \end{picture}   
    \end{center}
    \caption{The asymmetries $A^{000}(t)$ (full line) and $A^{+-0}(t)$ 
(dotted line). The plots have been obtained fixing
$ \eta^{000} =\eta^{+-0}= |\epsilon|e^{i\pi/4}$.}
    \protect\label{figksa}
\label{fig:interf} 
\end{figure}

By choosing appropriately $\ket{a}$ and $\ket{b}$, 
it is possible to construct interesting asymmetries. As an example,
a convenient choice to study $K\to 3\pi$ amplitudes  is 
given by $\ket{a}=\ket{3 \pi}$ and $\ket{b}=\ket{\pi l \nu}$ (as shown in 
sect.~\ref{sez:semil}, $|A(K_S\to \pi l \nu)|=|A(K_L\to \pi l \nu)|$),
which let us consider the following asymmetry\cite{handbook}
\beqa
A^{123}(t)
&=& \displaystyle{  \int\left[ I(\pi^1\pi^2\pi^3, l^+ \pi^- \nu; t) -
 I(\pi^1\pi^2\pi^3, l^- \pi^+ \nu; t) \right]d\phi_{3\pi}\, d\phi_{l\pi\nu}
\over \int\left[ I(\pi^1\pi^2\pi^3, l^+ \pi^- \nu; t) +
 I(\pi^1\pi^2\pi^3, l^- \pi^+ \nu; t) \right]d\phi_{3\pi} \, d\phi_{l\pi\nu}}, 
 \nonumber \\
        &=& \displaystyle{ 2(\Re \epsilon)e^{+{\Delta \Gamma  \over 2}t} 
-2\Re\left(\eta^{123}e^{+i \Delta m t}\right)  \over 
 e^{+{\Delta \Gamma  \over 2}t} +{\Gamma_S^{123} \over \Gamma_L^{123}} 
 e^{-{\Delta \Gamma  \over 2}t}  }, 
\label{apm0}
\eeqa
where $d\phi_{3\pi}$ and $d\phi_{l\pi\nu}$ 
indicate final-state phase--space elements.

The peculiarity\cite{Fukawa}  of  $A^{123}(t)$, with respect 
to analogous distributions measurable in different 
experimental set up, like fixed--target experiments, is the fact that
 $A^{123}(t)$ can be studied for  $t<0$. Events 
with $t<0$ are those where the  semileptonic decay occurs after 
the three--pion one, thus, as can be seen from Eq.~(\ref{apm0}) and
 fig.~\ref{fig:interf}, are much more sensible to the 
$CP$--violating $K_S \to 3\pi$ amplitude. Obviously the
statistics of these events is very low, and tends to zero 
for $t\ll 0$, but for small times ($|t|\lsim 5\tau_S$) 
the decrease of statistics does not compensate the increase of 
 sensibility.

The asymmetry  $A^{123}(t)$ is very useful to measure both  
$\eta_{000}$ and $\eta_{+-0}$. The measurement of  
$\eta^X_{+-0}$ is more difficult
since it requires an $X$-odd integration over the 
Dalitz Plot\cite{handbook} which drastically reduces the statistics.

At any rate, the sensitivity which should be reached on
$\eta_{000}$ and $\eta_{+-0}$ at KLOE is of the 
order of $10^{-3}$, still far form direct--$CP$--violating
effects expected in the  Standard  Model. Present bounds on 
$\eta_{+-0}$ are of the order of $10^{-2}$.\cite{E621eta,CPLEAReta}

\setcounter{equation}{0}
\setcounter{footnote}{0}
\section{$K\to \pi\pi\gamma$ decays.}
\label{cap:Kppg}

\subsection{Amplitude decomposition.}
\label{sez:Kppgdec}

The channels of
$K\to \pi\pi\gamma$ transitions are three:
\beq \ba{rcl} 
K^\pm &\to& \pi^\pm \pi^0 \gamma, \\
\Ko(\Kob) &\to& \pi^+ \pi^- \gamma, \\
\Ko(\Kob) &\to& \pi^0 \pi^0 \gamma, \ea
\eeq 
in any channel is possible to distinguish an electric
($E$) and a   magnetic  ($M$) amplitude.
The most general form, dictated by gauge and Lorentz 
invariance, for the transition amplitude
$K(p_K) \to \pi_1(p_1)\pi_2(p_2)\gamma(\epsilon, q)$
is given by:
\beq
A(K\to\pi\pi\gamma)= \epsilon_{\mu}\left[ E(z_i) (qp_1p_2^\mu - qp_2p_1^\mu)
+ M(z_i) \epsilon^{\mu\nu\rho\sigma}p_{1\nu}p_{2\rho}q_{\sigma}\right]
/M_K^3,  \label{kppgamp1} 
\eeq
where
\beq
z_i \doteq {p_iq \over M_K^2} \quad(i=1,2)\qquad \mbox{and} \qquad
z_3 \doteq z_1+z_2 = {p_K q \over M_K^2}. 
\eeq 
$E$ and $M$ thus defined 
are adimensional. Summing over the photon--helicity states, the 
differential width of the decay
is given by:
\beqa
{ \mbox{d}\Gamma \over \mbox{d}z_1\mbox{d}z_2 } &=& { M_K  \over 4 (4\pi)^3 }
\left( |E(z_i)|^2 + |M(z_i)|^2 \right) \no \\
&& \times \left[ z_1 z_2 (1-2z_3-r_1^2  -r_2^2)-r_1^2 z_2^2 - r_2^2
z_1^2 \right], \label{kppgw}      
\eeqa  
where $r_i=M_{\pi_i}/M_K$. Thus there is no interference 
among $E$ and $M$ if the photon helicity is not 
measured. 

In the  limit where the photon energy goes to zero, the 
electric amplitude is completely determined by 
Low theorem\cite{Low} which relates $\Gamma(K\to \pi\pi\gamma)$ to     
$\Gamma(K\to \pi\pi)$. 
For this reason is convenient to re--write $E$ in two parts:
the `bremsstrahlung' $E_{IB}$ and the `direct--emission' $E_{DE}$. 
The bremsstrahlung amplitude, fixed by  
Low theorem, diverges for $E_\gamma \to 0$
and corresponds in the classical limit
to the external--charged--particle radiation. 
If $eQ_i$ is the electric charge of the
pion $\pi_i$, we have 
\beq
E_{IB}(z_i) \doteq { e A(K\to \pi_1\pi_2) \over  M_K
z_3 }\left ( {Q_2 \over z_2}  
- {Q_1 \over z_1} \right).   
\eeq

\begin{table}[t] 
\[ \ba{|lr|c|c|} \hline
\multicolumn{2}{|c|}
{\mbox{decay}}  & BR(\mbox{bremsstrahlung}) & BR(\mbox{direct 
emission})  \\ \hline
K^\pm \to \pi^\pm \pi^0 \gamma\quad & _{ ( T_c^*=(55-90) MeV ) }  
& (2.57 \pm 0.16 ) \times 10^{-4} & (1.8 \pm 0.4 )\times 10^{-5} \\ \hline 
K_S \to \pi^+ \pi^- \gamma  & _{ ( E_\gamma^* > 50 MeV ) }  
& (1.78 \pm 0.05 )\times 10^{-3} & < 9 \times 10^{-5} \\ \hline 
K_L \to \pi^+ \pi^- \gamma    & _{ ( E_\gamma^* > 20 MeV ) }  
& (1.49 \pm 0.08 )\times 10^{-5} & (3.19 \pm 0.16 )  \times 10^{-5} \\ \hline 
K_S \to \pi^0 \pi^0 \gamma & & /  & 
- \\ \hline 
K_L \to \pi^0 \pi^0 \gamma & & /  & 
< 5.6 \times 10^{-6} \\ \hline \ea \]
\caption{Experimental values of  $K\to \pi\pi\gamma$  branching 
ratios\protect\cite{PDG} ($E_\gamma^*$ and $T_c^*$ are  
the photon energy and the $\pi^+$ kinetic energy in the 
kaon rest frame, respectively).}
\label{tab:Kppgexp} 
\end{table}  
\begin{table}[t] 
\[ \ba{|c|c|c|c|c|c|} \hline
\mbox{\rm process}  &  E_{IB}  &  E_1  &  M_1 &  E_2   &  M_2  \\ \hline
K^\pm \to \pi^\pm \pi^0 \gamma  & \omega & \cdot & \cdot & \cdot & \cdot 
  \\ \hline 
K_S \to \pi^+ \pi^- \gamma  & \cdot & \cdot &  CP  &  CP  & \cdot \\ \hline 
K_L \to \pi^+ \pi^- \gamma  & CP & CP  &  \cdot  &  \cdot  & CP \\ \hline 
K_S \to \pi^0 \pi^0 \gamma  & / & /
& /  & CP & \cdot \\ \hline
K_L \to \pi^0 \pi^0 \gamma  & / & /
& /  & \cdot & CP \\ \hline
\ea \]
\caption{Suppression factors for  $K\to \pi\pi\gamma$ amplitudes: 
$\cdot$ = allowed transitions, $CP$ = $CP$--violating transitions, 
$\omega$ = amplitudes suppressed by the $\Delta I=1/2$ rule, 
$/$ =  completely forbidden 
 amplitudes (by $Q_1=Q_2=0$ or by Bose symmetry). }
\label{tab:Kppgsopp} 
\end{table}   

\noindent
The electric direct emission amplitude is by definition
$E_{DE} \doteq E  - E_{IB}$ and, according to Low theorem,
we know that 
$E_{DE} = \mbox{cost.} + O(E_\gamma)$. The magnetic term by construction 
does not receive  bremsstrahlung contributions (is a pure direct--emission 
term) thus, analogously to the previous case,
$M = \mbox{cost.} + O(E_\gamma)$.
As can be noticed by table~\ref{tab:Kppgexp}, 
in the cases where the corresponding $K\to \pi\pi$ amplitude is not 
suppressed, the pole for  $E_\gamma \to 0$ naturally enhances 
the bremsstrahlung contribution respect  to the direct emission one.

The last decomposition which is convenient to  introduce is the 
so--called   multipole expansion for the 
direct--emission amplitudes
$E_{DE}$ and $M$:
\beqa
E_{DE}(z_i) &=& E_1 + E_2 (z_1-z_2) + O\left[(z_1-z_2)^2\right], \\
M(z_i) &=& M_1 + M_2 (z_1-z_2) + O\left[(z_1-z_2)^2\right]. 
\eeqa
This decomposition is useful essentially for two reasons: i)
since the phase space is limited ($|z_1-z_2|<0.2$) 
high--order multipoles are suppressed; ii) in the neutral 
channels even and odd multipoles have different $CP$--transformation 
properties:
$CP(E_J)=(-1)^{J+1}$,  $CP(M_J)=(-1)^{J}$.

\subsubsection{$CP$--violating observables.}
\label{sez:KppgCP}
              
As can be noticed by  tables~\ref{tab:Kppgexp} and~\ref{tab:Kppgsopp}, 
 $K_S \to \pi^+\pi^-\gamma$ and 
$K_{S,L} \to \pi^0\pi^0\gamma$  decays are not very interesting for 
the study of $CP$ violation. The first is dominated by the 
bremsstrahlung, which `hides' other contributions, 
whereas neutral channels are too suppressed 
to observe any kind of interference. The 
theoretical branching ratios for the latter\cite{ENP1}$^-$\cite{DI} 
are below $10^{-8}$.

Interesting channels for direct $CP$ violation
 are $K^\pm \to \pi^\pm \pi^0\gamma$
 and $K_L \to \pi^+ \pi^-\gamma$, where the 
 bremsstrahlung is suppressed and consequently it is easier to 
 measure interference between the latter and other amplitudes.
 If the photon polarization is not measured and the multipoles
 $E_2$ and $M_2$ are neglected, we can define only 
two observables which violate $CP$:
\beqa
\eta_{+-\gamma} &=& { A (K_L \to \pi^+\pi^-\gamma)_{E_{IB} +E_1} \over
A (K_S \to \pi^+\pi^-\gamma)_{E_{IB} +E_1} }, \label{k3pgeta} \\  
\delta\Gamma &=& { \Gamma (K^+ \to \pi^+ \pi^0 \gamma ) - 
\Gamma (K^- \to \pi^- \pi^0 \gamma ) \over  
\Gamma (K^+ \to \pi^+ \pi^0 \gamma ) + \Gamma (K^- \to \pi^- \pi^0 \gamma )}.
\label{k3pgdel}
\eeqa              
In the case where also $E_2$  and the photon polarization
are considered, it is possible to add other two 
$K_L \to \pi^+\pi^-\gamma$ observables, proportional to the 
interference of  $(E_{IB}+E_1)$ with 
 $E_2$ and $M_1$.
The first is the  Dalitz Plot asymmetry in the 
 $\pi^+\leftrightarrow
\pi^-$ exchange, the second is the  $\phi\to -\phi$ asymmetry, where
 $\phi$ is the angle between the $\gamma$--polarization plane and the 
 $\pi^+-\pi^-$ plane.
However, these  observables are less 
 interesting than  those of Eqs.~(\ref{k3pgeta}--\ref{k3pgdel}), 
because are not pure signals of 
direct $CP$ violation and  
are suppressed by the 
interference with higher order multipoles\cite{Sehgal,DI}. 
In the following we will not 
 consider them.

By the definition of  $\eta_{+-\gamma}$, using the 
identities 
\beqa
E_{IB}(K_L)&=&\eta_{+-}E_{IB}(K_S), \\ 
E_1(K_L)&=&\veps E_{1}(K_1)+E_{1}(K_2),
\eeqa
it follows
\beqa
\eta_{+-\gamma} &=& { E_{IB}(K_L)+E_{1}(K_L)  \over 
 E_{IB}(K_S)+E_{1}(K_S) } \no \\
		&=& \eta_{+-} +{
(\veps -\eta_{+-})E_1(K_1) + E_1(K_2) \over  E_{IB}(K_S)}\left[1+
O\left( { E_{1}(K_S)  \over E_{IB}(K_S) } \right) \right].
\eeqa
From the previous equation  we deduce that, 
contrary to the statement of Cheng\cite{Cheng1}, 
the difference 
\beq
\epsilon'_{+-\gamma} \doteq \eta_{+-\gamma}-  \eta_{+-}
\eeq
is an  index of direct $CP$ violation. 
Identifying in Eq.~(\ref{kppgamp1}) the $(p_1,p_2)$ pair with 
 $(p_+,p_-)$ and factorizing strong phases, we can write
\beqa
E_1(K_1) &=& e^{i\delta_n}  \Real E_n,  \label{kppg22} \\
E_1(K_2) &=& i e^{i\delta_n}  \Imm E_n,   \\
E_{IB}(K_S) &=& - e^{i\delta_0} \left({ e \sqrt{2} 
\Real A_0 \over M_K z_+z_-} \right)
\left[ 1+ O(\omega,\epsilon) \right], \label{kppg23}
\eeqa
where $E_n$  is a  complex amplitude 
which becomes real in the limit of  
$CP$ conservation. Using this decomposition we find 
\beqa                         
\epsilon'_{+-\gamma} &=& { e^{i(\delta_n -\delta_0)} M_K z_+z_- 
\Real E_n \over e \sqrt{2}  \Real A_0 } \left[ \epsp +i\left( 
{ \Imm A_0 \over \Real A_0 } - { \Imm E_n \over \Real E_n }  
\right)\right]\left(1+ O(\omega,\epsilon) \right) \no\\
& \simeq &  { i e^{i(\delta_n-\delta_0)} M_K z_+z_- 
\Real E_n  \over e \sqrt{2} \Real A_0 } 
 \left(  { \Imm A_0 \over \Real A_0 } -
{ \Imm E_n \over \Real E_n }  
\right),
\eeqa
where the second identity 
follows from the fact that the weak--phase difference 
between  $A_0$ and $E_n$
is not suppressed by $\omega$.

The observable $\delta \Gamma$ is a pure index of direct $CP$ violation.
Actually, analogously to the case of $K^\pm\to(3\pi)^\pm$ decays, 
instead of the width asymmetry is more convenient 
to consider the asymmetry of quantities which are directly 
proportional to interference terms
(like the  $g^\pm$ slopes in $K^\pm\to(3\pi)^\pm$). 
For this purpose is useful to 
consider the quantities $\Gamma^\pm_{DE}(E_\gamma^*)$, defined by
\beq
\Gamma^\pm_{DE}(E_\gamma^*) = \int_0^{E_\gamma^*} 
\mbox{d} E_\gamma  \left[ {\partial \Gamma(K^\pm \to \pi^\pm\pi^0 
\gamma) \over  \partial E_\gamma } - 
{\partial \Gamma(K^\pm \to \pi^\pm\pi^0 
\gamma)_{IB} \over  \partial E_\gamma }\right],   
\eeq
where  $\Gamma(K^\pm \to \pi^\pm\pi^0 \gamma)_{IB}$ is obtained  by
Eq.~(\ref{kppgw}) setting $E=E_{IB}$ and $M=0$. 
In the limit where the  magnetic term in Eq.~(\ref{kppgw}) 
is negligible, the expression of 
\beq
\delta \Gamma_{DE}= \frac{\Gamma_{DE}^+-\Gamma_{DE}^-}{
\Gamma_{DE}^++\Gamma_{DE}^-}
\eeq
is very simple: setting $(p_1,p_2)\equiv(p_\pm,p_0)$ and factorizing
strong phases analogously to 
Eqs.~(\ref{kppg22}--\ref{kppg23})
\beqa
E_1(K^\pm) &=&  e^{i \delta_c } E_c,  \label{kppg25} \\
E_{IB}(K^\pm) &=& - e^{i\delta_2} \left({ 3 e \Real A_2 \over 2 M_K z_\pm z_0
}\right), \label{kppg26}
\eeqa
we obtain:
\beq
\delta \Gamma_{DE}=\frac{ \Imm( A_2 E_c^*) \sin(
\delta_2-\delta_c) 
}{ \Real( A_2 E_c^*) \cos(\delta_2-\delta_c) }
\doteq \epsilon'_{+0\gamma} \tan (\delta_c-\delta_2).  
\label{kppg30}
\eeq
If the magnetic term is not negligible, 
Eq.~(\ref{kppg30})  is modified in  
\beq    
\delta \Gamma_{DE}=\frac{ \epsilon'_{+0\gamma} }{1+R}
 \tan ( \delta_c -\delta_2),  \label{kppg31}  
\eeq
where 
\beqa
R =\left\{ \int \mbox{d}z_+\mbox{d}z_0 
 \left[ z_+ z_0 (1-2z_3-r_+^2  -r_0^2)-r_+^2 z_0^2 - r_0^2
z_+^2 \right] |M(z_i)|^2   \right\}\times\qquad && \no\\
\times \left\{  2 \mbox{d}z_+\mbox{d}z_0  \left[ z_+ z_0 
(1-2z_3-r_+^2  -r_0^2)-r_+^2 z_0^2 - r_0^2
 z_+^2 \right] \Real(E_1^*(z_i)E_{IB}(z_i))  \right\}^{-1}.
\eeqa
Analogously to $\epsilon'_{+-\gamma}$, also
\beq
\epsilon'_{+0\gamma} = \left( 
{ \Imm E_c \over \Real E_c }  - { \Imm A_2 \over 
\Real A_2 } \right) = \left( 
{ \Imm E_c \over \Real E_c }  - { \Imm A_0 \over 
\Real A_0 } \right) -{\sqrt{2} |\epsp| \over \omega }
\eeq
is not suppressed  by the $\Delta I=1/2$ rule.

\subsubsection{$K\to\pi\pi\gamma$ amplitudes in CHPT.}
\label{sez:Kppgamp}

\begin{figure}[t]
    \begin{center}
       \setlength{\unitlength}{1truecm}
       \begin{picture}(10.0,3.0)
       \epsfxsize 10.  true cm       \epsfysize 3. true cm
       \epsffile{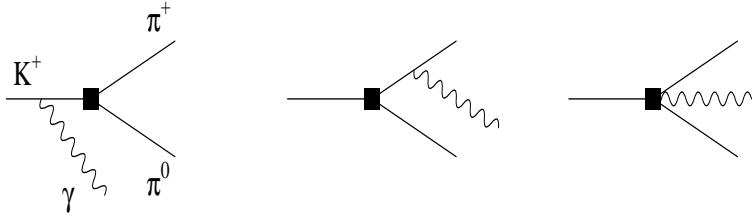}
       \end{picture}
    \end{center}
    \caption{Tree--level diagrams for the transition  
                $K^+ \to \pi^+\pi^0\gamma$.
The black box indicates the weak vertex.}
    \protect\label{fig:kppg1}
\end{figure}

\begin{figure}[t]
    \begin{center}
       \setlength{\unitlength}{1truecm}
       \begin{picture}(10.0,3.0)
       \epsfxsize 10.  true cm
       \epsfysize 3. true cm
       \epsffile{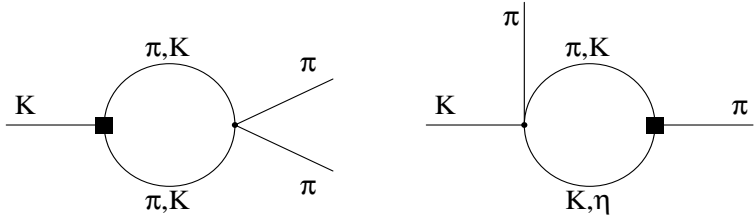}
       \end{picture}
    \end{center}
    \caption{One--loop diagrams relevant to the direct emission amplitudes 
in $K \to \pi\pi\gamma$ decays; for simplicity we have omitted the  
photon line, which has to be attached to any charged line and to 
any vertex. }
    \protect\label{fig:kppg2}
\end{figure}

The lowest order CHPT diagrams which contribute
 $K\to \pi\pi\gamma$ transitions are shown in 
fig.~\ref{fig:kppg1}. At this order only the 
bremsstrahlung amplitude is 
different from  zero. As can be easily deduced from Eq.~(\ref{kppgamp1}), 
is necessary to go beyond the lowest order to obtain 
non--vanishing contributions to  direct emission amplitudes. 
At order $p^4$ electric amplitudes receive contributions 
from both loops (\SSf\ref{fig:kppg2}) and 
counterterms, whereas the 
magnetic amplitudes receive contributions only by 
local operators\cite{ENP1}.\footnote{~In  $\pi^0\pi^0\gamma$ channels 
there is no contribution even at $O(p^4)$.} 

The complete $O(p^4)$ calculation of  electric direct--emission 
amplitudes, carried out in  Refs.\cite{ENP1,Sannino,DI},
give rise to two interesting results:
\begin{itemize}
\item{The loop contribution is finite both in 
 $\pi^+\pi^0\gamma$ and  $\pi^+\pi^-\gamma$. } 
\item{In both channels the counterterm combination is the same.}
\end{itemize}

Neglecting the small contribution of $\pi-K$ and $K-\eta$ loops,
the explicit  $O(p^4)$ expression of the weak amplitudes
$E_n$ and $E_c$ is:
\beqa
E_n &=& { e G_8  M^3_K \over 4 \pi^2 F_\pi }  \left[
{ 64 \pi^2 M_K^2 \over 1 + \rho_\pi } \Real \wt{C_{20}}(M_K^2,0)- 
N_{E_1}^{(4)} \right] \no \\
&\simeq& {e G_8  M_K^3 \over 4 \pi^2 F_\pi } 
[ 1.3  - N_{E_1}^{(4)} ], \label{kppg40}\\ 
E_c &=& {  e G_8  M_K^3  \over 8\pi^2  F_\pi } N_{E_1}^{(4)},  
\label{kppg41}
\eeqa                                
where the function $\wt{C_{20}}(x,y)$ is defined in the appendix,
$\rho_\pi = M^2_\pi /(M_K^2-M_\pi^2)$  and 
\beq
N_{E_1}^{(4)} = (4 \pi)^2 \left[ N_{14}-N_{15}-N_{16}-N_{17}\right]  
\eeq
is a $\mu$--independent combination  of  ${\cal L}_W^{(4)}$ 
coefficients (\SSp\ref{subsez:LW4}).  
%
Concerning strong phases, neglecting the small 
final--state interaction to $N_{E_1}$, we find:
\beqa
\delta_n &=& \arctan\left( {64\pi^2 M_K^2 \Imm \wt{C_{20}}(M^2_K,0) \over   
64\pi^2 M_K^2 \Real \wt{C_{20}}(M_K^2,0)- (1+\rho_\pi)\Real N_{E_1}^{(4)} }
\right)\no \\ &\simeq&  - \arctan\left( { 0.5 \over 1.3 -  
\Real N_{E_1}^{(4)} }\right), \\
\delta_c &=& 0.
\eeqa

Up to date it is impossible to determine 
the value of $\Real N_{E_1}^{(4)}$ using experimental data, 
since the available information on $K^+\to \pi^+\pi^0\gamma$ 
is not accurate enough 
to distinguish between electric and  magnetic amplitudes. To 
estimate $\Real N_{E_1}^{(4)}$ is necessary to
assume some theoretical model. In the framework of the 
factorization model discussed in sect.~\ref{sez:VMD}, which we 
expect give correct indications about sign and order of 
magnitude of counterterms, the result is
\beq
\Real N_{E_1}^{(4)} = -  k_f  { 8 \pi^2 F_\pi^2 \over M_V^2}
 = -(0.5 \div 1), \label{kppg39}
\eeq 
thus: 
\begin{itemize}
\item{In $K^+\to \pi^+\pi^0\gamma$  the interference between 
 $E_1$ and  $E_{IB}$ is positive (the loop contribution is negligible).}
\item{In $K_S\to \pi^+\pi^-\gamma$ loop and counterterm 
contributions are of the same order of the same sign and interfere 
destructively with the bremsstrahlung.}
\end{itemize}      

Regarding higher order electric multipoles, local $O(p^4)$
contributions to $E_2$ are forbidden by power--counting, 
but the kinematical dependence of the loop amplitudes
generate a non--vanishing contribution of this kind 
in $K^+\to \pi^+\pi^0\gamma$ and  
$K^0 \to \pi^+\pi^-\gamma$. However, the $O(p^4)$ prediction
for this higher order electric multipole is very small:\cite{ENP1,DI}
\beq
E_2^{(4)} (K_2) \simeq {  e G_8  M_K^3  \over 8\pi^2  F_\pi } [0.005  
(z_+ - z_-)]
\eeq
and we belive that the dominant contribution is generated 
only at $O(p^6)$, where counterterms are not forbidden. Indeed,
following Ref.\cite{DI} we can write
\beq
E_2^{(6)} (K_2) = {  e G_8  M_K^5  \over 48\pi^4  F^3_\pi } N_{E_2}^{(6)}  
(z_+ - z_-),
\eeq
and by power counting we expect $N_{E_2}^{(6)}\sim O(1)$.

For a detailed discussion about magnetic multipoles we refer the reader
the analysis of Ecker, Neufeld and Pich\cite{ENP1}.

\subsection{Estimates of $CP$ violation.}
\label{sez:kppgstimecp}
      
Using  Eqs.~(\ref{kppg40}--\ref{kppg41}) and the  
$O(p^2)$ expression of  $A_0$, is possible to relate each other
the direct--$CP$--violating parameters of $K\to\pi\pi\gamma$:
\beqa
\epsilon'_{+-\gamma}  &=&    { i e^{i(\delta_n-\delta_0)} M^2_K
\Real N_{E_1}^{(4)} z_+z_-  \over 8\pi^2 F_\pi^2 }   
{ \Imm N_{E_1}^{(4)} \over \Real N_{E_1}^{(4)} } 
[1+(\Omega_{IB},\omega,\rho_\pi)], \\
\epsilon'_{+0\gamma}  &=&  
{ \Imm N_{E_1}^{(4)} \over \Real N_{E_1}^{(4)} } -
{\sqrt{2} |\epsp| \over \omega } 
= - { i e^{i(\delta_0-\delta_n)}  F_\pi^2
 \over \Real N_{E_1}^{(4)} M^2_K z_+z_-  }  \epsilon'_{+-\gamma}
 - {\sqrt{2} |\epsp| \over \omega } \label{kppg43}
\eeqa
Eq.~(\ref{kppg43}) is the  analogous of 
Eq.~(\ref{LiWolfrel}), which relates
direct--$CP$--violating parameters
of $K\to2\pi$ and $K\to 3\pi$. However, since it 
does not imply $O(\omega)$ cancellation
among $\Delta I=1/2$ amplitudes, 
Eq.~(\ref{kppg43}) is definitively more stable than
 Eq.~(\ref{LiWolfrel})
with respect to next--order CHPT corrections. 

For what concerns  numerical estimates of 
$\epsilon'_{+0\gamma}$ and $\epsilon'_{+-\gamma}$, proceeding
similar to the $K\to 3 \pi$ case, i.e. assuming  that
all  weak phases are of the order of $\Imm A_0/\Real A_0$
and that interfere constructively, we find
\beq
|\epsilon'_{+-\gamma}| \lsim (3\times10^{-5})z_+z_-, 
\qquad \mbox{and} \qquad |\epsilon'_{+0\gamma}| \lsim  10^{-4}.
\label{kppg45}
\eeq                                        
Since the parameter $R$
introduced in Eq.~(\ref{kppg31}) is positive (due to the 
constructive interference
between  $E_{IB}$ and $E_1$ in $K^+\to \pi^+\pi^0\gamma$)
the limit on $|\epsilon'_{+0\gamma}|$ imply
\beq
|\delta \Gamma_{DE}| \lsim  10^{-4} \qquad \mbox{and} \qquad
|\delta \Gamma | \lsim  10^{-5}
\label{kppg46}
\eeq
in agreement with the estimates of Refs.\cite{McSanda,Cheng2,PaverS}.
                                
Actually, the four--quark--operator 
basis used for $K\to 2\pi$ and $K\to 3 \pi$ decays
is not complete for $K \to \pi\pi\gamma$ transitions. 
In this case we should add to ${\cal H}_{eff}^{|\Delta S|=1}$
the dimension--five electric--dipole operator\cite{Inami}:
\beq
{\cal H}_{eff}^{|\Delta S|=1; \gamma} =
{\cal H}_{eff}^{|\Delta S|=1} -
 {4 G_F \over  \sqrt{2}}  \left[   \lambda_t 
 C_{11}(\mu) O_{11}(\mu) +  \mbox{\rm h.c.}  \right],
\eeq
\beq
O_{11} = i (m_s \bar{s}_R \sigma_{\mu\nu} d_L + m_d
 \bar{s}_L \sigma_{\mu\nu} d_R ) F^{\mu\nu}.
\eeq
This operator generates a new short--distance contribution 
to the weak phases of $\Delta I=1/2$ amplitudes. 
However, the matrix elements of  
$O_{11}$  are suppressed with respect to those of 
$O_6$ (the dominant operator in the imaginary part of $A_0$), 
because are different from zero only at $O(p^6)$ in CHPT. 
Indeed, according to the chiral power counting exposed in
sect.~\ref{cap:CHPT},  we have $m_q \sim O(p^2)$, $F^{\mu\nu} \sim O(p^2)$
and $\bar{q}\sigma_{\mu\nu} q \sim \partial_{\mu}\phi \partial_{\nu}
\phi \sim O(p^2)$ (for an explicit chiral realization of  
$O_{11}$ see Ref.\protect\cite{DI}). Furthermore,
since the  Wilson coefficient of this 
operator is quite small\cite{Inami,Dib} $|C_{11}| < 2\times 10^{-2}$,
it is reasonable to expect that  limits
(\ref{kppg45}) are still valid.\footnote{~The value
of $\epsilon'_{+0\gamma}$ obtained by Dib and Peccei\protect\cite{Dib},
 that overcame the limit (\protect\ref{kppg46}), is overestimated,
as recently confirmed by one 
of the authors\protect\cite{Peccei}.}

Also in the $K\to \pi\pi \gamma$ case the situation is not very 
promising from the experimental point of view:
\begin{itemize}
\item{The parameter $\eta_{+-\gamma}$ has been recently 
measured at Fermilab\cite{FermiKL}, with an error $\sigma(\eta_{+-\gamma}) 
\sim 3 \times 10^{-4}$. In the next years
new high--statistics fixed--target experiments should 
reach $\sigma(\eta_{+-\gamma}) \gsim 10^{-5}$.}
\item{The asymmetry in the widths will be measured at 
 KLOE\cite{Alosio} with an error\cite{handbook}
$\sigma(\delta \Gamma_{DE})$ $\gsim 10^{-3}$. }
\end{itemize}

\setcounter{equation}{0}
\setcounter{footnote}{0}
\section{Decays with two photons in the final state.}
\label{cap:2gamma}

\subsection{$K  \to \gamma\gamma$.}

According to the 
photon polarizations, which can  be  
parallel  ($\sim F^{\mu\nu}F_{\mu\nu}$) or perpendicular 
($\sim \epsilon_{\mu\nu\rho\sigma} F^{\mu\nu}F^{\rho\sigma}$),
we can distinguish two channels in 
$\Ko(\Kob) \to \gamma\gamma$ transitions. 
The two channels transform  under $CP$ in such a way that 
the parameters 
\beqa
\eta_\parallel &=& { A(K_L \to 2\gamma_\parallel)  \over 
A(K_S \to 2\gamma_\parallel)  } = \epsilon + \epsilon'_\parallel, \\
\eta_\perp &=& { A(K_S \to 2\gamma_\perp) \over 
A(K_L \to 2\gamma_\perp)} = \epsilon + \epsilon'_\perp,  
\eeqa
measurable in  interference experiments,\footnote{~The 
need of interference experiments would drop if photon polarizations 
were directly measurable.}                  
would be zero if $CP$ was not violated\cite{deRafaelgg,Decker,Winstein}.

\begin{table}[t] 
\[ \ba{|l|c|} \hline
\mbox{decay}  & \mbox{branching ratio}  \\ \hline
K_L \to \gamma\gamma  & (5.73 \pm 0.27 ) \times 10^{-4} \\ \hline 
K_S \to \gamma\gamma  & (2.4 \pm 0.9 )\times 10^{-6} \\ \hline 
K_L \to \pi^0 \gamma\gamma  &  (1.70 \pm 0.28 )\times  10^{-6} \\ \hline 
K^\pm \to \pi^\pm \gamma\gamma  & \sim  10^{-6} \\ \hline 
  \ea \]  
\caption{Experimental data on
$K_{L,S} \to \gamma\gamma$ and  $K \to \pi \gamma\gamma$
decays\protect\cite{PDG,BarrKs,Shinkawa}.  }
\label{tab:Kgg} 
\end{table}  

\begin{figure}[t]
    \begin{center}
      \setlength{\unitlength}{1truecm}
       \begin{picture}(5.0,3.0)
       \epsfxsize 5. true  cm
       \epsfysize 3. true cm
       \epsffile{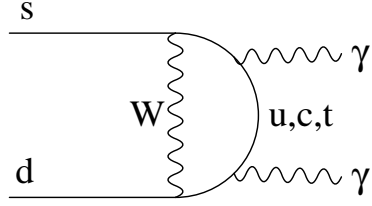}
       \end{picture}
    \end{center}
    \caption{Short--distance contribution to $K \to 
                    \gamma\gamma$ transitions. } 
  \label{fig:kgg1}
\end{figure}

It is useful to separate the   amplitude contributions 
into two classes:
the long-- and the short--distance ones. The first are 
generated by a non--leptonic transition ($K\to \pi$ or $K\to 2\pi$), 
ruled by ${\cal H}_{eff}^{|\Delta S|=1}$, 
followed by an electromagnetic process ($\pi\to\gamma\gamma$ 
or $\pi\pi\to\gamma\gamma$) which produces the two photons. The latter
are determined by new operators,  bilinear in the  quark fields,
like the electric--dipole operator (\SSp\ref{sez:kppgstimecp}) and 
the operator  generated by the box diagram of  
fig.~\ref{fig:kgg1}. By construction short--distance contributions,
recently analyzed by Herrlich  and Kalinowski\cite{Herrlichgg}, 
are either suppressed by 
the GIM mechanism or forbidden by the Furry theorem\cite{GL}. By comparing 
the short--distance calculation\cite{Herrlichgg}
with the experimental widths, we find:
\beq
\left| A_{short-d}(K\to \gamma\gamma) \over A_{long-d}(K\to \gamma\gamma) 
\right| < 10^{-4}. \label{shortkgg} 
\eeq

\begin{figure}[t]
    \begin{center}
       \setlength{\unitlength}{1truecm}
       \begin{picture}(10.0,7.0)
       \epsfxsize 10. true  cm
       \epsfysize 7.  true cm
       \epsffile{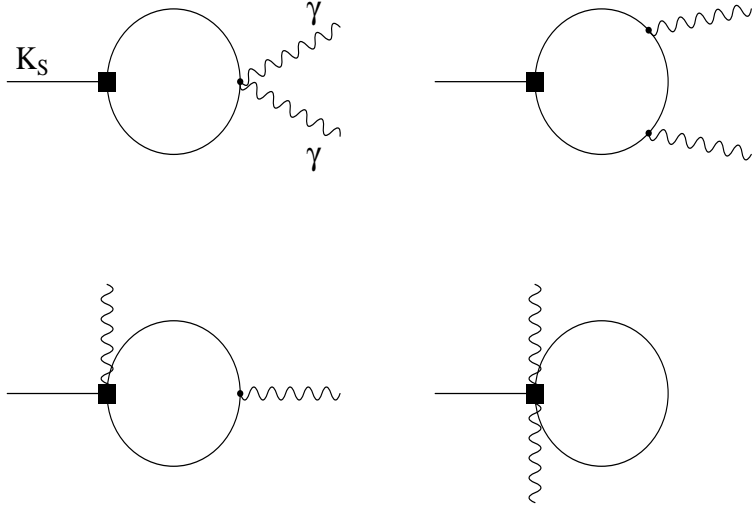}
       \end{picture}
    \end{center}
    \caption{One--loop diagrams for the transition $K_S \to \gamma
              \gamma$.}
    \label{fig:kggloop}
\end{figure}

In CHPT the first non--vanishing contribution to 
$K_S \to \gamma\gamma$ starts at  $O(p^4)$
and is  generated only by loop diagrams (\SSf\ref{fig:kggloop}).
The absence of counterterms, which implies the finiteness of
the loop calculation, leads to the  
unambiguous prediction\cite{Espriu,Goity}:
\beq
BR(K_S\to \gamma\gamma)^{O(p^4)} = 2.1 \times 10^{-6}.
\label{Kswidht}
\eeq
This result is in good agreement with the experimental data
(\SSt\ref{tab:Kgg}). Indeed, we expect that $O(p^6)$ contributions
in this channel are small because: 
i) are not enhanced by near--by resonance exchanges,
ii) unitarity correction to $\pi-\pi$ re--scattering are already 
included in the constant $G_8$.

\begin{figure}[t]
    \begin{center}
       \setlength{\unitlength}{1truecm}
       \begin{picture}(5.0,3.0)
       \epsfxsize 5.0  true  cm
       \epsfysize 3.  true cm
       \epsffile{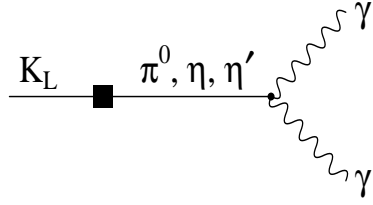}
       \end{picture}
    \end{center}
    \caption{Polar diagrams for the transition $K_L \to \gamma\gamma$.
The $P\to \gamma\gamma$ vertices of order $p^4$ are  
generated by the anomalous--functional $Z_{WZW}$.}
    \label{fig:kggpoles}
\end{figure}

If $CP$ is conserved then $K_L \to \gamma
\gamma$ does not receive any contribution at $O(p^4)$: at 
this order the pole diagrams with $\pi^0$ and $\eta$ exchange 
(\SSf\ref{fig:kggpoles}) cancel each other. 
Due to the large branching ratio of the process,  
this cancellation implies that, contrary to the
$K_S\to \gamma\gamma$ case,  $O(p^6)$ operators have to generate large
effects. Since the $CP$--violating phase of these operators contribute 
to  $\eta_\perp$, it is reasonable assume
\beq
|\epsp_\parallel |  < |\epsp_\perp |. 
\eeq
However, since\cite{Buccella} $|\epsp_\perp |^{(4)} \sim |\epsp|$,
we expect that also  $|\epsp_\perp |$ is dominated by local 
$O(p^6)$ contributions. 

Neglecting for the moment short distance effects, 
analogously to   $K\to 3 \pi$
and $K\to \pi\pi\gamma$ cases, we find
\beq
|{\epsp_\perp}|  \lsim  \left| { \Imm A_0 \over  \Real A_0} \right|.
\label{cpgg2}
\eeq
For what concerns  short  distance contributions, due to the 
suppression (\ref{shortkgg}), even if the new operators 
had a $CP$--violating phase of order one, their effect  on
${\epsp_\perp}$ and ${\epsp_\parallel}$ could not overcome the limit
(\ref{cpgg2}). Results near to this limit  
have been obtained for instance in Refs.\cite{Buccella,Chaugg}. 

\subsection{$K_L \to \pi^0 \gamma\gamma$.}
\label{subsez:klpgg}  

$\Ko(\Kob) \to \pi^0 \gamma\gamma$ transitions  
are not very interesting by themselves for the study of
$CP$ violation. However, the   process 
$K_L \to \pi^0 \gamma\gamma$ has an important role as 
 intermediate state in the 
decay $K_L \to \pi^0 e^+e^-$, that is very interesting for 
the study
$CP$ violation (\SSp\ref{sez:kpff}). 

The  $CP$--invariant decay amplitude of   
$K_L \to \pi^0 \gamma\gamma$  can be 
decomposed in the following way:
\beq
M(K_L(p)\rightarrow \pi^0(p')\gamma (q_1,\varepsilon_1)
 \gamma (q_2,\varepsilon_2))=
\varepsilon_{1\mu}\varepsilon_{2\nu}M^{\mu\nu}(p,q_1,q_2)~,
\label{eq:klpggaa}
\eeq
where
\beqa
M^{\mu\nu}&=&{A(y,z) \over M_K^2}(q_2^{\mu}q_1^{\nu}-q_1q_2g^{\mu\nu})
\nonumber \\
& &+{2B(y,z) \over M_K^4}
(-pq_1 pq_2g^{\mu\nu}-q_1 q_2 p^\mu p^\nu +
p q_1 q_2^\mu p^\nu + p q_2 p^\mu q_1^\nu)
\label{deco1klpgg}
\eeqa
and the variables $y$ and $z$  are defined   by
\beq
y=p (q_1-q_2)/M_K^2\quad\qquad \mbox{and}\quad 
\qquad z=(q_1+q_2)^2/M_K^2~.
\label{eq:yzdef}
\eeq
Due to Bose symmetry $A(y,z)$ and $B(y,z)$ must be
symmetric for $q_1 \leftrightarrow q_2$ and consequently
depend only on $y^2$.

The physical region in the dimensionless variables y and z
is given by the inequalities
\beq
|y|\le {1\over2} \lambda^{1/2}(1,z,r_\pi^2)~,\qquad\qquad
0\le z \le (1-r_\pi)^2 ~,
\label{eq:phsp}
\eeq
where $r_\pi = M_\pi / M_K$ and $\lambda(a,b,c)$ is a kinematical
function defined by
\beq
\lambda(x,y,z) = x^2 + y^2 + z^2 - 2 (xy + yz + zx)~.
\label{lambdaxyz}
\eeq 
 From (\ref{eq:klpggaa}) and (\ref{deco1klpgg})
we obtain the double differential decay rate for unpolarized photons:
\beq
 {d^2 \Gamma \over dy \ dz} = {M_K \over 2^9\pi^3} \left
\{z^2 \vert A+B \vert ^2 +
\left[y^2-{1\over 4}\lambda(1,z,r_\pi^2) \right]^2
\vert B \vert ^2 \right\}.
\label{eq:klpggc}
\eeq

We remark that, due to the different tensor structure in (\ref{deco1klpgg}),
the $A$ and
$B$ parts of the amplitude give rise to contributions to the differential decay
rate which have different dependence on the two--photon invariant mass $z$.
In particular, the second term in (\ref{deco1klpgg})  gives a non--vanishing
contribution to $\displaystyle{{d \Gamma(K_L \rightarrow \pi^0 \gamma \gamma)
\over dz}}$ in the limit $z \rightarrow 0$. Thus the kinematical region with
collinear photons is important to extract the $B$ amplitude, that
plays a crucial role in $K_L \to \pi^0 e^+e^-$ 
 (\SSp\ref{sez:kpff}). 

\begin{figure}[t]
    \begin{center}
       \setlength{\unitlength}{1truecm}
       \begin{picture}(7.0,9.0)
       \put(-3.5,-6.0){\includegraphics{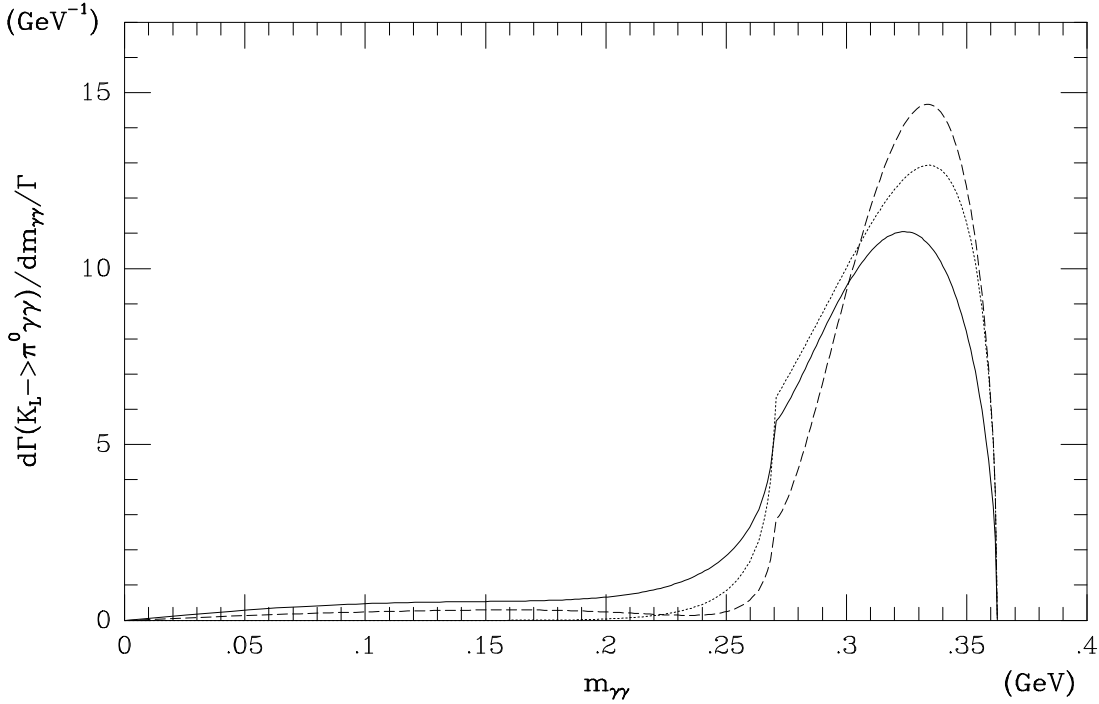}}
       \end{picture}
    \end{center}
    \caption{Theoretical predictions for the 
 width of  $K_L \rightarrow \pi^0 \gamma\gamma$
 as a function of the two--photon invariant mass. The dotted curve is the 
 $O(p^4)$ contribution,  dashed and full lines
 correspond to the  $O(p^6)$ estimates\protect\cite{Cohen,DP8} for 
 $a_V=0$ and  $a_V=-0.8$,  respectively.
 The three distributions are normalized to  
 the $50$ unambiguous events of NA31 (\SSf\protect\ref{fig:Klpgg2}).}
\label{fig:Klpgg1}
\end{figure}

\begin{figure}[t]
    \begin{center}
       \setlength{\unitlength}{1truecm}
       \begin{picture}(7.0,9.0)
       \put(-3.5,-6.0){\includegraphics{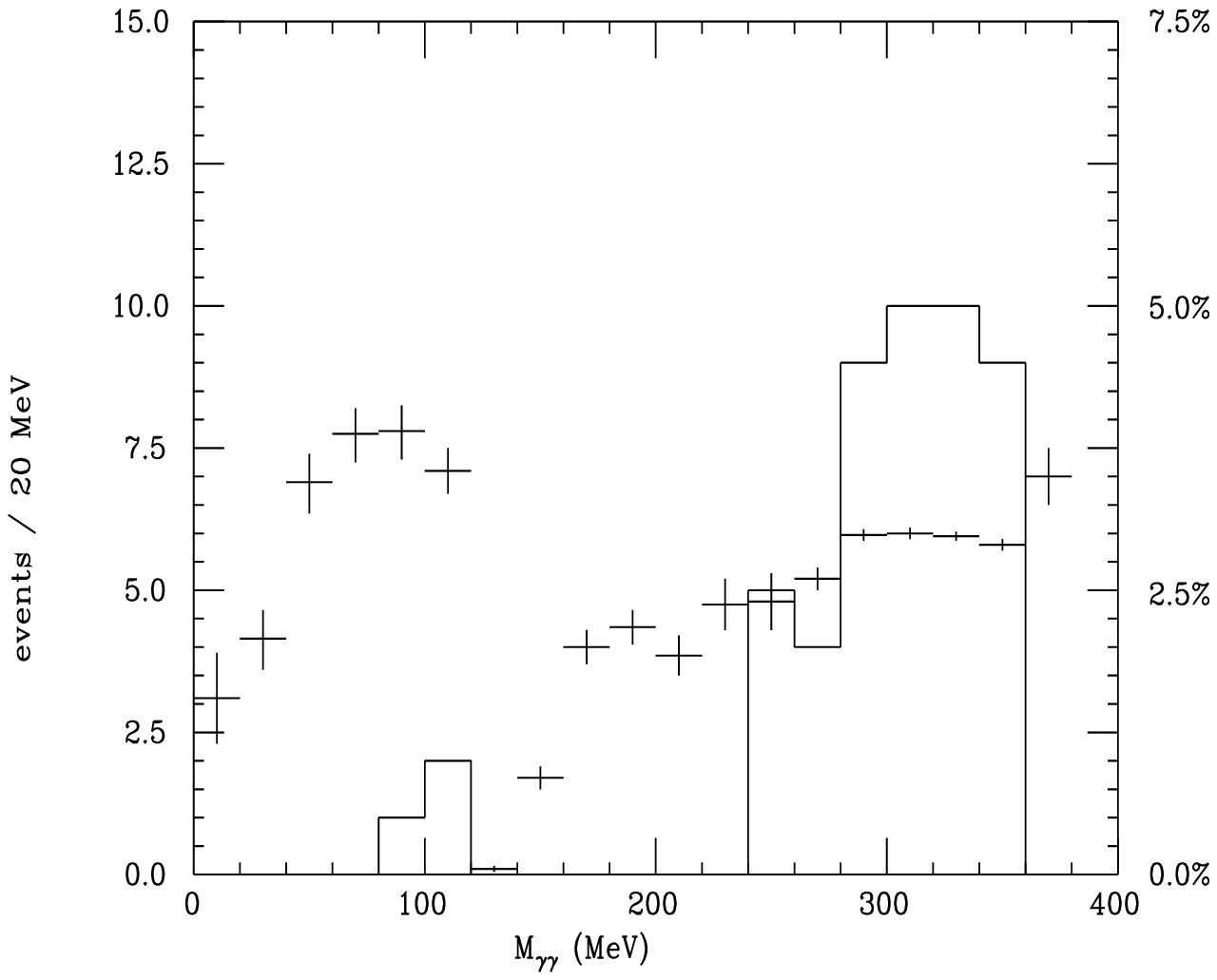}}
       \end{picture}
    \end{center}
    \caption{Distributions of the 50 unambiguously 
$K_L \rightarrow \pi^0 \gamma\gamma$ events reconstructed 
by NA31\protect\cite{NAKL} 
(histograms). Crosses indicate 
the experimental acceptance (scale on the right).}
\label{fig:Klpgg2}
\end{figure}

 Analogously to  
$K_S \to \gamma\gamma$, also 
$K_L \to \pi^0 \gamma\gamma$  receive $O(p^4)$ contributions 
only by loops, which thus are finite\cite{EPRkl,Cappiello}  and generate
only an $A$-type amplitude. The diagrams are very similar to the ones
of $K_S \to \gamma\gamma$ (\SSf\ref{fig:kggloop}) in the diagonal basis of 
Ref.\cite{EPR2}.  The shape of 
 the  photon spectrum at
 $O(p^4)$ (\SSf\ref{fig:Klpgg1}), determined by the cut
 $K_L \to 3\pi \to \pi \gamma\gamma$,  
is in perfect agreement with the data
(\SSf\ref{fig:Klpgg2}), however the branching ratio 
\beq
BR(K_L\to \pi^0 \gamma\gamma)^{O(p^4)} = 0.61 \times 10^{-6},
\label{Klwidht}
\eeq
is definitely underestimated (\SSt\ref{tab:Kgg}).
This implies that  $O(p^6)$ effects are not
negligible, nevertheless the $B$-type contribution should be small.
Though the full  ${\cal O}(p^6)$  calculation 
is still missing, several authors have
 considered some ${\cal O}(p^6)$ contributions 
(see, e.g.  Ref.\cite{Eckerrep} and 
references cited therein). 
At this order  there are counterterms and loops.

Similarly to the strong sector
one can assume that  nearby resonances generate the
bulk of the local contributions,
however we do not know the weak coupling of resonances and we have to rely
on models. 
A useful parametrization
of the local $O(p^6)$ contributions generated by vector resonances 
was introduced in Ref.\cite{EPR0},  by means of an
effective coupling $a_V$ (of order one):
\beq
A={G_8 M_K^2\alpha\over \pi } a_V(3-z+r_{\pi}^2) ~,\qquad\qquad
B=-{2G_8 M_K^2\alpha\over \pi } a_V \label{klppge}~.
\eeq
Thus, in general vector exchange can  generate a $B$ amplitude
changing the $O(p^4)$ spectrum, particularly in the region
of  small $z$, and contributing to the CP conserving part of $K_L\to\pi^0
e^+e^-$. 

Also non-local contributions play a crucial role. Indeed, 
the ${\cal O}(p^2)$ $K\to 3\pi$ vertex from (\ref{lagr2w}), used in 
the $K_L\rightarrow \pi^0 \gamma \gamma$ loop amplitude 
does not take into account the quadratic slopes of $K\to 3\pi$ 
(\S~Eq.~(\ref{slopes})) and describes the 
linear ones with 20$\%$--$30\%$ errors (\SSt\ref{tab:k3piconf}).
Only at ${\cal O}(p^4)$ the full physical $K\to 3\pi$ amplitudes are 
recovered\cite{KMWlett}. Using the latter as an effective $K\to 3\pi$ vertex
for $K_L\rightarrow \pi^0 \gamma \gamma$ leads to
a $40\%$ increase in the width and a change in the spectrum\cite{CDM,Cohen},
due to the quadratic slopes which generate a $B$ amplitude
(\SSf\ref{fig:Klpgg1}). 

Including both local and non--local effects, one can
choose appropriately $a_V$ ($a_V\sim - 0.9 $)
to reproduce the experimental spectrum and the
experimental branching ratio\cite{Cohen}.
Finally, a more complete unitarization of $\pi-\pi$ intermediate 
states 
(Khuri--Treiman treatment) and the inclusion of the experimental
$\gamma\gamma\to \pi^0\pi^0$ amplitude\cite{KaHo} increases the
$K_L\to \pi^0\gamma\gamma$ width by another 10\% and the resulting
spectrum (\SSf\ref{fig:Klpgg1}) requires a smaller 
 $a_V$ ($a_V\sim -0.8$)\cite{DP8}.
The general framework for  weak vector meson exchange to
 $K_L \rightarrow \pi^0 \gamma \gamma $ and to 
$K_L \rightarrow  \gamma \gamma^* $ has been studied in
Ref.~\cite{DP8} and the value for the slope to 
$K_L \rightarrow  \gamma \gamma^* $ has been connected 
to $a_V$. Agreement with phenomenology is met in two 
factorization models (FM and FMV). 
The factorization model with vectors
(FMV) seems to give a more complete and predictive picture\cite{DP8}.
In particular, the phenomenological 
value for the weak coupling appearing in this model
is consistent with the perturbative 
value of $C_-$ in (3.27).

Experiments  test the presence
of a $B$ amplitude  by studying
the spectrum of $K_L\rightarrow\pi^0 \gamma \gamma$ at low $z$. Since 
NA31\cite{NAKL} (\SSf\ref{fig:Klpgg2})  reports no evidence
of a $B$ amplitude, this  implies, as we shall 
see in sect.~\ref{sez:kpff}, very interesting
consequences for $K_L \to \pi^0 e^+e^-$.          
In the next section we shall see how the 
the relative role of unitarity corrections 
and vector meson contributions
can be tested\cite{DP6} also in $K^\pm \to \pi^\pm
\gamma\gamma$.

\subsection{Charge asymmetry in $K^\pm \to \pi^\pm \gamma\gamma$.}

Analogously to $K\to \gamma\gamma$ transitions, also  
$K^\pm \to \pi^\pm \gamma\gamma$ is
dominated by long--distance effects and receive the first non--vanishing 
contribution at $O(p^4)$.  However, since in this case 
the final state is not a  $CP$ eigenstate and 
contains a charged pion, 
$K^\pm \to \pi^\pm \gamma\gamma$ receive contributions not only 
from loops but also from 
$Z_{WZW}$ and non--anomalous counterterms.

The $O(p^4)$ decay amplitude can be decomposed in the following way:
\beqa
&&M(K^+(p) \rightarrow \pi^+(p') \gamma(q_1,\epsilon_1) \gamma(q_2,\epsilon_2)
 ) = \no \\ \qquad
&&\quad=\epsilon_\mu(q_1) \epsilon_\nu(q_2)
\left[ A(y,z)
\dis{ (q_2^{\mu}q_1^{\nu}-q_1 q_2g^{\mu\nu}) \over  M_K^2} +
 C(y,z)\varepsilon^{\mu\nu\alpha\beta}
\dis{q_{1\alpha}q_{2\beta}  \over M_K^2}
\right],\qquad \label{pippolippo}
\eeqa
where $C(y,z)$ is the anomalous contribution.  The variables
$y$ and $z$ and their relative phase space
 are defined in (\ref{eq:yzdef}) and (\ref{eq:phsp}).
The $O(p^4)$ result for $A(y,z)$ and $C(y,z)$ is\cite{EPR2}:
\beqa
A(y,z) &=& {G_8 M_K^2 \alpha \over 2 \pi z}\left[(r_\pi^2-1-z)
F\left(
{z \over r_\pi^2}\right) +(1 - r_\pi^2 -z) F\left(z\right)
+ \hat{c}z\right], \label{eq:kppgga}\\
C(y,z) &=& {G_8 M_K^2 \alpha \over \pi }
\left[{z-r_\pi^2 \over z - r_\pi^2
+ir_\pi{\Gamma_{\pi^0} \over M_K}} -  {z-{2+r_\pi^2 \over 3} \over
 z - r_\eta^2}\right], \label{eq:kppggb}
\eeqa
where  $r_i = M_i / M_K$ ($i=\pi,\eta$), 
$F(z)$ is defined in the appendix and 
$\hat{c}$ is a finite combination of counterterms:
\beq
\hat{c} ={ {128 \pi^2}\over{3}}[3(L_9+L_{10}) +N_{14}-N_{15}-2N_{18}] ~.
\label{eq:c-hat}
\eeq

\begin{figure}[t]
    \begin{center}
       \setlength{\unitlength}{1truecm}
       \begin{picture}(7.0,9.0)
       \put(-3.5,-6.0){\includegraphics{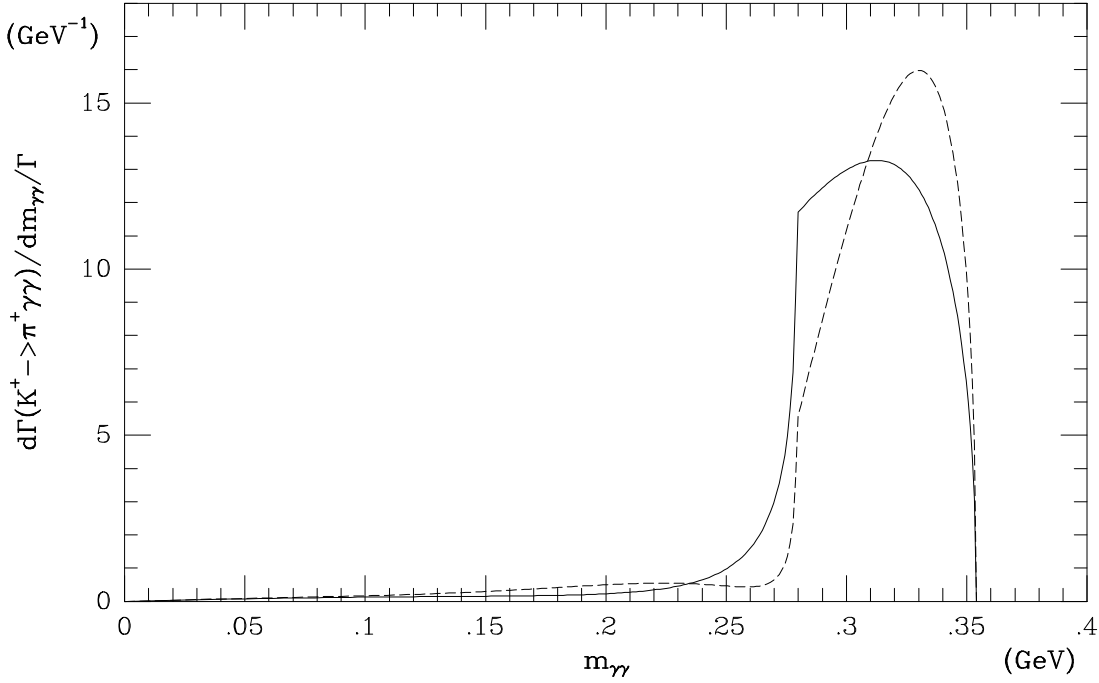}}
       \end{picture}
    \end{center}
    \caption{Theoretical predictions for the 
 normalized width of  $K^+ \rightarrow \pi^+ \gamma\gamma$
as a function of the two--photon invariant mass for
$\hat{c}=-2.3$ (NF, dashed line) and $\hat{c}=0$ (WDM, full line)
 \protect\cite{DP6}.} 
\label{fig:Kppgg1}
\end{figure}

\begin{figure}[t]
    \begin{center}
       \setlength{\unitlength}{1truecm}
       \begin{picture}(7.0,9.0)
       \put(-3.5,-6.0){\includegraphics{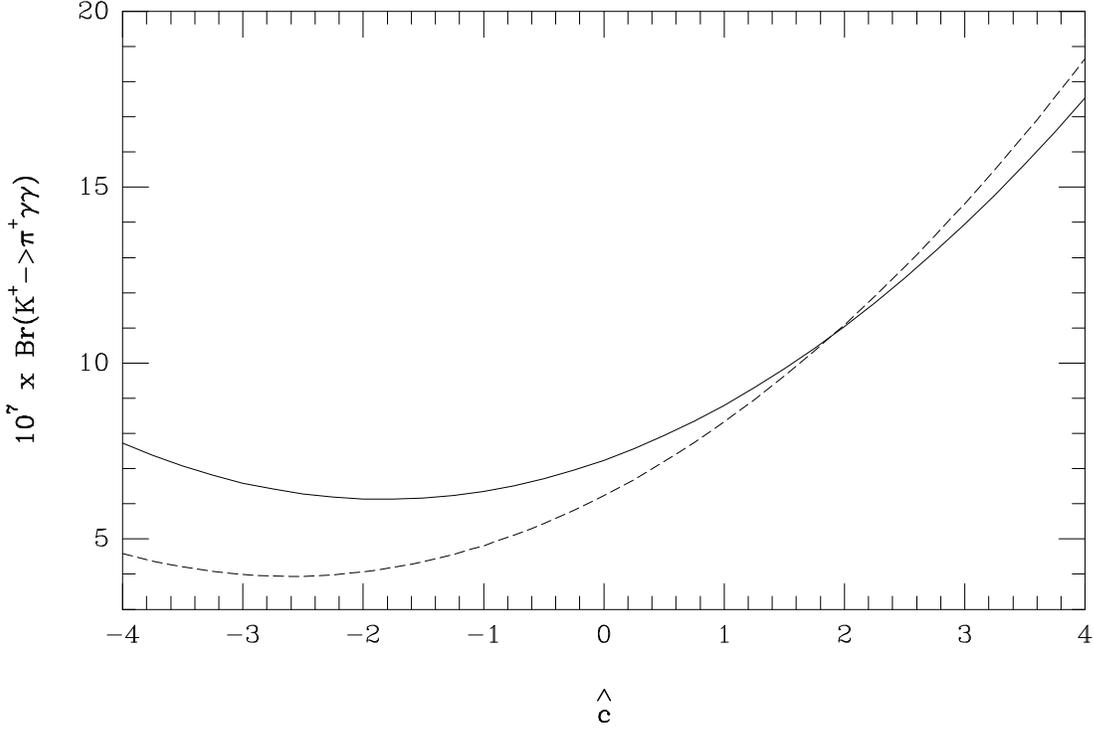}}
       \end{picture}
    \end{center}
    \caption{ $BR(K^+ \rightarrow \pi^+ \gamma\gamma)$ 
as a function of $\hat{c}$.
The dashed line corresponds to the ${\cal O}(p^4)$ CHPT amplitude. The
full line corresponds to the amplitude including the evaluated 
${\cal O}(p^6)$ corrections\protect\cite{DP6}.}
\label{fig:Kppgg2}
\end{figure}

Very similarly to $K_L \rightarrow \pi^0\gamma\gamma$ (\SSp\ref{subsez:klpgg}),
at $O(p^6)$ there are i) unitarity corrections from the inclusion
of the physical  $K \rightarrow 3\pi$ vertex in the loops, and ii)
corrections generated by 
vector meson exchange\cite{DP6,DP8}. Differently from 
 $K_L \rightarrow \pi^0\gamma\gamma$ one expects\cite{EPR0,DP6,DP8} 
that the $O(p^6)$
vector meson exchange is negligible. However, unitarity
corrections are large here too and generate a B-amplitude 
(see Eq.~(\ref{deco1klpgg})) 
as in  $K_L \rightarrow \pi^0\gamma\gamma$. The resulting 
diphoton spectrum
is shown in fig.~\ref{fig:Kppgg1} for two values of $\hat{c}$: 0.0 and
-2.3, corresponding to the theoretical predictions
of the weak deformation model (WDM) and of the naive 
factorization (NF), respectively. 
In fig.~\ref{fig:Kppgg2} is shown the
 $BR(K^+ \rightarrow \pi^+\gamma\gamma)$ as function of
$\hat{c}$. Brookhaven\cite{Shinkawa}
has actually now 30 candidates for this channel 
with a tendency to  $\hat{c}=0$.

Since loops generate an  absorptive contribution, if 
$\hat{c}$  has a non--vanishing phase, the condition {\bf [2]} 
of sect.~\ref{cap:kksystem} is satisfied and is possible to observe
direct $CP$ violation. Indeed, 
from Eqs.~(\ref{pippolippo}-\ref{eq:kppgga}) it follows\cite{EPR2}:
\beqa                                        
&& \Gamma(K^+ \rightarrow \pi^+\gamma\gamma)-
  \Gamma(K^- \rightarrow \pi^-\gamma\gamma) 
= \dis{\Imm\hat{c}|G_8 \alpha|^2 M_{K^+}^5 \over 2^{10} \pi^5 } \times
\no \\ &&\qquad\qquad
\times \int_{4 r_\pi^2}^{(1-r_\pi)^2} dz \lambda^{1 \over 2}(1,z,r_\pi^2) 
(r_\pi^2-1-z)z  \Imm F (z / r_\pi^2),
\eeqa
where $\lambda(a,b,c)$ is the kinematical function defined 
in (\ref{lambdaxyz}).
Unitarity corrections to this formula have been taken into account
in Ref.\cite{CDM} and lead to
\beq
|\delta \Gamma| =
{ | \Gamma(K^+ \to \pi^+ \gamma\gamma ) - \Gamma(K^- \to \pi^-  \gamma\gamma
 )| \over \Gamma(K^+ \to \pi^+  
\gamma\gamma ) + \Gamma(K^- \to \pi^-  \gamma\gamma
 ) } \lsim |\Imm {\hat c}|. 
\label{cpasimmkgg}
\eeq

For what concerns the estimate of $\Imm {\hat c}$, 
the situation is completely similar to the 
$K\to \gamma\gamma$ case.
Since short--distance contributions\cite{Bruno,Herrlichgg} 
are suppressed at least by a factor $10^{-4}$, we 
argue 
\beq
{ | \Gamma(K^+ \to \pi^+  \gamma\gamma ) - \Gamma(K^- \to \pi^-  \gamma\gamma
 )| \over \Gamma(K^+ \to \pi^+ 
 \gamma\gamma ) + \Gamma(K^- \to \pi^-  \gamma\gamma ) } < 10^{-4}. 
\label{cpasimmkgg2}
\eeq
Since $BR(K^+ \to \pi^+ \gamma\gamma )\lsim  10^{-6}$,
the above result implies that also this asymmetry  is far from the 
near--future experimental sensitivities.

\setcounter{equation}{0}
\setcounter{footnote}{0}
\section{Decays with two leptons in the final state.}
\label{cap:fbarf}

\subsection{$K \to \pi f {\bar f}$.}
\label{sez:kpff}
                                               
$K \to \pi f {\bar f}$ decays can be divided in 
two categories:
\beq
\mbox{(a)}\ K \to \pi l^+ l^- \qquad\qquad  \mbox{and} \qquad\qquad
\mbox{(b)}\ K \to \pi \nu \bar{\nu},  \label{kff1}
\eeq
where $l=e,\mu$.
Even if the
branching ratios of these processes  (\SSt\ref{tab:Kff})
 are very small compared to those considered before, the different 
role between short-- and long--distance contributions
make them very interesting for the study of  $CP$ violation.

\begin{figure}[t]
    \begin{center}
       \setlength{\unitlength}{1truecm}
       \begin{picture}(14.0,4.0)
       \epsfxsize 14.  true cm
       \epsfysize 4.  true cm
       \epsffile{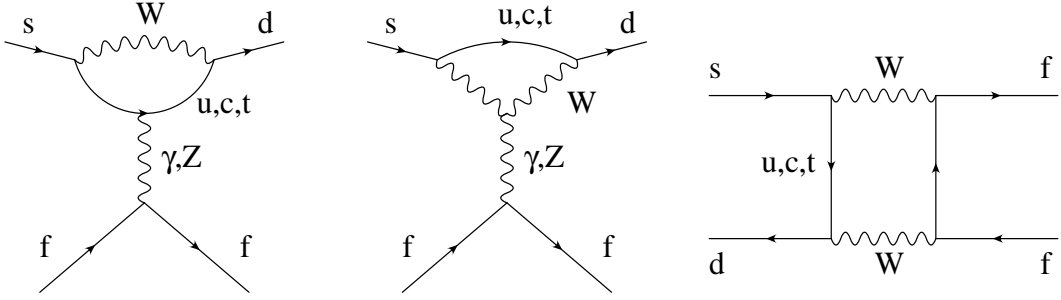}
       \end{picture}
    \end{center}
    \caption{Short--distance contributions to 
                 $K \to \pi f {\bar f}$ decays. }
    \protect\label{fig:shortdist}
\end{figure}

Short--distance contributions
are generated by the
loop diagrams in  fig. \ref{fig:shortdist},  
which give rise to the following local operators:
\beqa
O^V_{f\bar{f}} &=& \bar{s}_L \gamma_\mu d_L \bar{f} \gamma^\mu f, \\
O^A_{f\bar{f}} &=& \bar{s}_L \gamma_\mu d_L \bar{f} \gamma^\mu\gamma_5 f.
\eeqa
Due to the 
GIM suppression, the dominant contribution to the 
Wilson coefficients of these operators is generated by the 
quark top and is proportional to $\lambda_t$. Thus 
short--distance contributions carry a large $CP$--violating phase.

There are two kinds of long--distance contributions. First of all
$K\to\pi\gamma^*(Z^*)$ transitions, ruled by
the non--leptonic weak hamiltonian (\ref{ciuham}).  
Secondly, but only for case (a), 
$K\to\pi\gamma\gamma$ transitions followed by  
$\gamma\gamma \to l^+ l^-$ re--scattering. 

Both  short--distance and
$K\to\pi\gamma^*(Z^*)$ contributions  produce the 
lepton pair in a $J^{CP}=1^{--}$ or  $1^{++}$ state,\cite{BGlashow}
so that $CP|\pi^0 f {\bar f}\rangle= +|\pi^0 f {\bar f}\rangle$.
As a consequence, in $K_L \to \pi^0 l^+ l^- (\nu {\bar \nu})$  
these two contributions violate $CP$. Since the phase  of
$\lambda_t$ is of order one and the phase
of the weak hamiltonian (\ref{ciuham}) is very small
($\sim \Imm A_0/\Real A_0$), the short--distance contribution is
essentially a  direct $CP$ violation whereas the   
long--distance contribution is dominated by indirect $CP$ violation.
Only the re--scattering $\gamma\gamma \to l^+ l^-$,
that is however very suppressed,
generates  a $CP$--invariant contribution.

$K_L \to \pi^0 l^+ l^- (\nu {\bar \nu})$ decays 
have not been observed yet and certainly have very small
branching ratios (\SSt \ref{tab:Kff}). However, if the  short--distance 
contribution was dominant then an observation of these decays 
would imply the  evidence of direct $CP$ violation\cite{LittenbergKL}.
In the following we will try to analyze  under 
which conditions this is true.

\begin{table}[t] 
\[ \ba{|l|c|} \hline
\mbox{decay}  & \mbox{branching ratio}  \\ \hline
K^\pm \to \pi^\pm e^+e^-  & (2.74 \pm 0.23 )\times 10^{-7} \\ \hline 
K^\pm \to \pi^\pm \mu^+\mu^-  & < 2.3 \times 10^{-7} \\ \hline 
K^\pm \to \pi^\pm \nu \bar{\nu}  & < 5.2 \times 10^{-9} \\ \hline  
K_L \to \pi^0 e^+e^-  & < 4.3 \times 10^{-9} \\ \hline 
K_L \to \pi^0 \mu^+\mu^-  & < 5.1 \times 10^{-9} \\ \hline 
K_L \to \pi^0 \nu \bar{\nu}  & < 2.2 \times 10^{-4} \\ \hline 
K_S \to \pi^0 e^+e^-  & < 1.1 \times 10^{-6} \\ \hline  \ea \]  
\caption{Experimental data on 
$K \to \pi f {\bar f}$ decays\protect\cite{PDG}.  }
\label{tab:Kff} 
\end{table}  
                                                                 
\subsubsection{Direct $CP$ violation in $K_L \to \pi^0 f \bar{f}$.} 
\label{sez:gim}
                           
The effective hamiltonian describing  short distance effects 
in these decays is\cite{Inami}:
\beq
{\cal H}_{eff}^{|\Delta S|=1;f \bar{f} } = {2 G_F \alpha_{em} \over \sqrt{2} }
\sum_{q=u,c,t}
\lambda_q  (\bar{s}_L\gamma_\mu d_L )  \bar{f} \gamma^\mu \left( 
V^q_{f \bar{f}} + A^q_{f \bar{f}}\gamma_5 \right) f  
 + \mbox{h.c.}  \label{ffham}
\eeq
As anticipated, in spite of the $\lambda_t$ suppression,
the dominant contribution in (\ref{ffham}) is obtained for $q=t$.
The coefficients  $V^t_{f \bar{f}}$ and 
 $A^t_{f \bar{f}}$ have been calculated including 
next--to--leading--order  QCD corrections\cite{BurasKl}, 
for $\mu \sim 1$ GeV the result is
\beq
V^t_{l \bar{l}} = 3.4 \pm  0.1 \qquad
V^t_{\nu \bar{\nu}} = -A^t_{\nu \bar{\nu}} = {1\over 2} 
A^t_{l \bar{l}} = 1.6 \pm 0.2.
\eeq             
Differently that in  (\ref{ciuham}), in  this case 
the  $\mu$--dependence and the uncertainties related to 
$\alpha_s$ are quite small: the error is dominated by the 
uncertainty on $m_t$.

The hadronic part of $O^V_{ll}$ and $O^A_{ll}$  matrix elements   
is well known because is related, by
isospin symmetry, to the matrix element  of 
$K^+ \to \pi^0 e^+ \nu_e$:\footnote{~It is 
possible to derive the same result also by means of CHPT 
(\SSp \ref{subsez:LS}), but is necessary to    
keep also $O(p^4)$ contributions to obtain the   
correct form--factor  behaviour at $q^2\not=0$.}
\beq
\bra \pi^0(p_\pi) | \bar{d} \gamma_\mu s \ket{\Ko(p_K)}  = 
{ f_+(q^2) \over \sqrt{2} } (p_K+p_\pi)_\mu +
{ f_-(q^2) \over \sqrt{2} } (p_K-p_\pi)_\mu; \label{kl3elem}
\eeq  
where
\beq
f_+(q^2) = 1+ \lambda { q^2 \over M^2_{\pi^+} }\qquad  \mbox{and } \qquad
 \lambda = (0.030 \pm 0.002).
\eeq
Using the previous equations, in the limit $m_f=0$, we 
find:\footnote{~The contribution of $f_-(q^2)$ 
is proportional to $m_f$ and thus negligible in the case of 
$e^+e^-$ and  $\nu\bar{\nu}$ pairs.} 
\beqa
A(K_2 \to \pi^0 f\bar{f} ) &=& i G_F \alpha_{em} 
\Imm \lambda_t f_+(q^2) \times \no\\
&& \times (p_K+p_\pi)_\mu
\bar{u}(k) \gamma^\mu \left[ V^t_{f \bar{f}} + A^t_{f \bar{f}}\gamma_5 
\right] v(k'), \label{elemento000}
\eeqa
which implies 
\beq
BR_{CP-dir}(K_L \to \pi^0 f \bar{f} ) = (1.16 \times 10^{-5}) \left[ 
(V^t_{f\bar{f}})^2+(A^t_{f\bar{f}})^2 \right] ( A^2 \lambda^5 \eta)^2.
\eeq
Using for $A$, $\lambda$ and $\eta$ the values of tab.~\ref{tab:CKMpar} 
and summing over the three neutrino families, we finally 
obtain
\beq \ba{rclcl}
1.3 &<& 10^{12} \times BR_{CP-dir}(K_L \to \pi^0 e^+ e^- ) &<&   5.0,  \\
0.8 &<& 10^{12} \times BR_{CP-dir}(K_L \to \pi^0 \nu \bar{\nu} ) &<&   3.3, \\
0.3 &<& 10^{12} \times BR_{CP-dir}(K_L \to \pi^0 \mu^+ \mu^- ) &<&   1.0. \\
\ea \label{cpdir}
\eeq
                                                                 
\subsubsection{Indirect $CP$ violation in $K_L \to \pi^0 f \bar{f}$.}    
\label{sez:cpind}                                 

Neglecting the interference pieces among the different terms,
the indirect--$CP$--violating contribution to the branching ratio
is given by
\beqa
BR_{CP-ind}(K_L \to \pi^0 f \bar{f} ) &=& |\eps|^2 
{\Gamma_S \over \Gamma_L } BR(K_S \to \pi^0 f \bar{f} )  \no \\
 &=& 3\times 10^{-3} \times BR(K_S \to \pi^0 f \bar{f} ). \label{cpind1}
\eeqa
Unfortunately, present limits 
on $K_S \to \pi^0 f\bar{f}$ branching ratios
(\SSt \ref{tab:Kff}) are not sufficient to establish the   
relative weight between Eq.~(\ref{cpind1})  and Eq.~(\ref{cpdir}).
Thus we need to estimate theoretically  
the  $K_S \to \pi^0 f \bar{f}$ width.

\begin{figure}[t]
    \begin{center}
       \setlength{\unitlength}{1truecm}
       \begin{picture}(8.0,4.0)
       \epsfxsize 8.  true cm
       \epsfysize 4.  true cm
        \epsffile{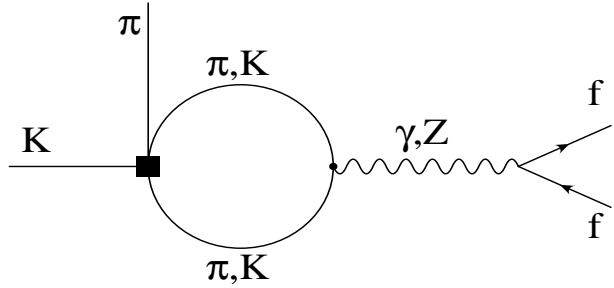}
	\end{picture}
    \end{center}
    \caption{One--loop diagram (in the diagonal basis of
Ecker, Pich and de Rafael\protect\cite{EPR2})  
relevant to $K\to\pi f\bar{f}$ transitions.}
    \protect\label{fig:longdist}
\end{figure} 

This process receives contributions both from the effective hamiltonian 
in Eq.~(\ref{ffham}) and from the one in Eq.~(\ref{ciuham}). 
The former is negligible since generates 
a width of the same order of
$\Gamma_{CP-dir}(K_L \to \pi^0 f \bar{f} )$
estimated in the previous subsection.
The long--distance contribution generated by ${\cal H}_{eff}^{|\Delta S|=1}$ 
can be evaluated in the CHPT framework. The lowest order
result  vanishes both in the $K_S\to\pi^0 \gamma^*$ case\cite{EPR1} 
and in the $K_S\to\pi^0 Z^*$ one\cite{Geng}. The first non--vanishing 
contribution arises at $O(p^4)$. 

Before going on with the calculation, we note that the one--loop diagram 
of fig.~\ref{fig:longdist} with a $Z^*$  is heavily suppressed 
($\sim (M_K/M_Z)^2$) respect to the corresponding one with 
$\gamma^*$, thus 
\beq
{ BR_{long-d}(K_S \to \pi^0 \nu \bar{\nu} ) \over
  BR_{long-d}(K_S \to \pi^0 e^+ e^- )  } \lsim
\left( M_K^2 \alpha_{em} \over M_Z^2 \alpha_{W} \right) <
10^{-7}. 
\eeq
This result, together with the experimental limit 
on $K_S\to \pi^0 e^+e^-$ (\SSt\ref{tab:Kff}),
let us state that   
{\it the dominant contribution to 
$K_L \to \pi^0 \nu \bar{\nu}$  is generated by direct $CP$ 
violation}.
 Unfortunately the observation of this process 
 is very difficult: it requires an extremely good $\pi^0$--tagging and an
hermetic detector able to eliminate the background coming 
from $K_L \to \pi^0 \pi^0 $ (with two missing photons),
that has a branching ratio eight orders of magnitude 
larger. The KTeV\cite{KTev}  expected sensitivity
for $K_L \to \pi^0 \nu \bar{\nu}$ is  $\sim 10^{-8}$.

Coming back to the $O(p^4)$ calculation, the 
result for $K_S \to \pi^0 e^+e^-$ is\cite{EPR1}: 
\beqa
&& A^{(4)}( K_S \to \pi^0 e^+e^- ) =  {G_8 \alpha_{em} \over 4\pi } 
\left[ 2 \varphi  
\left({q^2\over M_K^2}\right) +\omega_S \right]  \times  \no \\
&&\qquad\quad 
\times \left\{ (p_K+p_\pi)_\mu - (M^2_K-M^2_\pi){q^\mu \over q^2} \right\}
\bar{u}(k) \gamma^\mu v(k'),
\eeqa
where  $\varphi(z)$ is defined in the appendix and
\beq
\omega_S = {2\over 3} (4\pi)^2 \left[ 2N_{14}^r(\mu) +N_{15}^r(\mu)
\right] -{1\over 3}\log\left(M_K^2\over \mu^2\right).
\label{omegasdef}
\eeq
Also in this case the counterterm combination is not  
experimentally known.\footnote{~Differently to the cases analyzed before,
the counterterm combination which appears in $K_S \to \pi^0 e^+e^- $ 
is not finite. However, the constant  $\omega_S$ of 
Eq.~(\protect\ref{omegasdef})
is by construction 
$\mu$--independent.}

Expressing $K_S \to \pi^0 e^+e^- $ 
branching ratio as a function of
$\omega_S$, leads to\cite{Eckerrep}  
\beq
BR(K_S \to \pi^0 e^+e^- ) = \left[ 3.07 - 18.7\omega_S +28.4\omega_S^2 
\right] \times 10^{-10}.\label{brksff}
\eeq 
Within the  factorization model (\SSp\ref{sez:VMD}) is
possible to relate  $\omega_S$  to the 
counterterm combination that appears in   
$K^+ \to \pi^+ e^+e^-$ (\SSp \ref{sez:kppff}),
the only $K\to \pi f\bar{f}$ channel till now 
observed\cite{EckerWyler,Eckerrep}:  
\beq
\omega_S = (0.5 \pm 0.2) + 4.6(2 k_f -1). \label{omega_skf} 
\eeq
In the factorization model earlier proposed in Ref.\cite{EPR1},
the choice $k_f=1/2$ was adopted.
This value of $k_f$ essentially minimize Eq.~(\ref{brksff})
and for this reason in the literature has been sometimes claimed
that indirect $CP$ violation is negligible in
 $K_L \to \pi^0 e^+e^-$ (see e.g. Ref.\cite{Pich}). Though supported 
by a calculation done in a different framework\cite{Portoles},  
 this statement is very model dependent 
(as can be easily deduced from Eq.~(\ref{omega_skf})). Indeed the choice
$k_f>0.7$, perfectly consistent from the theoretical point of view, 
implies $BR(K_S\to \pi^0 e^+e^-)> 10^{-8}$, i.e. 
$BR_{CP-ind}(K_L \to \pi^0 e^+e^-)>3\times 10^{-11}$. 
In our opinion, the only model independent statement that
can be done now is:
\beq \ba{rcccl}
  5 \times 10^{-10} & < & BR(K_S \to \pi^0 e^+ e^- ) &<&
  5 \times 10^{-8}, \\
  1.5 \times 10^{-12} & <& BR_{CP-ind} (K_L \to \pi^0 e^+ e^- ) &<&
  1.5 \times 10^{-10}, \\
\ea \eeq
thus {\it today is not possible to establish if   
 $K_L \to \pi^0 e^+ e^-$ is  dominated by direct or
indirect $CP$ violation.} The only possibility to solve the 
question is a direct measurement of 
$\Gamma(K_S \to \pi^0 e^+ e^-)$,  or an upper limit on it at the 
level of $10^{-9}$, within the 
reach of KLOE\cite{Alosio,handbook}.
We stress that this question is of great relevance 
since  the sensitivity on  $K_L \to \pi^0 e^+ e^-$
which should be reached at  KTeV  is 
$\sim 10^{-11}$ (for a discussion about backgrounds and related cuts in 
 $K_L \to \pi^0 e^+ e^-$ see Ref.\cite{Greenlee}). 

\subsubsection{$CP$--invariant contribution of 
$K_L \to \pi^0 \gamma\gamma$ to $K_L \to \pi^0 e^+ e^-$.}

As anticipated, the decay $K_L \to \pi^0 e^+ e^-$ receives also 
a $CP$--invariant contribution form the 
two--photon  re--scattering in $K_L \to \pi^0 \gamma\gamma$ 
(contribution that has been widely discussed in the literature,
see e.g. Refs.\cite{SehgalKL}$^-$\cite{Gabbiani,Cappiello,Cohen}).
                    
As we have seen in sect.~\ref{subsez:klpgg},  the 
$K_L \to \pi^0 \gamma\gamma$ amplitude can be 
decomposed in to two parts:
 $A$ and $B$ (\S~Eq.~(\ref{deco1klpgg})).   
When the two photons interact creating an 
$e^+e^-$ pair the contribution of the  $A$ amplitude is negligible
being proportional to $m_e$.

In the parametrization of Ecker, Pich and de Rafael\cite{EPR0}
(\S~Eq.~(\ref{klppge}))
the absorptive contribution generated by on--shell photons 
coming from the  $B$ term is given by\cite{Cohen,Gabbiani,DP8}:
\beqa
BR_{CP-cons}(K_L \to \pi^0e^+e^-)\big|_{abs}
 &=& 0.3 \times 10^{-12}, \qquad\quad  a_V=0,\label{av1} \\
BR_{CP-cons}(K_L \to \pi^0e^+e^-)\big|_{abs}
 &=& 1.8 \times 10^{-12}, \qquad\quad  a_V=-0.9.
\label{av2}\eeqa
Using these results we can say that 
the $CP$--invariant contribution should be smaller than 
the direct--$CP$--violating one.
At any rate, even in this case a more precise determination of 
 $\Gamma(K_L \to \pi^0 \gamma\gamma)$ at small $z$
(definitely within the reach of KLOE)  could help to 
evaluate better the situation.  

Another important question is the dispersive contribution
generated by off--shell photons, which is more complicated since 
the dispersive integral is in general not convergent.
A first estimate of this integral was done in Ref.\cite{Gabbiani},
introducing opportune form factors which suppress 
the virtual--photon couplings at high $q^2$. The 
dispersive contribution estimated in this way is of the 
same order of the absorptive  one, but is quite model dependent. A
more refined analysis is in progress\cite{donop}.

Finally, we remark that the different $CP$-conserving and
$CP$-violating contributions to $K_L \to \pi^0e^+e^-$ could be partially
disentangled if the asymmetry in the 
electron--positron energy distribution\cite{Gabbiani} and the 
time--dependent  interference\cite{Littenberg2,Paschos}
of  $K_{L,S} \to \pi^0e^+e^-$ would be measured in addition to the 
total width. 

\subsubsection{$K^\pm \to \pi^\pm l^+ l^-$.}
\label{sez:kppff}
      
Another $CP$--violating observable that can be studied in  
$K\to \pi f \bar{f}$ decays is the 
charge asymmetry of $K^\pm \to \pi^\pm e^+ e^-$ widths.
As we have seen in sect.~\ref{cap:kksystem}, 
in order to have a non--vanishing charge asymmetry is necessary 
to consider  processes with non--vanishing  re--scattering 
phases. This happens in 
$K^\pm \to \pi^\pm e^+ e^-$ decays due to the absorptive 
contribution of the loop diagram in fig. \ref{fig:longdist}.

Analogously to the $K^0\to \pi^0 e^+e^-$ case,
the $O(p^4)$ amplitude of 
$K^+\to \pi^+ e^+e^-$ is given by\cite{EPR1}: 
\beqa
&& A^{(4)}( K^+ \to \pi^+ e^+e^- ) =  {G_8 \alpha_{em} \over 4\pi } 
\left[ - \varphi  
\left({q^2\over M_K^2}\right)
- \varphi \left({q^2\over M_\pi^2}\right)
 - \omega_+ \right]  \times  \no \\
&&\qquad\quad 
\times \left\{ (p_K+p_\pi)_\mu - (M^2_K-M^2_\pi){q^\mu \over q^2} \right\}
\bar{u}(k) \gamma^\mu v(k'),  \label{ampCHPTkpff}
\eeqa
where 
\beq
\omega_+ = {4\over 3} (4\pi)^2 \left[ N_{14}^r(\mu) -N_{15}^r(\mu)
+3L_9^r(\mu)
\right] -{1\over 3}\log\left(M_K^2\over \mu^2\right).
\eeq
Thus the charge asymmetry is given by:
\beqa
&& \Gamma(K^+ \to \pi^+ e^+e^- ) - \Gamma(K^- \to \pi^- e^+e^- )
= \Imm w_+ {|G_8|^2 \alpha^2_{em} 
M_K^5 \over 192 \pi^5  } \times \no \\ 
&& \qquad 
\times \int_{4r_e^2}^{(1-r_\pi^2} \mbox{d}z \lambda^{3/2}(1,z,r^2_\pi)
\sqrt{ 1-4{r_e^2  \over z} }\left( 1+2{r_e^2  \over z}  \right)
\Imm \varphi\left({z \over r_\pi^2} \right), \label{asimkff}
\eeqa
where $z=q^2/ M_K^2$, $r_i=m_i/ M_K$ and
$\lambda(x,y,z)$ is defined in Eq.~(\ref{lambdaxyz}).
Integrating Eq.~(\ref{asimkff}) and using the experimental value of
$\Gamma (K^+ \to \pi^+ e^+e^-)$ we finally get
\beq
{ \Gamma(K^+ \to \pi^+ e^+e^- ) - \Gamma(K^- \to \pi^- e^+e^- ) \over
 \Gamma(K^+ \to \pi^+ e^+e^- ) + \Gamma(K^- \to \pi^- e^+e^- ) }
\simeq 10^{-2}  \Imm w_+.
\eeq

The real part of  $w_+$ is fixed by the experimental
information on the width and the spectrum of the decay\cite{Alliegro}:  
\beq
\Real w_+ = 0.89^{+0.24}_{-0.14}.
\eeq 
On the other hand, we expect that $\Imm w_+$
is theoretically determined by short distance contributions.
Comparing Eq.~(\ref{ampCHPTkpff}) with the amplitude obtained
by ${\cal H}_{eff}^{|\Delta S|=1;f \bar{f} }$, we find:
\beq
|\Imm w_+| \simeq 4\pi A^2 \lambda^5 \eta V^t_{f \bar{f}},  
\eeq
which implies 
\beq
|\delta \Gamma |= 
{ | \Gamma(K^+ \to \pi^+ e^+e^- ) - \Gamma(K^- \to \pi^- e^+e^- )| \over
 \Gamma(K^+ \to \pi^+ e^+e^- ) + \Gamma(K^- \to \pi^- e^+e^- ) }
\simeq 2 \times 10^{-4} \times A^2 \eta. \label{cpasimmkff} 
\eeq
Unfortunately, due to  the small branching ratio of the process,
the statistics necessary to test this interesting prediction 
is beyond  near--future experimental programs.

Apart from the asymmetry of $K^\pm$ widths, 
in $K^+ \to \pi^+ \mu^+ \mu^-$ (and in $K^- \to \pi^- \mu^+ \mu^-$) 
it is possible to measure also
asymmetries which involve muon polarizations. These can be useful both 
to study $T$ violation\cite{Agrawal} and to 
provide valuable information about CKM matrix elements\cite{Savage}. 
However, these measurements are not easy from the experimental point
of view and
thus we will not discuss them 
(for accurate analyses see 
Refs.\cite{Agrawal}$^-$\cite{Buraspolar}).
 
\subsubsection{$K^+ \to \pi^+ \nu \bar{\nu}$.}
\label{sez:kppnn}

As in  the $K_L \to \pi^0 \nu \bar{\nu}$ case, also this decay is by far 
dominated by short distance. This process is not directly 
interesting for $CP$ violation but
is one of the best channels to put constraints on
CKM parameters $\rho$ and $\eta$. 

Using the short distance hamiltonian (\ref{ffham}) we find, analogously to 
Eq. (\ref{elemento000}),
\beq
A(K^+ \to \pi^+ \nu\bar{\nu} ) =  G_F \alpha_{em} 
f_+(q^2) \sum_q  \lambda_q    V^q_{\nu \bar{\nu}} 
 (p_K+p_\pi)_\mu \bar{u}(k) \gamma^\mu \left[ 1  - \gamma_5 
\right] v(k'). \label{elemento001}
\eeq
As in the previous cases the top contribution is dominant, 
however, since  $K^+ \to \pi^+ \nu\bar{\nu}$ 
is a $CP$ conserving amplitude, in this case
the charm effect is not completely negligible. 
Following Buras et al.\cite{BurasVtd}     
we can parametrize the
branching ratio in the following way:
\beq
BR(K^+ \to \pi^+ \nu \bar{\nu} ) = 2 \times 10^{-11} A^4 \left[ 
\eta^2 +{2\over 3}(\rho-\rho^e)^2+
{1\over 3}(\rho-\rho^\tau)^2 \right]\left({m_t\over M_W}\right)^{2.3},
\label{Vtddeterm}
\eeq
where $\rho^e$ and $\rho^\tau$ differ from unity because of the presence
of the charm contribution; using 
$\overline{m}_c(m_c)=1.30 \pm 0.05$ GeV from 
Ref.\cite{BurasVtd} we find
\beq
1.42 \leq \rho^e \leq 1.55 \qquad \qquad 
1.27 \leq \rho^\tau \leq 1.38.
\eeq

\begin{figure}[t]
    \begin{center}
       \setlength{\unitlength}{1truecm}
       \begin{picture}(10.0,6.0)
       \epsfxsize 10.0  true cm
       \epsfysize 6.0 true cm
       \epsffile{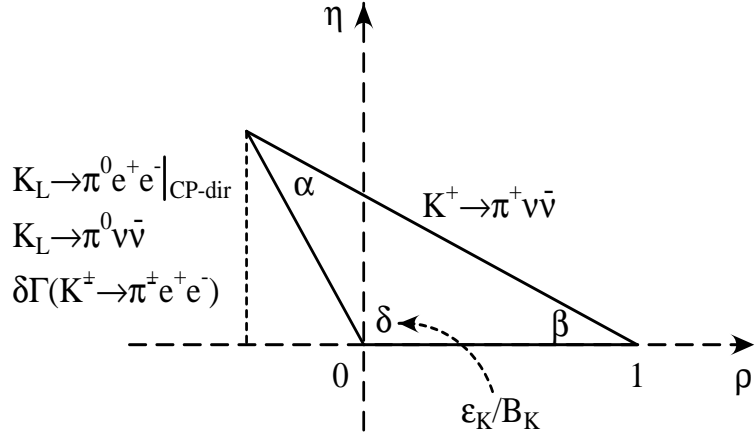}
       \end{picture}
    \end{center}
\caption{Kaon decays and the unitarity triangle.} 
\label{fig:triangle2}   
\end{figure}

Present limits on this decay (\SSt\ref{tab:Kff}) are still
more than one order of magnitude far from the Standard Model value,
which however should be reached in the near future.\cite{Shinkawa}

As shown in fig.~\ref{fig:triangle2}, $K \to \pi f\bar{f}$ measurements
would allow in principle a complete determination of sizes and angles of
the unitarity triangle introduced in sect.~\ref{sez:Bdecays}. Even if 
these measurements are not easy to perform, it is important to stress that 
could compete for completeness and cleanliness with those in the $B$ sector. 
 
\subsection{$K \to l^+l^-$}

The decay amplitude  $A(\Ko \to l^+l^-)$ 
can be generally written as
\beq
A(\Ko \to l^+l^-) = \bar{u}(k) (i B +A \gamma_5) v(k'),
\eeq
then 
\beq
\Gamma (\Ko \to l^+l^-)  = {M_K \beta_{(l)} \over 8 \pi} 
\left( |A|^2 +\beta_{(l)}^2 |B|^2 \right), \qquad \quad 
\beta_{(l)}=\left(1-{4 m_l^2 \over M_K^2}\right)^{1/2}.
\eeq
Up to now, only the $K_L \to \mu^+\mu^-$
decay has been observed (\SSt \ref{tab:Kll}).

\begin{table}[t] 
\[ \ba{|l|c|} \hline
\mbox{decay}  & \mbox{branching ratio}  \\ \hline
K_L \to  \mu^+\mu^- &  (7.4 \pm 0.4) \times 10^{-9} \\ \hline 
K_L \to  e^+e^-     & < 4.1 \times 10^{-11} \\ \hline 
K_S \to  \mu^+\mu^- & < 3.2 \times 10^{-7} \\ \hline 
K_S \to  e^+e^-     & < 1.0 \times 10^{-5} \\ \hline  \ea \]  
\caption{Experimental data\protect\cite{PDG} on
$K_{L,S} \to  l^+l^-$.  }
\label{tab:Kll} 
\end{table}  

Analogously to previous decays, also in 
in this case it is convenient to decompose 
amplitude   contributions 
in short-- and long--distance ones. In both channels
($K_L$ and $K_S$) the dominant contribution is the long--distance one,
generated by the two--photon re--scattering in
$K_L (K_S) \to \gamma\gamma$ transitions. 
Since $A(\gamma\gamma \to l^+l^-)$ 
is proportional to $m_l$, this explains why 
only the  $K_L \to \mu^+\mu^-$ has been observed till now.
 
$A$ and $B$ amplitudes have different transformation
properties under $CP$: if $CP$ is conserved
$K_2\to l^+l^-$ receives a contribution only 
from $A$ whereas $K_1\to l^+l^-$ only form $B$.  Thus
$CP$ violation in $K \to l^+l^-$ decays can be observed 
trough the asymmetry
\beq
\langle P^{(l)} \rangle
 = { N_+ (l^-) - N_-(l^-) \over N_+ (l^-) + N_-(l^-) } 
\propto \Imm (A^*B),
\eeq
where $N_\pm (l^-)$ indicates the number of $l^-$ emitted with 
positive or negative helicity. In the 
$K_L$ case we obtain:
\beq
\left| \langle P^{(\mu)}_L \rangle  \right| \simeq \beta_{(\mu)} \left| 
\Imm \left( {B_2 +\veps B_1 \over A_2} \right) \right|,
\label{klmmasimm}
\eeq
where the subscripts of $A$ and $B$ indicate if the amplitudes 
belong to $K_1$ or $K_2$.

The amplitude $B_2$, responsible of direct $CP$ violation, is 
generated in the Standard Model 
by the short--distance contribution of the 
effective operator\cite{Botella} $\bar{s}d H$, where
$H$ is the physical Higgs field. Its effect
is completely negligible with respect to the
indirect--$CP$--violating one.
 
The $CP$--invariant  amplitude $A_2$ has a
large imaginary part ($|\Real A_2| \ll |\Imm A_2|$),  
because the absorptive contribution to $K_L \to \gamma
\gamma \to \mu^+\mu^-$ essentially saturates the 
experimental value of $\Gamma (K_L \to \mu^+\mu^-)$:
\beq
BR(K_L \to \mu^+\mu^-)\big|_{abs} = (6.85 \pm 0.32)\times 10^{-9}.
\eeq
As a consequence, if we neglect both direct $CP$ violation and 
$\Real A_2/\Imm A_2$, and we assume $\eps=|\eps|e^{i\pi/4}$, 
then Eq.~(\ref{klmmasimm}) becomes:
\beq
\left| \langle P^{(\mu)}_L \rangle 
 \right| \simeq { \beta_{(\mu)} |\eps| \sqrt{2}
\over | \Imm A_2 | } \left| \Imm B_1 -\Real B_1 \right|.
\label{klmmasimm2}
\eeq

The amplitude $B_1$ can be calculated unambiguously in 
CHPT at the lowest non--vanishing order ($K_S \to \gamma\gamma\ O(p^4)$ +
$\gamma\gamma \to l^+l^-\ O(e^2)$) since, as in
 $K_S \to \gamma\gamma$, the loop calculation is 
finite\cite{EckerPich}. The results
thus obtained are
\beqa
\Imm B_1 &=& +0.54 \times 10^{-12},  \\
\Real B_1 &=& -1.25 \times 10^{-12},  
\eeqa
and imply\cite{EckerPich}
\beq
\left| \langle P^{(\mu)}_L \rangle  \right| \simeq 2 \times 10^{-3}.
\eeq

The measurement of $ \langle P^{(\mu)}_L \rangle $ is not useful 
for the study of direct $CP$ 
violation within the Standard Model. However,
the above result tell us that 
a measurement of $  \langle P^{(\mu)}_L \rangle $ at the level of  $10^{-3}$ 
could be very useful  to exclude new 
$CP$--violating mechanisms with additional scalar fields\cite{Pich}.

Another interesting question in this channel is the short 
distance\cite{BurasKLmm}
contribution to $\Real A_2$, which depends on the CKM matrix element
$V_{td}$. Thus $K_L \to \mu^+\mu^-$ could in principle add new 
information about the unitarity triangle of fig.~\ref{fig:triangle2}.
However, $\Real A_2$ receives also a long--distance (model dependent)
contribution from the dispersive integral $K_L \to \gamma^\ast
\gamma^\ast \to \mu^+\mu^-$. The smallness of $\Real A_2$ implies a 
 cancellation among these two terms. The extent of this cancellation
and the accuracy with which one can evaluate the dispersive integral,
determine the sensitivity to $V_{td}$.
$K_L \to \gamma  l^+l^- $ 
and $K_L \to  e^+e^- \mu^+\mu^-$ 
decays  could bring some information on the relevant 
form factor (see Refs.\cite{BMS,Ecker78,DP8} 
and references therein), however the result is still model dependent.

\subsection{$K \to \pi l \nu.$}
\label{sez:Kmu3}

Analogously to the previous case, also the 
transverse muon polarization  
in $K\to \pi \mu\nu$ decays ($\langle P_\perp^{(\mu)}\rangle$)
is very sensitive to  new $CP$--violating mechanisms with additional
scalar fields (for an update
discussion see Ref.\cite{Peccei}). 
$\langle P_\perp^{(\mu)}\rangle$  is a measurement of the 
muon polarization perpendicular to the decay plane 
and, by construction,  is related  to the correlation   
\beq 
 \langle \svect{s}_{(\mu)} \cdot ( \svect{p}_{(\mu)}\times \svect{p}_\pi)
\rangle
\eeq
which violates $T$ in absence of final--state interactions. 
In the $K_L \to \pi^- \mu^+ \nu $ case, with two charged 
particles in the final state, electromagnetic interactions 
can generate\cite{kmu31} 
$\langle P_\perp^{(\mu)}\rangle_{FSI} \sim \alpha/\pi \sim 10^{-3}$.
However, in $K^+ \to \pi^0 \mu^+ \nu $ this effect is much 
smaller\cite{kmu32} ($\langle P_\perp^{(\mu)}\rangle_{FSI} \sim 10^{-6}$) and 
$T$--violation could be dominant.

In the framework of the Standard Model and in any model where
the  $K^+ \to \pi^0 \mu^+ \nu $ decay is mediated by vector meson
exchanges, $\langle P_\perp^{(\mu)}\rangle$ is zero at  tree-level and is 
expected to be very small\cite{kmu33}. On the other hand, interference between
$W$ bosons and $CP$--violating  scalars can produce a large effect. 
Writing the effective amplitude as\cite{PDG} 
\beq
A(K^+ \to \pi^0 \mu^+ \nu  ) \propto f_+(q^2)  \left[  (p_K + p_\pi)_\mu 
\bar{\mu} \gamma^\mu  (1-\gamma_5 ) \nu + \xi(q^2)  m_{(\mu)}
\bar{\mu}  (1-\gamma_5 ) \nu \right], \label{elemento002}
\eeq
neglecting the $q^2$ dependence of $\xi(q^2)$ and 
averaging over the phase space, leads to\cite{CabibboM}:
\beq
\langle P_\perp^{(\mu)}\rangle \simeq 0.2 (\Imm \xi).
\eeq 

The present experimental determination of $\Imm \xi$ is\cite{PDG}:
\beq
\Imm \xi  = -0.0017 \pm 0.025,
\eeq 
but an on--going experiment at KEK (E246)
should improve soon this limit by an order of magnitude\cite{E246}. 
From the theoretical point of view, 
it is interesting to remark that present
limits on multi--Higgs models, coming from 
neutron electric dipole moment and 
$B\to X \tau \nu_\tau$, do not exclude a
value of $\Imm \xi$  larger 
than the sensitivity 
achievable at KEK\cite{Peccei}. A large 
$\Imm \xi$ is expected in some SUSY models\cite{Ng96}
and an eventual evidence 
at the level of $10^{-3}$ would imply interesting 
consequences for the next--generation of experiments in 
the $B$ sector.

\subsection{$K \to  \pi\pi l^+l^-.$}
\label{sez:kppll}

The last channels  we are going to discuss are 
$K\to \pi\pi l^+l^-$ transitions. 
The dynamics of these processes is essentially the same of
$K \to \pi\pi\gamma$ transitions (\SSp~\ref{cap:Kppg}),
with the difference that the photon is virtual.
For this reason we shall discuss these decays only briefly.

With respect to $K \to \pi\pi\gamma$ transitions, 
$K \to \pi\pi l^+l^-$ decays have the disadvantage
that the branching ratio is sensibly smaller
(obviously the $e^+e^-$ pair is favoured with respect to the 
$\mu^+\mu^-$ one). Nevertheless, there is also an advantage:
the lepton plane furnishes a measurement of the photon polarization
vector. This is particularly useful in the case of
$K_L \to  \pi^+ \pi^-  e^+e^-$,\footnote{~This decay 
has not been observed yet, the theoretical branching 
ratio\protect\cite{Eckerrep} is
 $BR(K_L \to  \pi^+ \pi^-  e^+e^-)=
2.8 \times 10^{-7}$.} 
because let us to measure the  $CP$--violating interference 
between electric and magnetic amplitudes (\SSp \ref{sez:KppgCP})
also in experimental apparata where  
photon polarizations are not directly accessible. 
The observable proportional to this interference
is the $\phi_{\pi/e}$--distribution, where $\phi_{\pi/e}$ 
is the angle between $e^+e^-$ and 
$\pi^+ \pi^-$ planes. 

The asymmetry in the $\phi_{\pi/e}$--distribution has been recently 
estimated in Refs.\cite{Sehgal,Wise} and turns out to be quite large
($\sim 10\%$), within the reach of KLOE. However, 
since the electric amplitude of
$K_L \to \pi^+\pi^-\gamma$ is dominated by the  bremsstrahlung of 
$K_L \to \pi^+\pi^-$, this effect is essentially 
an indirect $CP$ violation. Elwood et al.\cite{Wise} 
 have shown 
how to construct an asymmetry which is essentially 
an index of direct $CP$ violation. In this case, however, 
the prediction is of the order of $10^{-4}$, 
far from experimental sensitivities.
 

\setcounter{equation}{0}
\setcounter{footnote}{0}
\section{Conclusions.}

In table~\ref{tab:finale} we report the Standard Model
predictions discussed in this review 
for the direct--$CP$--violating observables of  $K\to 2\pi$,  
$K\to 3 \pi$, 
$K\to 2\pi\gamma$, $K\to \pi f{\bar f}$, $K\to \gamma\gamma$
and $K\to \pi \gamma\gamma$ decays.

In many cases, due to the uncertainties  of next--to--leading order
CHPT corrections, it has not been 
possible to make definite predictions but only 
to put some upper limits. However, this analysis 
is still very useful 
since an experimental evidence beyond these limits would 
imply the existence of  new $CP$ violating mechanisms.

\begin{table}
\[ \ba{|c|rcl|lr|} \hline  
\mbox{channel} & \multicolumn{1}{|r}{ \mbox{observable}} &
\multicolumn{1}{c}{ - } & \multicolumn{1}{l|}{ \mbox{prediction}} 
& \multicolumn{2}{|c|}{ \sigma } \\ \hline & & & & & \\   
 (2\pi)^0 & \left\vert {\epsp\over\eps}\right\vert  &< &  10^{-3} 
& \gsim 10^{-4}&  \\ 
& & & & & \\ \hline & & & & & \\
 (3\pi)^\pm   & |\delta g| &<&  10^{-5} 
& \gsim 10^{-4}&  \\  
 (3\pi)^0  &  |\epsilon^X_{+-0}|,|\epsp_{+-0}| &<& 5 \times 10^{-5} 
& \gsim 10^{-4}&  \\ 
& & & & & \\ \hline  & & & & & \\
 \pi^+\pi^-\gamma  &  |\eta_{+-\gamma}-\eta_{+-}| &<&  5\times 10^{-6} 
& \gsim 10^{-5}&  \\  
 \pi^\pm\pi^0\gamma  &  |\delta \Gamma_{DE} | &<& 10^{-4} 
& \gsim 10^{-3}&  \\ & & & & & \\ \hline & & & & & \\ 
 \gamma\gamma  & |\epsp_\perp|,|\epsp_\parallel|  &<&
10^{-4} & \gsim 10^{-3}&   \\  
 \pi^\pm \gamma\gamma  & |\delta \Gamma|  &<&
10^{-4} & \gsim 10^{-2} &   \\ & & & & & \\ \hline & & & & & \\ 
 \pi^0\nu\bar{\nu}  & BR(K_L \to \pi^0\nu\bar{\nu})  &=&
(0.8 \div 3.3 )\times 10^{-12} & \gsim 10^{-8}& \\  
 \pi^0e^+e^-  & BR(K_L \to \pi^0e^+e^-)  &=&
(1.3 \div 5.0 )\times 10^{-12}\  ^{(*)} &  10^{-11}&   \\
 \pi^\pm e^+e^-  & |\delta \Gamma|  &=&
(0.4 \pm 0.2 )\times 10^{-4} & \gsim 10^{-2}&  \\  
& & & & & \\ \hline
\ea \]
\caption{Standard Model predictions for
direct--$CP$--violating observables of $K$ decays. 
In the third column we report a rough estimate of the 
expected sensitivities, achievable by combining
KTeV\protect\cite{KTev}, NA48\protect\cite{NA48} and 
KLOE\protect\cite{Alosio} future results.
The `$*$' in $K_L \to \pi^0e^+e^-$ concerns the 
question of the indirect--$CP$--violating contribution
 (\SSp\protect\ref{sez:cpind}). }
\label{tab:finale} 
\end{table}  

The essential points of our analysis can be summarized as follows:
\begin{itemize}
\item{In 
$K\to 3 \pi$, $K\to 2\pi\gamma$, $K\to \gamma\gamma$ 
and $K\to \pi\gamma\gamma$, the presence of
several $\Delta I=1/2$ amplitudes generally let to
overcome the $\omega = \Real A_2/\Real 
A_0$ suppression which depresses direct $CP$ violation in $K\to 2 \pi$.
Thus in the above decays  direct--$CP$--violating observables are 
usually larger than $\epsp$ of about one order of magnitude. 
Nevertheless, due to the small branching ratios,
the experimental  sensitivities
achievable in these decays are well below those of  $K\to 2\pi$.

With respect to some controversial 
questions in the literature, we stress  that 
charge asymmetries in
$K^\pm \to (3\pi)^\pm$, as well as 
those in $K^\pm \to \pi^\pm\pi^0\gamma$, 
cannot exceed $10^{-5}$.  }

\item{In $K_L\to \pi^0 f \bar{f}$ decays, the absence or the 
large suppression of  $CP$--invariant contributions
implies that $CP$ violation plays a fundamental role:
the question is whether the direct $CP$ violation dominates over
the other contributions.

In $K_L\to \pi^0 \nu \bar{\nu}$ the direct $CP$ violation is
certainly dominant, but the observation of this decay is 
extremely difficult.

In $K_L\to \pi^0 e^+ e^-$, due to the uncertainties of both 
$O(p^4)$ effects in $K_S\to \pi^0 e^+ e^-$ and 
the dispersive contribution from $K_L\to \pi^0\gamma\gamma$, 
is impossible to establish without model dependent assumptions 
which is the dominant amplitude. Only a measurement of 
$BR(K_S\to \pi^0 e^+ e^-)$ (or an upper limit on it
at the level of $10^{-9}$) together with a more precise 
determination of the dispersive contribution form
$K_L\to \pi^0\gamma\gamma$   
  could solve the question.}
\end{itemize}

\vglue .5 true cm                         
\noindent
To conclude this analysis, we can say that 
in the near future there is a realistic hope 
to observe direct $CP$ violation only in 
$K\to 2\pi$ and $K_L\to \pi^0 e^+e^-$, but even in
these decays a positive result is not guaranteed.
Nevertheless, 
a new significant insight in the study of this  interesting phenomenon
will certainly start in few years with the next--generation of 
experiments on $B$ decays\cite{NakadaP} and, possibly, with new
 rare kaon decay experiments\cite{LittenbergP}.
                                          
\vglue .5 true cm                         
\subsection*{Acknowledgments.}
We are grateful to G. Ecker and L. Maiani for enlightening
discussions and valuable comments on the manuscript.
We thank also M. Ciuchini, E. Franco, G. Martinelli, H. Neufeld, 
N. Paver, J. Portoles and A. Pugliese for interesting discussions 
and/or collaborations on some of the subjects presented here.          
G.D. would like to thank the hospitality of the Particle Theory Group
at MIT where part of this work was done.  

\appendix
\section{Loop functions.}
\label{appendb}
\newcounter{zahler}
\renewcommand{\theequation}{\Alph{zahler}.\arabic{equation}}
\setcounter{zahler}{1}
\setcounter{equation}{0}

The function $\wt{C_{20}}(x,y)$, 
which appear at one loop in $K\to\pi\pi\gamma$ direct--emission
amplitudes, is defined as\cite{DEIN}
\beq
\wt{C_{20}}(x,y) = \frac{C_{20}(x,y) - C_{20}(x,0)}{y}, \label{Cdiff}
\eeq
in terms of the three--propagator one--loop function
$C_{20}(p^2, kp)$ for $k^2 = 0$:     
\beqa
&&\int {\mbox{d}^d l \over (2\pi)^d}\  {l^\mu l^\nu  \over [l^2 - M_\pi^2] 
[(l+k)^2 - M_\pi^2] [(l+p)^2 - M_\pi^2] }   \no \\
&&\qquad\quad 
= i g^{\mu\nu} C_{20}( p^2, kp )\ +O(p^\mu,k^\mu).
\eeqa
The explicit expression for $x,\ x-2y > 4 M_\pi^2$ is
\beqa
(4\pi)^2 \Real \wt{C_{20}}(x,y) &=& {x \over 8 y^2}\left\lbrace 
\left(1-2{y\over x}\right)\left[\beta\log\left({1+\beta
\over 1-\beta}\right) -\beta_0\log\left({1+\beta_0
\over 1-\beta_0}\right)\right] \right. \no\\ 
&& \left. \qquad + 
{M_\pi^2\over x} \left[\log^2\left({1+\beta_0 \over 1-\beta_0}
\right)-\log^2\left({1+\beta \over 1-\beta }\right) \right]
+ 2{y\over x} \right\rbrace, \\
(16\pi) \Imm \wt{C_{20}}(x,y) &=& - {x \over 8 y^2}\left\lbrace 
\left(1-2{y\over x}\right)\left[\beta-\beta_0\right] \right. \no\\ 
&& \left. \qquad + 
{2 M_\pi^2 \over x} \left[\log\left({1+\beta_0 \over 1-\beta_0}
\right)-\log\left({1+\beta \over 1-\beta }\right) \right]
+ 2{y\over x} \right\rbrace,
\eeqa
where 
\beq
\beta_0=\sqrt{1-{4M_\pi^2\over x }}\qquad\mbox{and }\qquad
\beta=\sqrt{1-{4M_\pi^2\over ( x-2y )}}.
\eeq

The other one--loop functions introduced in the text are\cite{EPR2}: 
\beq
 F(z) = \Biggl\{
   \ba{ll}
      1 - \dis\frac{4}{z} \arcsin^{2}{(\sqrt{z}/2)}
           \qquad  & z\le 4 \\
      1 + \dis\frac{1}{z}\left(\log\dis\frac{1-\sqrt{1-4/z}}{1+ \sqrt{1-4/z}}+
       i\pi \right)^2  & z\ge 4
   \Biggr. \ea
\eeq
and
\beq
\varphi(z) = {5\over 18} - {4\over 3 z} - {1\over 3}\left(1 - {4\over z}
\right) G(z),
\eeq
where
\beq
 G(z) = \Biggl\{
   \ba{ll}
      \sqrt{4/z -1} \arcsin{(\sqrt{z}/2)} \qquad & z\le 4 \\
      -{1\over 2}\sqrt{1-4/z}
      \left( \log \dis\frac{1-\sqrt{1-4/z}}{1+\sqrt{1-4/z}} + i\pi \right)
       & z\ge 4
   \Biggr. \ea
\eeq


\end{document}